\documentclass[12pt,letterpaper,a4paper]{article}
\usepackage[includeheadfoot,
            marginratio={1:1,2:3},
            width=480pt,
            height=786pt,]{geometry}
\usepackage{amsmath}
\usepackage{amsfonts}
\usepackage{amssymb}
\usepackage{graphicx}
\usepackage{cancel}

\usepackage{empheq}
\usepackage{color}
\usepackage{hyperref}

\usepackage{float}
\restylefloat{table}
\restylefloat{figure}

\newcommand{\nc}{\newcommand}
\nc{\beq}{\begin{equation}}
\nc{\eeq}{\end{equation}}
\nc{\bea}{\begin{eqnarray}}
\nc{\eea}{\end{eqnarray}}

\def\ov{\overline}

\allowdisplaybreaks[2]
\numberwithin{equation}{section}

\begin{document}

\vspace{0.0cm}
\begin{center}
{\LARGE
Revisiting the two formulations of Bianchi identities and \vskip0.2cm  their implications on moduli stabilization}
\vspace{0.4cm}
\end{center}

\vspace{0.35cm}
\begin{center}
Pramod Shukla \footnote{Email: shukla.pramod@ictp.it}
\end{center}

\vspace{0.1cm}
\begin{center}
{ICTP, Strada Costiera 11, Trieste 34151, Italy}
\vskip0.4cm
\end{center}

\vspace{1cm}

\begin{abstract}
In the context of non-geometric type II orientifold compactifications, there have been two formulations for representing the various NS-NS Bianchi-identities. In the first formulation, the standard three-form flux ($H_3$), the geometric flux ($\omega$) and the non-geometric fluxes ($Q$ and $R$) are expressed by using the real six-dimensional indices (e.g. $H_{ijk}, \omega_{ij}{}^k, Q_i{}^{jk}$ and $R^{ijk}$), and this formulation has been heavily utilized for simplifying the scalar potentials in toroidal-orientifolds. On the other hand, relevant for the studies beyond toroidal backgrounds,  a second formulation is utilized in which all flux components are written in terms of various involutively even/odd $(2,1)$- and $(1,1)$-cohomologies of the complex threefold. In the lights of recent model building interests and some observations made in \cite{Ihl:2007ah,Robbins:2007yv}, in this article, we revisit two most commonly studied toroidal examples in detail to illustrate that the present forms of these two formulations are not completely equivalent. To demonstrate the same, we translate all the identities of the first formulation into cohomology ingredients, and after a tedious reshuffling of the subsequent constraints, interestingly we find that all the identities of the second formulation are embedded into the first formulation which has some additional constraints.  In addition, we look for the possible solutions of these Bianchi identities in a detailed analysis, and we find that some solutions can reduce the size of scalar potential very significantly, and in some cases are too strong to break the no-scale structure completely. Finally, we also comment on the influence of imposing some of the solutions of Bianchi identities in studying moduli stabilization. 

\end{abstract}

\clearpage

\tableofcontents



\section{Introduction}
\label{sec_intro}
In recent years, non-geometric flux compactification has received a great amount of attention towards model building applications in the superstring compactification frameworks \cite{Derendinger:2004jn,Grana:2012rr,Dibitetto:2012rk, Danielsson:2012by, Blaback:2013ht, Damian:2013dq, Damian:2013dwa, Hassler:2014mla, Ihl:2007ah, deCarlos:2009qm, Danielsson:2009ff, Blaback:2015zra, Dibitetto:2011qs}. The basic argument on the origin of non-geometric fluxes lies in the process of a successive application of T-duality on the three form $H$-flux of the type II orientifold theories, where a chain with geometric and non-geometric fluxes appears as \cite{Shelton:2005cf},
\bea
\label{eq:Tdual}
& & H_{ijk} \longrightarrow \omega_{ij}{}^k  \longrightarrow Q^{jk}{}_i  \longrightarrow R^{ijk} \, .
\eea
Apart from these set of fluxes ($H, \omega, Q$ and $R$), the modular completion arguments for the four dimensional effective potentials of type IIB superstring compactifications demand to introduce a new kind of non-geometric $P$-flux which is S-dual to the non-geometric $Q$-flux \cite{Aldazabal:2006up, Aldazabal:2008zza,Font:2008vd,Guarino:2008ik, Hull:2004in, Kumar:1996zx, Hull:2003kr}. The appearance of various kinds of fluxes as a set of parameters in the generalized superpotential induces an effective scalar potential which can generically stabilized all moduli at the tree level. This could be considered as one of the most attractive features for motivating the model building efforts in a non-geometric framework of type IIB superstring compactifications as one may (at least in principle) stabilize {\it all} moduli at tree level, and therefore without paying as much attention to the infinite series of (un-)known perturbaive and non-perturbative corrections as in the standard flux compactification. Note that such subleading corrections have been also found to be crucially useful in the conventional flux compactification without any non-geometric flux, where the compex structure moduli are stabilized at tree level along with the axion-dilaton while the no-scale structure preserves the flatness in the K\"ahler moduli directions. The same are lifted once some additional corrections are included; for example, the non-perturbative corrections to the superpotential \cite{Kachru:2003aw, Balasubramanian:2005zx, Grana:2005jc, Blumenhagen:2006ci, Douglas:2006es, Denef:2005mm, Blumenhagen:2007sm}. Moreover, there have been enormous amount of continuous efforts made in realizing de-Sitter vacua in the standard flux compactification scenarios; for example see \cite{Kachru:2003aw, Burgess:2003ic, Westphal:2006tn, Burgess:2006cb, Achucarro:2006zf, Parameswaran:2006jh, Cremades:2007ig, Krippendorf:2009zza, Louis:2012nb, Cicoli:2012fh, Cicoli:2013cha, Blaback:2013qza, Kallosh:2014wsa, Braun:2015pza, Retolaza:2015nvh, Guarino:2015gos,Bergshoeff:2015jxa, Cicoli:2015ylx,Garcia-Etxebarria:2015lif}. Further, the presence of many non-geometric fluxes of different couplings has been also found to be helpful in appropriate sampling of flux parameters on the way of easing the moduli stabilization process, and it sometimes also creates the possibility of realizing de-Sitter vacua in non-geometric flux-compactification framework \cite{Blaback:2013ht,Damian:2013dq, Damian:2013dwa, Hassler:2014mla, Blaback:2015zra, deCarlos:2009fq, Blumenhagen:2015xpa}. 

Being simple and suitable for performing explicit computations, toroidal orientifolds have been utilized as toolkits in the conventional approach of studying 4D type II effective theories in a non-geometric flux compactification framwork. In fact most of the studies have been centered around a ${\mathbb T}^6/({\mathbb Z}_2 \times {\mathbb Z}_2)$ orientifold. Toroidal setups have not only helped in understanding model building aspects \cite{Aldazabal:2008zza,Font:2008vd,Guarino:2008ik,deCarlos:2009fq, Danielsson:2012by,Damian:2013dq, Damian:2013dwa} but also have demonstrated their utilities in  taking the initial steps for invoking the ten-dimensional origin of the 4D effective type IIB potentials \cite{Blumenhagen:2013hva, Gao:2015nra, Shukla:2015rua, Shukla:2015bca, Shukla:2015hpa,Shukla:2016hyy}. Some very significant steps have been also taken towards exploring the form of non-geometric 10D action via Double Field Theory (DFT) \cite{Andriot:2013xca, Andriot:2011uh, Blumenhagen:2015lta} as well as supergravity \cite{Villadoro:2005cu, Blumenhagen:2013hva, Gao:2015nra, Shukla:2015rua, Shukla:2015bca} \footnote{See \cite{Andriot:2012wx, Andriot:2012an,Andriot:2014qla,Blair:2014zba} also.}.  Moreover, apart from the direct  model building motivations, the interesting relations among the ingredients of superstring flux-compactifications and those of the gauged supergravities have boosted significant interests in understanding both the sectors as fluxes in one setting are related to the gauging in the other one  \cite{ Derendinger:2004jn, Derendinger:2005ph, Shelton:2005cf, Aldazabal:2006up, Dall'Agata:2009gv, Aldazabal:2011yz, Aldazabal:2011nj,Geissbuhler:2011mx,Grana:2012rr,Dibitetto:2012rk, Villadoro:2005cu}. 

The very recent progress made in \cite{Hassler:2014mla, Blumenhagen:2015qda, Blumenhagen:2015kja,Blumenhagen:2015jva, Blumenhagen:2015xpa,  Li:2015taa} regarding the formal developments along with applications towards moduli stabilization, searching de-Sitter vacua as well as  building inflationary models have boosted the interests in setups beyond toroidal example, say Calabi Yaus. As the explicit form of the metric for a generic Calabi Yau threefold is not known, the same has led to explore other possibilities of expressing the effective 4D potentials in a framework where one could bypass the need of knowing the Calabi Yau metric. In this regard, earlier attempts of studying the close connections between the symplectic geometry and effective potentials of type II supergravity theories \cite{Ceresole:1995ca, D'Auria:2007ay, Taylor:1999ii} have been recently extended to include non-geometric fluxes in \cite{Shukla:2015hpa,Shukla:2016hyy}. These proposals provide some alternative and compact ways of representing the scalar potential in terms of some new peculiar flux combinations, (proposed in cohomology language in \cite{Shukla:2015rua},) using which imposing the NS-NS Bianchi identities could be easier as we elaborate in one of the examples in this article. 

Although the complexity induced by introducing many flux parameters of various kinds facilitates a possibly easier samplings of parameters to fit the values, for example cosmological parameters (such as scalar/tensor power spectrum and spectral indices etc.)   while building an inflationary model, however the process as a whole does not remain simple and clean, and it enforces some inevitably hard challenges. A couple of those are enumerated as under,
\begin{itemize}
 \item{To figure of what/which kind of, and how many fluxes can be simultaneously turned-on in a given consistent compactification setup still needs a clear answer.}
 \item{The resulting 4D scalar potentials are very often so huge in concrete examples (say in Type IIB on ${\mathbb T}^6/({\mathbb Z}_2\times{\mathbb Z}_2)$ orientifold) that even it gets hard to analytically solve the extremization conditions as the same demands to solve very high degree polynomials.}
 \item{The difficulty in dealing with the extremization conditions is so much involved that one has to look either for simplified ansatz by switching-off certain flux components at a time, or else one has to opt for an involved numerical analysis \cite{Aldazabal:2008zza,Font:2008vd,Guarino:2008ik,Danielsson:2012by,Damian:2013dq, Damian:2013dwa}.}
 \item{To consistently impose all the NS-NS Bianchi identities and tadpole cancellation conditions while performing moduli stabilization and de-Sitter vacua search is quite challenging.}
\end{itemize}
These points suggest that even though one starts with a large number of fluxes of distinct nature and that too having different types of superpotential couplings (in the sense that some fluxes only couple to $\tau$ while some others couple to the odd moduli $G^a$ or the $T_\alpha$ moduli only), the additional flux constraints may not allow as much freedom to play with flux parameters as one could have naively thought of. In this regard we will explore the possibility of reducing the flux parameter space by looking at the various possible solutions of NS-NS Bianchi identities.

\subsubsection*{Motivation and main goals}
Let us mention at the outset that there are two main formulations of Bianchi identities which are being utilized in the literature for simplifying the type IIB effective potential. One formulation involves fluxes which are denoted in terms of real six dimensional indices (e.g. $H_{ijk}, \omega_{ij}{}^k, Q_i^{jk}, R^{ijk}$) while in the later one, all flux components are written out using cohomology indices (e.g. $H_\Lambda, \omega_{a \Lambda}$ etc. where $\Lambda \in h^{21}_-(CY), a \in h^{11}_-(CY)$). The first formulation has been heavily utilized for simplifying the scalar potential of toroidal examples \cite{Aldazabal:2008zza,Font:2008vd,Guarino:2008ik,Danielsson:2012by,Damian:2013dq, Damian:2013dwa} while the recent interests in building models beyond toroidal example \cite{Blumenhagen:2015qda, Blumenhagen:2015kja,Blumenhagen:2015jva, Blumenhagen:2015xpa,  Li:2015taa} have utilized the identities of the second formulation. In this article,  
\begin{itemize} 
 \item{motivated by some observations of \cite{Ihl:2007ah,Robbins:2007yv}, we plan to demonstrate that the two formulations of identities are not equivalent in their present forms. In fact, the flux constraints of the second formulations are already contained in the first formulation which has some additional ones that cannot be obtained from the known version of the second formulation.}
 \item{we present a detailed investigation of the possible solutions of Bianchi identities and their effects on moduli stabilization in two concrete examples.}
\end{itemize}
The article is organized as follows: In section \ref{sec_setup} we provide some relevant preliminaries of type IIB non-geometric flux compactification. Subsequently, in section \ref{sec_two_formulation}, we will present the two known formulations of the NS-NS Bianchi identities which are utilized for simplifying the scalar potentials of toroidal as well as beyond toroidal examples. Section \ref{sec_convertBIs} is devoted to demonstrate that the two formulations are not equivalent (at least in their presently known versions). This has been done by converting all the complicated flux constraints of the first formulation into cohomology ingredients in the context of two concrete examples. Followed by the same,  in section \ref{sec_solBIs2} and section \ref{sec_solBIs1}, we perform a detailed analysis for looking at the possible solutions of Bianchi identities in the two examples, and subsequently we explore on their implications towards moduli stabilization. 
The section \ref{sec_conclusion} presents an overall conclusion with future directions.  Moreover, we include two appendices where we present a short derivation of Bianchi identities of the second formulation, followed by another one where we discuss the effect of including non-geometric $P$-flux on the Bianchi identities and their solutions for one of the examples.

\section{Basic preliminaries}
\label{sec_setup}
Let us consider Type IIB superstring theory compactified on an orientifold of a Calabi-Yau threefold $X$. To describe the four dimensional effective theory we need several ingredients, and let us start with fixing the conventions. The two classes of possible orientifold projections are described by their action on the
K\"ahler form $J$ and the holomorphic three-form $\Omega_3$ of
the Calabi-Yau, and are given as under \cite{Grimm:2004uq}:
\begin{eqnarray}
\label{eq:orientifold}
 {\cal O}= \begin{cases}
                       \Omega_p\, \sigma  & : \, 
                       \sigma^*(J)=J\,,  \, \, \sigma^*(\Omega_3)=\Omega_3 \, ,\\[0.1cm]
                       (-)^{F_L}\,\Omega_p\, \sigma & :\, 
        \sigma^*(J)=J\,, \, \, \sigma^*(\Omega_3)=-\Omega_3\, ,
\end{cases}
\end{eqnarray}
where $\Omega_p$ is the world-sheet parity, $F_L$ is the left-moving space-time fermion number, and $\sigma$ is a holomorphic, isometric involution. The first choice leads to orientifold with $O5/O9$-planes
whereas the second choice to $O3/O7$-planes. {Let us fix our conventions as those of \cite{Robbins:2007yv} where,
\begin{itemize}
\item{we denote the bases  of even/odd two-forms as $(\mu_\alpha, \, \nu_a)$, four-forms as $(\tilde{\mu}_\alpha, \, \tilde{\nu}_a)$ while the three-forms in two even/odd symplectic pairs as $(a_K, b^J)$ and $({\cal A}_\Lambda, {\cal B}^\Delta)$. We denote the zero- and six- even forms as ${\bf 1}$ and $\Phi_6$ respectively.}
\item{Here the various indices run into their respective cohomology dimensions as; for example, $\alpha\in \left\{1, 2, .., h^{1,1}_+(X)\right\}, \, a\in \left\{1,2,..., h^{1,1}_-(X) \right\}$, $K\in \{1, ..., h^{2,1}_+(X)\}$ and $\Lambda\in \{0, ..., h^{2,1}_-(X)\}$ for the $O3/O7$-cases. For $O5/O9$-planes which is not relevant for our current studies in this work, one has $K\in \{0, ..., h^{2,1}_+(X)\}$ and $\Lambda\in \{1, ..., h^{2,1}_-(X)\}$.}
\item{The definitions of integration over the intersection of various cohomology bases are,
\bea
\label{eq:intersection}
& & \hskip-1.7cm \int_X \Phi_6 = f, \, \, \int_X \, \mu_\alpha \wedge \tilde{\mu}^\beta = \hat{d}_\alpha^{\, \, \, \beta} , \, \, \int_X \, \nu_a \wedge \tilde{\nu}^b = {d}_a^{\, \, \,b}, \quad \int_X \, \mu_\alpha \wedge \mu_\beta \wedge \mu_\gamma = k_{\alpha \beta \gamma}, \\
& & \hskip-0.7cm  \int_X \, \mu_\alpha \wedge \nu_a \wedge \nu_b = \hat{k}_{\alpha a b}, \quad \int_X a_K \wedge b^J = \delta_K{}^J, \, \, \, \, \, \int_X {\cal A}_\Lambda \wedge {\cal B}^\Delta = \delta_\Lambda{}^\Delta \nonumber
\eea
Note that if four-form bases are chosen to be dual to the two-form bases, one will have $\hat{d}_\alpha^{\, \, \, \beta} = \hat{\delta}_\alpha^{\, \, \, \beta}$ and ${d}_a^{\, \, \,b} = {\delta}_a^{\, \, \,b}$. However here we follow  a bit more generic case as in \cite{Robbins:2007yv}.}
\end{itemize}
The massless states in the four dimensional effective theory are in one-to-one correspondence with harmonic forms which are either  even or odd under the action of $\sigma$, and these do generate the equivariant  cohomology groups $H^{p,q}_\pm (X)$. Now, the various field ingredients can be expanded in appropriate bases of the equivariant cohomologies. For example, the K\"{a}hler form $J$, the
two-forms $B_2$,  $C_2$ and the R-R four-form $C_4$ can be expanded as \cite{Grimm:2004uq}
\bea
\label{eq:fieldExpansions}
& & \hskip-1.6cm J = t^\alpha\, \mu_\alpha, \quad  B_2= b^a\, \nu_a, \quad C_2 =c^a\, \nu_a, \quad C_4 = D_2^{\alpha}\wedge \mu_\alpha + V^{K}\wedge a_K + U_{K}\wedge b^K + {\rho}_{\alpha} \, \tilde\mu^\alpha 
\eea
where $t^\alpha$ is string-frame two-cycle volume moduli, while $b^a, \, c^a$ and $\rho_\alpha$ are various axions.  Further, ($V^K$, $U_K$) forms a dual pair of space-time one-forms
and $D_2^{\alpha}$ is a space-time two-form dual to the scalar field $\rho_\alpha$. Moreover, the overall volume of the complex threefold is given as $ {\cal V}_E = \frac{1}{6} \,{k_{\alpha \beta \gamma} \, t^\alpha\, t^\beta \, t^{\gamma}}$.
Now, we consider a complex multi-form of even degree $\Phi_c^{even}$ defined as \cite{Benmachiche:2006df},
\bea
& & \hskip-1cm \Phi_c^{even} = e^{B_2} \wedge C_{RR} + i \, e^{-\phi} Re(e^{B_2+i\, J}) \equiv \tau + G^a \, \,\nu_a + T_\alpha \, \, \tilde{\mu}^\alpha \, ,
\eea
which suggests the following forms for the Einstein-frame chiral variables appearing in ${\cal N}=1$ 4D-effective theory,
\bea
\label{eq:N=1_coords}
& & \hskip-1.6cm \tau = C_0 + \, i \, e^{-\phi}, \quad G^a= c^a + \tau \, b^a,\quad T_\alpha= \left({\rho}_\alpha +  \hat{\kappa}_{\alpha a b} c^a b^b + \frac{1}{2} \, \tau \, \hat{\kappa}_{\alpha a b} b^a \, b^b \right)  -\frac{i}{2} \, \kappa_{\alpha\beta\gamma} t^\beta t^\gamma\,,
\eea
where $\kappa_{\alpha\beta\gamma}=(\hat{d^{-1}})_\alpha^{ \, \,\delta} \, k_{\delta\beta\gamma}$ and $\hat{\kappa}_{\alpha a b} = (\hat{d^{-1}})_\alpha^{ \, \,\delta} \, \hat{k}_{\delta a b}$ in our conventions. 

\subsection*{Four dimensional scalar potential}
The dynamics of low energy effective supergravity action is encoded in three building blocks; namely a K\"{a}hler potential ($K$), a holomorphic superpotential ($W$) and a holomorphic gauge kinetic function ($\hat{\cal G}$) written in terms of appropriate chiral variables. Subsequently, the total ${\cal N}=1$ scalar potential is computed via
\bea
\label{eq:Vtot}
& & \hskip-1cm V=e^{K}\Big(K^{i\bar\jmath}D_i W\, D_{\bar\jmath} \ov W-3\, |W|^2\Big) + \frac{1}{2} (Re \, \, \, \hat{\cal G})^{{-1}{JK}} \, D_J D_K + \frac{1}{2} (Re \, \, \, \hat{\cal G})^{-1}{}_{{JK}} \, D^J D^K.
\eea
Let us briefly elaborate on the ingredients of the total scalar potential.
\subsubsection*{The K\"ahler potential ($K$)}
Using appropriate chiral variables, a generic form of the tree level K\"{a}hler potential can be written as a sum of three pieces motivated from their underlying ${\cal N}=2$ special K\"ahler and quaternionic structures, and the same is give as under,
\bea
\label{eq:K}
& & \hskip-2.2cm K = -\ln\left(i\int_{X}\Omega_3\wedge{\bar\Omega_3}\right) - \ln\left(-i(\tau-\ov\tau)\right) -2\ln\left({\cal V}_E\,(\tau, G^a, T_\alpha; \ov \tau, \ov G^a, \ov T_\alpha)\right) \,.
\eea
Here, the involutively-odd holomorphic three-form $\Omega_3$ generically depends on the complex structure moduli ($z^k$) and can be written out in terms of period vectors, 
\bea
& &  \Omega_3\, \equiv  {\cal X}^\Lambda \, {\cal A}_\Lambda - \, {\cal F}_{\Lambda} \, {\cal B}^\Lambda \,  
\eea
via using a generic tree level pre-potential  as under,
\bea
\label{eq:prepotential}
& & {\cal F} = (X^0)^2 \, \, f({z^i}) \,, \quad \quad  f({z^i}) = \frac{1}{6} \,{\hat{l}_{ijk}} \, z^i\, z^j \, z^k +  \frac{1}{2} \,a_{ij} \, z^i\, z^j +  \,b_{i} \, z^i +  \frac{i}{2} \,\gamma\,.
\eea 
Here special coordinates $z^i =\frac{\delta^i_\Lambda \, X^\Lambda}{X^0}$ are used, and $\hat{l}_{ijk}$ are triple intersection numbers on the Mirror Calabi Yau. Further,  the quantities $a_{ij}, b_i$ and $\gamma$ are real parameters \cite{Hosono:1994av,Arends:2014qca}.  In general, $f({z^i})$ will have an infinite series of non-perturbative contributions (say ${\cal F}_{\rm inst.}(z^i)$), however  for the current purpose, we are assuming the large complex structure limit to suppress the same.

\subsubsection*{Distribution of fluxes for inducing superpotential ($W$) and D-terms ($D^K, D_K$)}
Turning on various fluxes on the internal background induces a non-trivial flux superpotential \cite{Taylor:1999ii}. To construct a generic form of the superpotential, one has to understand the splitting of various geometric as well as non-geometric fluxes into the suitable orientifold even/odd bases. Moreover, it is important to note that in a given setup, all flux-components will not be generically allowed under the full orietifold action ${\cal O} = \Omega_p (-)^{F_L} \sigma$. For example, only geometric flux $\omega$ and non-geometric flux $R$ remain invariant under  $(\Omega_p (-)^{F_L})$, while the standard fluxes $(F, H)$ and non-geometric flux $(Q)$ are anti-invariant \cite{Blumenhagen:2015kja, Robbins:2007yv}. Therefore, under the full orientifold action, we can only have the following flux-components 
\bea
\label{eq:allowedFluxes}
& & \hskip-0.00cm F\equiv \left(F_\Lambda, F^\Lambda\right), \quad  H\equiv \left(H_\Lambda, H^\Lambda\right), \quad \omega\equiv \left({\omega}_a{}^\Lambda, {\omega}_{a \Lambda} , \hat{\omega}_\alpha{}^K, \hat{\omega}_{\alpha K}\right),\nonumber\\
& &  R\equiv \left(R_K, R^K \right), \quad Q\equiv \left({Q}^{a{}K}, \, {Q}^{a}{}_{K}, \, \hat{Q}^{\alpha{}\Lambda} , \, \hat{Q}^{\alpha}{}_{\Lambda}\right), 
\eea
For writing a general flux-superpotential, one needs to define a twisted differential operator, ${\cal D}$ involving the actions from all the NS-NS (non-)geometric fluxes as \cite{Robbins:2007yv}, 
\bea
\label{eq:twistedD}
& & {\cal D} = d + H \wedge.  + \omega \triangleleft . + Q \triangleright. + R \bullet \, \, .
\eea
The action of operators $\triangleleft, \triangleright$ and $\bullet$ on a $p$-form changes it into a $(p+1)$, $(p-1)$ and $(p-3)$-form respectively \footnote{The details of various flux-actions on the non-trivial cohomology basis elements  are given in eqns. (\ref{eq:action1})-(\ref{eq:action2}) while on a generic $p$-form are given in (\ref{eq:action0}).}. With these ingredients in hand, a generic form of flux superpotential is given as under,
\bea
\label{eq:W1}
& & \hskip-0.5cm W =- \int_X \biggl[F+ {\cal D} \Phi_c^{even}\biggr]_3 \wedge \Omega_3  = - \int_{X} \biggl[{F} +\tau \, {H} + \, \omega_a {G}^a + \, {\hat Q}^{\alpha} \,{T}_\alpha \biggr]_3 \wedge \Omega_3. 
\eea
This generic flux superpotential $W$ can be equivalently written as,
\begin{eqnarray}
\label{eq:W_gen}
& & W = e_\Lambda \, {\cal X}^\Lambda + m^\Lambda \, {\cal F}_\Lambda,
\end{eqnarray} 
where
\begin{eqnarray}
\label{eq:eANDm}
& &  \hskip-0.75cm e_\Lambda = \left({F}_\Lambda + \tau \, H_\Lambda + \omega_{a\, \Lambda}\, G^a+ \hat{Q}^\alpha{}_{\Lambda} \, T_\alpha \right), \quad m^\Lambda = \left({F}^\Lambda + \tau \, {H}^\Lambda + {\omega}_a{}^{\Lambda}\, G^a + \hat{Q}^{\alpha \,\Lambda} \, T_\alpha \right).
\end{eqnarray} 
Note that, among the new fluxes only the $\omega_a$ and $\hat{Q}^\alpha$ components are allowed by the choice of involution to contribute into the superpotential, and in order to turn-on the non-geometric $R$-fluxes, one has to induce the following D-terms via implementing a non-trivial even sector of $H^{2,1}(X)$-cohomology \cite{Robbins:2007yv,Shukla:2015bca,Shukla:2015rua},
\bea
\label{eq:DtermOld}
& & \hskip-1.2cm \quad D_K = \frac{1}{2\, s\,{\cal V}_E}\, \biggl[ \frac{{\mathbb R}_K}{f} \, \left({\cal V}_E -\frac{s}{2}\hat{k}_{\alpha a b} t^\alpha b^a b^b\right) + s\, (d^{-1})_b{}^a Q^b{}_K \, \hat{k}_{\alpha a c} t^\alpha b^c - s\, t^\alpha \, \,\hat{\omega}_{\alpha K}\, \biggr]\\
& & \hskip-1.2cm \quad D^K = -\frac{1}{2\, s\,{\cal V}_E}\, \biggl[ \frac{{\mathbb R}^K}{f} \, \left({\cal V}_E -\frac{s}{2}\hat{k}_{\alpha a b} t^\alpha b^a b^b\right) + s\, (d^{-1})_b{}^a Q^{b K} \, \hat{k}_{\alpha a c} t^\alpha b^c - s\, t^\alpha \, \,\hat{\omega}_{\alpha}{}^{K}\, \biggr]\nonumber
\eea

\subsubsection*{New generalized flux orbits}
A closer investigation of the symplectic vectors $(e_\Lambda, m^\Lambda)$ and ($D_K, D^K$), which are responsible for generating $F$-term and $D$-term contributions to the scalar potential, suggests for defining some peculiar flux combination as  {\it new generalized flux orbits} \cite{Shukla:2015bca,Shukla:2015rua}. The flux orbits in NS-NS sector with orientifold odd-indices $\Lambda \in h^{2,1}_-(X)$ are given as,
\bea
\label{eq:OddOrbitA}
& &  {\mathbb H}_\Lambda = H_\Lambda + \omega_{a\Lambda} \, {b}^a + \hat{Q}^\alpha{}_\Lambda \, \left(\frac{1}{2}\, ({\hat{d}^{-1}})_\alpha^{ \, \,\delta} \, \hat{k}_{\delta a b}\, b^a b^b\right) \nonumber\\
& &  {\mathbb H}^\Lambda = H^\Lambda + \omega_{a}{}^{\Lambda} \, {b}^a + \hat{Q}^{\alpha \Lambda} \, \left(\frac{1}{2}\, ({\hat{d}^{-1}})_\alpha^{ \, \,\delta} \, \hat{k}_{\delta a b}\, b^a b^b\right)\\
& & \hskip-2cm {\mathbb\mho}_{a\Lambda} = \omega_{a\Lambda} + \hat{Q}^\alpha{}_\Lambda \, \left(({\hat{d}^{-1}})_\alpha^{ \, \,\delta} \, \hat{k}_{\delta a b}\, b^b\right), \quad {\mathbb\mho}_{a}{}^{\Lambda} = \omega_{a}{}^{\Lambda} + \hat{Q}^{\alpha \Lambda} \, \left(({\hat{d}^{-1}})_\alpha^{ \, \,\delta} \, \hat{k}_{\delta a b}\, b^b\right) \nonumber\\
& & \hskip2cm \hat{\mathbb Q}^\alpha{}_\Lambda = \hat{Q}^\alpha{}_\Lambda , \quad \hat{\mathbb Q}^{\alpha \Lambda} = \hat{Q}^{\alpha \Lambda} \nonumber
\eea
and the RR three-form flux orbits are generalized in the following form, 
\bea
& &  \hskip-0.9cm {\mathbb F}_\Lambda = F_\Lambda + \omega_{a\Lambda} \, {c}^a + \hat{Q}^\alpha{}_\Lambda \, \left({\rho}_\alpha + \hat{\kappa}_{\alpha a b} c^a b^b\right), \\
& & \hskip2cm \quad  {\mathbb F}^\Lambda = F^\Lambda + \omega_a{}^\Lambda \, {c}^a + \hat{Q}^{\alpha \Lambda} \, \left({\rho}_\alpha + \hat{\kappa}_{\alpha a b} c^a b^b\right)\,. \, \nonumber
\eea
Further, the flux components of even-index $K\in h^{2,1}_+(X)$ are given as, 
\bea
\label{eq:OddOrbitB}
& & \hat{\mho}_{\alpha K} = \hat{\omega}_{\alpha K}\, - (d^{-1})_b{}^a  \, Q^{b}{}_{K} \, \, \left(\hat{k}_{\alpha a c} \, b^c \right) + f^{-1} \, \, R_K \, \left(\frac{1}{2}\hat{k}_{\alpha a b} \, b^a \,b^b\right)\nonumber\\
& & \hat{\mho}_{\alpha}{}^{K} =\hat{\omega}_{\alpha}{}^{K}\, - (d^{-1})_b{}^a \, \, Q^{b K} \, \left(\hat{k}_{\alpha a c} \, b^c\right)+ f^{-1} \, \, R^K \, \left(\frac{1}{2}\hat{k}_{\alpha a b} \, b^a \,b^b\right)\\
& & \hskip-1.5cm {\mathbb Q}^{a}{}_{K} =  {Q}^{a}{}_{K} - f^{-1} \,\,d_b{}^a \, (R_K \, b^b), \quad \quad \quad \quad \quad {\mathbb Q}^{a{}K} = {Q}^{a{}K} - f^{-1} \,\, d_b{}^a  \, (R^K \, \, b^b), \nonumber\\
& & \hskip3cm {\mathbb R}_K = R_K, \quad {\mathbb R}^K = R^K \, .\nonumber
\eea
Using these flux orbits, symplectic vectors $(e_\Lambda, m^\Lambda)$ and ($D_K, D^K$) are compactly written as,
\bea
& &  \hskip-1.5cm e_\Lambda = \, \left({\mathbb F}_\Lambda +C_0 {\mathbb H}_\Lambda \right) + i \, \left(s \, {\mathbb H}_\Lambda \right)- i\, \left(\hat{\mathbb Q}^{\alpha}{}_{\Lambda}\, \,\sigma_\alpha \right), \nonumber\\
& & \quad  \, m^\Lambda =\,\left({\mathbb F}^\Lambda + C_0 \, {\mathbb H}^\Lambda \right)+ i \, \left(s \, {\mathbb H}^\Lambda \right)- i\, \left(\hat{\mathbb{Q}}^{\alpha \Lambda}\, \, \sigma_\alpha \right), 
\eea
and
\bea
\label{eq:D-termCompact}
& & \hskip-1.5cm  D_K  = \frac{1}{2\, s\,{\cal V}_E}\, \biggl[ f^{-1} R_K \, {\cal V}_E - s\, t^\alpha \, \hat{\mho}_{\alpha K} \biggr]\, , \\
& & \hskip1.0cm   D^K  = -\frac{1}{2\, s\,{\cal V}_E}\, \biggl[ f^{-1} R^K \, {\cal V}_E - s\, t^\alpha \,\hat{\mho}_{\alpha}{}^{K} \biggr] \, , \nonumber
\eea
where the symbol $\sigma_\alpha$ represents Einstein-frame four-cycle volume given as: $\sigma_\alpha = \frac{1}{2}\,{\kappa}_{\alpha \beta \gamma} t^\beta t^\gamma$.

\section{Two formulations of the NS-NS Bianchi identities}
\label{sec_two_formulation}
For studying moduli stabilization and any subsequent phenomenology, a very crucial step to follow is to impose the constraints from various NS-NS Bianchi identities as well as RR tadpoles to get the {\it true} non-vanishing contribution to the effective four dimensional scalar potential. We have two formulations for representing the (NS-NS) Bianchi identities, and we emphasize here that both sets of Bianchi identities have their own advantages and limitations.  Let us elaborate on it as under, 
\subsection{First formulation}
One set of identities is the one in which all fluxes are expressed as $H_{lmn}, \omega_{lm}{}^n, Q^{lm}{}_n$ and $R^{lmn}$ where $l,m,n$ are indices corresponding to the internal real six dimensional coordinates, and this formulation has five classes of identities given as under, 
\bea
\label{eq:bianchids1}
        {H}_{m[\underline{ij}} {\omega}_{\underline{kl}]}{}^{m}&=0  \nonumber\\
      {H}_{m[\underline{ij}} \, {Q}^{ml}{}_{\underline{k}]} -  {\omega}_{[\underline{ij}}{}^{m}{}  \, {\omega}_{\underline{k}]m}{}^{l}{}&=0\nonumber\\
     {H}_{ijm} \, {R}^{klm} + {\omega}_{{ij}}{}^{m}{} \, {Q}^{{kl}}{}_{m}{}-
     4\, {\omega}_{m[\underline{i}}{}^{[\underline{k}}{} \, {Q}^{\underline{l}]m}{}_{\underline{j}]}{} &=0\, \\
      {\omega}_{mi}{}^{[\underline{j}}{}\,\, {R}^{\underline{kl}]m} \, \,
  - {Q}^{[\underline{jk}}{}_{m}{}  \, {Q}^{\underline{l}]m}{}_{i}{}&=0\nonumber\\
     {Q}^{[\underline{ij}}{}_{m}{}\, {R}^{\underline{kl}]m} \, &=0  , \, \nonumber
\eea 
There have been several ways of deriving these sets of constraints; for example see \cite{Ihl:2007ah, Shelton:2005cf, Blumenhagen:2013hva, Aldazabal:2006up, Blumenhagen:2012pc, Geissbuhler:2013uka, Andriot:2014uda}. We do not intend to provide the detailed derivation, however let us sketch a couple of routes to arrive at these constraints,
\begin{itemize}
\item{One way to derive these identities is via the Jacobi identities of the following Lie brackets for NS-NS fluxes \cite{Aldazabal:2006up},
\bea  
& & \left[Z_i, Z_j  \right] \, \, \, \, = \omega_{ij}{}^k\, Z_k - \, H_{ijk}\, X^k, \quad \nonumber\\
& & \left[ Z_i, X^j \right] \, = -\omega_{ik}{}^j\, X^k - \, Q_i{}^{jk} Z_k, \quad \\
& & \left[ X^i, X^j\right] = Q_k{}^{ij} X^k - \, R^{ijk} Z_k \,, \nonumber
\eea
where $Z_i$ and $X^i$'s are generators of the gauge transformations corresponding to the two gauge groups consisting of two set of $d$-dimensional vectors obtained, from the metric and the $B$-field respectively, via the reduction of  type IIB superstring theory on a $d$-dimensional torus.}
\item{Another route to derive these identities is via considering the nilpotency of the twisted differential operator ${\cal D}$. For that, we consider the following flux-actions on a generic $p$-form, $X_p = \frac{1}{p!} X_{i_1 ....i_p} dx^1 \wedge dx^2 ....\wedge dx^{p}$ being defined as under \cite{Robbins:2007yv,Shelton:2006fd},
\bea
\label{eq:action0}
& & (\omega \triangleleft X)_{i_1i_2...i_{p+1}} = \left(\begin{array}{c}p+1\\2\end{array}\right) \, \, \omega_{[\underline{i_1 i_2}}{}^{j} X_{j|\underline{i_3.....i_{p+1}}]} + \frac{1}{2} \left(\begin{array}{c}p+1\\1\end{array}\right) \, \, \omega_{j[\underline{i_1}}{}^{j} X_{\underline{i_2 i_3.....i_{p+1}}]}\nonumber\\
& & (Q \triangleright X)_{i_1i_2...i_{p-1}} = \frac{1}{2}\left(\begin{array}{c}p-1\\1\end{array}\right) \, \, Q^{jk}{}_{[\underline{i_1}} X_{jk|\underline{i_2.....i_{p-1}}]} + \frac{1}{2} \left(\begin{array}{c}p-1\\0\end{array}\right) \, \, Q^{jk}{}_{j} X_{k |i_1 i_2.....i_{p+1}}\nonumber\\
& & (R \bullet X)_{i_1i_2...i_{p-3}} = \frac{1}{3!}\left(\begin{array}{c}p-3\\0\end{array}\right) \, \, R^{jkl} X_{jkl|i_1.....i_{p-3}]} \, ,
\eea 
where underlined indices are anti-symmetrized. Here, one can notice that the action of (non-)geometric-fluxes via $\triangleleft$, $\triangleright$ and $\bullet$ on a $p$-from changes the same into a $(p+1)$-form, a $(p-1)$-form and a $(p-3)$-form respectively. Then the set of Bianchi identities in eqn. (\ref{eq:bianchids1}) can be derived from the nilpotency of twisted differential operator ${\cal D}$ via ${\cal D}^2 A_p = 0$. Moreover, the action of ${\cal D}^2$ on a generic p-form $A_p$ results in the following additional constraints \cite{Ihl:2007ah},
\bea
\label{eq:bianchids1additional}
&  H_{kl[\ov i} Q^{kl}{}_{\ov j]}-\frac{1}{2}\, Q^{kl}{}_k\,H_{lij}-\,\frac{1}{2}\, \omega_{kl}{}^k\, \omega_{ij}{}^l &=0 , \nonumber\\
& H_{kli}\, R^{klj} - Q^{kl}{}_i\, \omega_{kl}^j  -\omega_{kl}{}^k\, Q^{lj}{}_i-\, Q^{kl}{}_k\,\omega_{li}{}^j &= 0 ,\nonumber\\
& \omega_{kl}^{[\ov i} \, R^{kl \ov j]}  +\frac{1}{2}\,\omega_{kl}{}^k\, R^{lij} + \frac{1}{2}\,Q^{kl}{}_k\,Q^{ij}{}_l &  = 0, \nonumber\\
&  2\, H_{klm}\, R^{klm} - 3\, \omega_{kl}{}^k\, Q^{lm}{}_m &= 0 . 
\eea
However, a closer look ensures that the first three of these additional identities in eqn. (\ref{eq:bianchids1additional}) can be obtained by contracting two indices from the respective main identities in eqn. (\ref{eq:bianchids1}), while the last one in eqn. (\ref{eq:bianchids1additional}) generically holds by the orientifold construction itself. Thus, the additional identities in eqn. (\ref{eq:bianchids1additional}) are effectively not the new ones to worry about. Nevertheless, we will explain their relevance in a different sense while we compare the two formulations in explicit examples later on.}
\end{itemize}
Let us mention that this formulation of Bianchi identities has the following benefits and limitations,
\begin{itemize}
\item{While studying a setup based on toroidal orientifolds, one can write down all the fluxes and moduli in terms of components with real six-dimensional indices, and for such cases, one can directly utilize the Bianchi identities (\ref{eq:bianchids1}) of the first formulation to simplify the scalar potential. }
\item{However, even for the simple toroidal setups such as ${\mathbb T}^6/({\mathbb Z}_2 \times {\mathbb Z}_2)$-orientifold, the total scalar potential has huge number of terms, specially when fluxes are written in terms of real six dimensional indices, e.g. see \cite{Blumenhagen:2013hva,Gao:2015nra,Shukla:2015bca}. Practically speaking, this is more often too huge to impose the Bianchi identities in a clean manner, and moreover performing moduli stabilization demands to solve high degree polynomial constraints, and one is forced to consider simplified flux-ansatz such as taking isotropic limit or switching-off certain fluxes at a time. }
\item{Moreover, these quadratic flux-constraints (\ref{eq:bianchids1}) will not be directly useful for generic setups beyond toroidal backgrounds (such as Calabi Yau orientifolds) as for simplifying scalar potential of those setups, one needs to write down all the fluxes/moduli with indices counted by various even/odd cohomology bases.}
\end{itemize}

\subsection{Second formulation}
On the lines of motivations raised in the last point at the end of the first formulation, in the second formulation, all the Bianchi identity constraints consist of fluxes written with cohomology indices such as $(H_\Lambda, H^\Lambda), \left({\omega}_a{}^\Lambda, {\omega}_{a \Lambda} , \hat{\omega}_\alpha{}^K, \hat{\omega}_{\alpha K}\right), \left({Q}^{a{}K}, \, {Q}^{a}{}_{K}, \, \hat{Q}^{\alpha{}\Lambda} , \, \hat{Q}^{\alpha}{}_{\Lambda}\right)$ and $\left(R_K, R^K \right)$ where $\Lambda\in h^{2,1}_-(CY)$, $K\in h^{2,1}_+(CY)$, $\alpha\in h^{1,1}_+(CY)$ and $a\in h^{1,1}_-(CY)$. 
Now the relevant flux actions represented in the cohomology language are given as under \cite{Robbins:2007yv},
\bea
\label{eq:action1}
& & \hskip1cm H = {H}^\Lambda {\cal A}_\Lambda + H_\Lambda \, \,{\cal B}^\Lambda, \, \, \, \, \, F = {F}^\Lambda {\cal A}_\Lambda + F_\Lambda \, \,{\cal B}^\Delta,\nonumber\\
& & \hskip-1cm \omega_a \equiv (\omega \triangleleft \nu_a) = {\omega}_a{}^\Lambda \, {\cal A}_\Lambda + \omega_{a{}\Lambda} {\cal B}^\Lambda, \quad  \hat{Q}^{\alpha}\equiv (Q \triangleright {\tilde\mu}^\alpha) = \hat{Q}^{\alpha{}\Lambda} {\cal A}_\Lambda + \hat{Q}^{\alpha}{}_{\Lambda} {\cal B}^\Lambda \\
& & \hskip-1cm \hat{\omega}_\alpha \equiv (\omega \triangleleft \mu_\alpha) = \hat{\omega}_\alpha{}^K a_K + \hat{\omega}_{\alpha{}K} b^K, \quad  {Q}^{a}\equiv (Q \triangleright \tilde{\nu}^a) = {Q}^{a{}K} \, a_K + Q^{a}{}_{K} b^K, \nonumber\\
& & \hskip3cm R\bullet \Phi = R^K a_K + R_K b^K \, . \nonumber
\eea
and
\bea
\label{eq:action2}
& & \hskip-0.6cm H \wedge {\cal A}_\Lambda = - f^{-1} H_\Lambda \,\, \Phi_6, \quad \quad \quad \, \, \, \, \quad H \wedge {\cal B}^\Lambda = f^{-1} H^\Lambda \, \,\Phi_6 \\
& & \hskip-0.4cm \omega\triangleleft {\cal A}_\Lambda= - \left({d}^{-1}\right)_a{}^b \,{\omega}_{b \Lambda} \, \tilde{\nu}^a, \, \quad  \quad \quad \omega\triangleleft {\cal B}^\Lambda=\left({d}^{-1}\right)_a{}^b \, {\omega}_{b}{}^{\Lambda} \, \tilde{\nu}^a\nonumber\\
& & \hskip-0.4cm Q\triangleright {\cal A}_\Lambda=-\left(\hat{d}^{-1}\right)_\alpha{}^\beta \,\hat{Q}^\alpha_{\Lambda} \, {\mu}_\beta, \, \quad \quad  \, Q\triangleright {\cal B}^\Lambda=\left(\hat{d}^{-1}\right)_\alpha{}^\beta \, \hat{Q}^{\alpha \Lambda} \,{\mu}_\beta ,\nonumber\\
& & \hskip-0.4cm R\bullet a_K = - f^{-1} \, R_K \, {\bf 1}, \quad  \quad \quad \quad \,\,\, \quad R \bullet b^K = f^{-1} \, R^K \,{\bf 1}\nonumber\\
& & \hskip-0.4cm \omega\triangleleft a_K= - \left(\hat{d}^{-1}\right)_\alpha{}^\beta \hat{\omega}_{\beta K} \, \tilde{\mu}^\alpha,  \, \, \, \quad \quad \omega\triangleleft b^K=\left(\hat{d}^{-1}\right)_\alpha{}^\beta \hat{\omega}_{\beta}{}^{K} \, \tilde{\mu}^\alpha\nonumber\\
& & \hskip-0.4cm Q\triangleright a_K=-\left({d}^{-1}\right)_a{}^b {Q}^a_{K} \,{\nu}_b,\quad  \quad \, \quad Q\triangleright b^K=\left({d}^{-1}\right)_a{}^b \,{Q}^{a K} \, {\nu}_b \, .\nonumber
\eea
The following NS-NS Bianchi identities are obtained via demanding the nilpotency (${\cal D}^2 =0$) of the twisted differential operator on the harmonic forms \cite{Robbins:2007yv}, 
\bea
\label{eq:BIs1}
& & \hskip1.0cm H^\Lambda \, \hat{Q}_\Lambda{}^\alpha - H_\Lambda \hat{Q}^{\alpha \Lambda} = 0, \quad \quad \quad H_\Lambda \, \omega_{a}{}^{\Lambda} - H^\Lambda \, \omega_{\Lambda a} = 0 \\
& & \hskip-1cm \hat{Q}^{\alpha\Lambda} \hat{Q}^\beta{}_{k} - \hat{Q}^{\beta \Lambda} \hat{Q}^\alpha{}_{\Lambda} = 0, \quad  \omega_{a}{}^{\Lambda} \omega_{b \Lambda} - \omega_{b}{}^{\Lambda} \omega_{a \Lambda} =0 , \quad \omega_{a \Lambda} \hat{Q}^{\alpha \Lambda} - \omega_{a}{}^{\Lambda} \hat{Q}^\alpha{}_{\Lambda} = 0\nonumber\\
& & \hskip1.0cm R^K \, \hat{\omega}_{\alpha K} - R_K \hat{\omega}_{\alpha}{}^{K} = 0, \quad \quad \quad R_K \, Q^{a K} - R^K \, Q^{a}{}_{K} = 0 \nonumber\\
& & \hskip-1cm \hat{\omega}_{\alpha}{}^{K} \hat{\omega}_{\beta K} - \hat{\omega}_{\beta}{}^{K} \hat{\omega}_{\alpha K} = 0, \quad  Q^{a K} Q^{b}{}_{K} -Q^{b K} Q^{a}{}_{K} =0 , \quad Q^{a K} \hat{\omega}_{\alpha K} - Q^{a}{}_{K} \hat{\omega}_{\alpha}^{K} = 0\nonumber
\eea
and
\bea
\label{eq:BIs2}
& & \hskip1.0cm f^{-1} \, H_\Lambda \, R_K + (d^{-1})_a{}^b \, \omega_{b \Lambda} \, Q^a{}_K + (\hat{d}^{-1})_\alpha{}^\beta \, \hat{Q}^\alpha{}_\Lambda \, \hat{\omega}_{\beta K} = 0 \\
& & \hskip1cm f^{-1} \, H^\Lambda \, R_K + (d^{-1})_a{}^b \, \omega_{b}{}^{ \Lambda} \, Q^a{}_K + (\hat{d}^{-1})_\alpha{}^\beta \, \hat{Q}^{\alpha{}\Lambda} \, \hat{\omega}_{\beta K} = 0\nonumber\\
& & \hskip1.0cm f^{-1} \, H_\Lambda \, R^K +(d^{-1})_a{}^b \, \omega_{b \Lambda} \, Q^{a{}K} + (\hat{d}^{-1})_\alpha{}^\beta \, \hat{Q}^\alpha{}_\Lambda \, \hat{\omega}_{\beta}{}^{K} = 0\nonumber\\
& & \hskip1.0cm f^{-1} \, H^\Lambda \, R^K + (d^{-1})_a{}^b \, \omega_{b}{}^{ \Lambda} \, Q^{a K} + (\hat{d}^{-1})_\alpha{}^\beta \, \hat{Q}^{\alpha{}\Lambda} \, \hat{\omega}_{\beta}{}^{K} = 0\nonumber
\eea
A direct route of arriving at these quadratic flux constraints is given in the appendix \ref{sec_BIs1} with some detail. Moreover, these identities can also be read-off from the flux-constraints of \cite{Grana:2006hr} via implementing the full orientifold projections on various flux components, and taking care of appropriate normalization of forms.  Let us specifically provide some of the utilities of the second formulation enumerated as under,
\begin{itemize}
\item{The best thing is that all fluxes are written with cohomology indices, and so easily applicable/extendable for simplifying the potentials beyonds toroidal examples such as Calabi Yaus. This is quite promising !}
\item{{\bf Special solutions satisfying the flux constraints:} \\
The constraints given in eqns. (\ref{eq:BIs1}) have easy-to-guess solutions which we call ``special solutions''. One of the same is given as,
\bea
\label{eq:symplecticRotation}
& & \hskip-2cm H^\Lambda =0 , \quad \hat{Q}^{\alpha \Lambda} = 0, \quad \omega_a{}^{\Lambda}= 0, \quad R^{K} =0, \quad Q^{aK} = 0, \quad \omega_\alpha^K =0 \, .
\eea
Or equivalently one could consider fluxes with lower even/odd (2,1)-cohomology indices to be zero. Subsequently, the only constraint to worry about, among those in eqn. (\ref{eq:BIs2}) of the second formulation of Bianchi identities, is $$f^{-1} \, H_\Lambda \, R_K + (d^{-1})_a{}^b \, \omega_{b \Lambda} \, Q^a{}_K + (\hat{d}^{-1})_\alpha{}^\beta \, \hat{Q}^\alpha{}_\Lambda \, \hat{\omega}_{\beta K} = 0$$ Moreover, if one is dealing with an involution such that $h^{2,1}_+(CY) =0$ which one does very often in phenomenological application, then all the relations in eqns. (\ref{eq:BIs2}) are automatically trivial. An equivalent amount of simplification of scalar potential terms by looking at  the first formulation identities would not have been this much easier. These simplifications are quite significant as they can reduce a huge number of terms from the scalar potential! }
\end{itemize} 
Let us also point out here that the existence of the possibility for setting half of the fluxes to zero is not merely a random choice of simplification. It is rooted into the fact that  constraints given in eqns. (\ref{eq:BIs1}) have a nice symplectic structure. For example, the flux-pairs $({H}_\Lambda, {H}^\Lambda), (\omega_{a \Lambda}, {\omega}_a{}^\Lambda)$ and $({\hat Q}^\alpha_\Lambda, {\hat Q}^{\alpha\Lambda})$ form orthogonal vectors in sympelctic basis $({\cal A}_\Lambda, {\cal B}^\Lambda)$, and similarly the flux-pairs $(R_K, R^K), (\hat{\omega}_{\alpha K}, \hat{\omega}_\alpha{}^K)$ and $(Q^a{}_K, Q^{aK})$ are orthogonal vectors in sympelctic basis $(a_K, b^K)$. Now given a set of orthogonal symplectic vectors, there always exits a symplectic map which can rotate away half of the components (say those with upper $h^{2,1}$ indices), and so one can switch-off half of the NS-NS fluxes assuming that the fluxes are quantized to integral values \cite{Robbins:2007yv}, which we will assume throughout this work. 
\subsubsection*{An interesting remark}
Using the new generalized flux orbits as given in (\ref{eq:OddOrbitA})-(\ref{eq:OddOrbitB}) and the Bianchi identities of the second formulation in eqns. (\ref{eq:BIs1})-(\ref{eq:BIs2}), one finds that the following holds,
\bea
\label{eq:newOrbitsBIs1}
& & \hskip1.0cm {\mathbb H}^\Lambda \, \hat{{\mathbb Q}}_\Lambda{}^\alpha - {\mathbb H}_\Lambda \hat{{\mathbb Q}}^{\alpha \Lambda} = 0, \quad \quad \quad {\mathbb H}_\Lambda \, {\mathbb \mho}_{a}{}^{\Lambda} - {\mathbb H}^\Lambda \, {\mathbb \mho}_{\Lambda a} = 0 \\
& & \hskip-1cm \hat{{\mathbb Q}}^{\alpha\Lambda} \hat{{\mathbb Q}}^\beta{}_{k} - \hat{{\mathbb Q}}^{\beta \Lambda} \hat{{\mathbb Q}}^\alpha{}_{\Lambda} = 0, \quad  {\mathbb \mho}_{a}{}^{\Lambda} {\mathbb \mho}_{b \Lambda} - {\mathbb \mho}_{b}{}^{\Lambda} {\mathbb \mho}_{a \Lambda} =0 , \quad {\mathbb \mho}_{a \Lambda} \hat{{\mathbb Q}}^{\alpha \Lambda} - {\mathbb \mho}_{a}{}^{\Lambda} \hat{{\mathbb Q}}^\alpha{}_{\Lambda} = 0\nonumber\\
& & \hskip1.0cm {\mathbb R}^K \, \hat{{\mathbb \mho}}_{\alpha K} - {\mathbb R}_K \hat{{\mathbb \mho}}_{\alpha}{}^{K} = 0, \quad \quad \quad {\mathbb R}_K \, {\mathbb Q}^{a K} - {\mathbb R}^K \, {\mathbb Q}^{a}{}_{K} = 0 \nonumber\\
& & \hskip-1cm \hat{{\mathbb \mho}}_{\alpha}{}^{K} \hat{{\mathbb \mho}}_{\beta K} - \hat{{\mathbb \mho}}_{\beta}{}^{K} \hat{{\mathbb \mho}}_{\alpha K} = 0, \quad  {\mathbb Q}^{a K} {\mathbb Q}^{b}{}_{K} -{\mathbb Q}^{b K} {\mathbb Q}^{a}{}_{K} =0 , \quad {\mathbb Q}^{a K} \hat{{\mathbb \mho}}_{\alpha K} - {\mathbb Q}^{a}{}_{K} \hat{{\mathbb \mho}}_{\alpha}^{K} = 0\nonumber
\eea
and
\bea
\label{eq:newOrbitsBIs2}
& & \hskip1.0cm f^{-1} \, {\mathbb H}_\Lambda \, {\mathbb R}_K + (d^{-1})_a{}^b \, {\mathbb \mho}_{b \Lambda} \, {\mathbb Q}^a{}_K + (\hat{d}^{-1})_\alpha{}^\beta \, \hat{{\mathbb Q}}^\alpha{}_\Lambda \, \hat{{\mathbb \mho}}_{\beta K} = 0 \\
& & \hskip1cm f^{-1} \, {\mathbb H}^\Lambda \, {\mathbb R}_K + (d^{-1})_a{}^b \, {\mathbb \mho}_{b}{}^{ \Lambda} \, {\mathbb Q}^a{}_K + (\hat{d}^{-1})_\alpha{}^\beta \, \hat{\mathbb Q}^{\alpha{}\Lambda} \, \hat{{\mathbb \mho}}_{\beta K} = 0\nonumber\\
& & \hskip1.0cm f^{-1} \, {\mathbb H}_\Lambda \, {\mathbb R}^K + (d^{-1})_a{}^b \, {\mathbb \mho}a_{b \Lambda} \, {\mathbb Q}^{a{}K} + (\hat{d}^{-1})_\alpha{}^\beta \, \hat{{\mathbb Q}}^\alpha{}_\Lambda \, \hat{{\mathbb \mho}}_{\beta}{}^{K} = 0\nonumber\\
& & \hskip1.0cm f^{-1} \, {\mathbb H}^\Lambda \, {\mathbb R}^K + (d^{-1})_a{}^b \, {\mathbb \mho}_{b}{}^{ \Lambda} \, {\mathbb Q}^{a K} + (\hat{d}^{-1})_\alpha{}^\beta \, \hat{{\mathbb Q}}^{\alpha{}\Lambda} \, \hat{{\mathbb \mho}}_{\beta}{}^{K} = 0\nonumber
\eea
Thus we find that new generalized flux orbits in eqns. (\ref{eq:OddOrbitA})-(\ref{eq:OddOrbitB}) respect the same structure of the flux constraints as to those of the old flux orbits given in eqns. (\ref{eq:BIs1})-(\ref{eq:BIs2}). This looks quite interesting !  The reason for this to happen is the fact that these relations in (\ref{eq:newOrbitsBIs1})-(\ref{eq:newOrbitsBIs2}) can directly be derived from the nilpotency of a `new' generalized twisted differential (${\mathfrak D}^2 =0$), where $\mathfrak D$ being defined as ${\mathfrak D} = (d + {\mathbb H} \wedge.  + {\mathbb \mho} \triangleleft . + {\mathbb Q} \triangleright. + {\mathbb R} \bullet .)$ in a similar fashion as the usual ${\cal D}$ is defined in eqn. (\ref{eq:twistedD}). This new observation on Bianchi identities illustrates more use and relevance of the cohomological version of the `new' generalized flux orbits proposed  in \cite{Shukla:2015rua}, and found useful in many subsequent studies in \cite{Blumenhagen:2015lta, Shukla:2015hpa, Shukla:2016hyy}. 

\subsection{Two approaches for simplifying the 4D scalar potential}
As we have mentioned in the list of benefits and challenges about using both of the formulations in simplifying the 4D scalar potential, there are basically two possible approaches to follow, 
\begin{itemize}
\item{Option one is to compute the total scalar potential by converting all fluxes appearing in the superpotential as well as the D-terms into real index components and use the identities of the first formulation. For that purpose one has to re-express the flux dependent three-form symplectic components ($e_\Lambda, m^\Lambda$) given in eqn. (\ref{eq:eANDm}) as well as D-terms ($D_K, D_K$) in eqn. (\ref{eq:D-termCompact})  in real six dimensional indices. This strategy has been adopted in most of the previous phenomenology oriented studies based on toroidal setups.}
\item{Option two is to directly use the second formulation constraints for scalar potential pieces obtained via the superpotential (\ref{eq:W_gen}) and D-term (\ref{eq:D-termCompact}) written in cohomology ingredients. Or equivalently one should convert the first formulation identities given in eqn. (\ref{eq:bianchids1}) into the ones with cohomology basis, and subsequently use them directly. }
\end{itemize}
Now, there is a subtlety between the two formulations of the Bianchi identities. For some particular toroidal examples in type IIA and type IIB compactifications on the orientifolds of ${\mathbb T}^6/{\mathbb Z}_4$ sixfold, it has been observed in \cite{Ihl:2007ah, Robbins:2007yv} that the two sets of Bianchi identities are not the same !  In fact, as we will elaborate later on in two different and concrete toroidal orientifolds, one has to supplement some additional constraints in the second formulations to have a match with the first one. 

For checking the inequivalence of the two formulations, one has to work in a setup in which one can do computations for both formulations. Fortunately, for toroidal setups, one can follow both approaches as one can easily switch from one set of fluxes into the other one, and so one can convert Bianchi identities (\ref{eq:bianchids1}) into cohomology based fluxes. Moreover, one can also convert the superpotential given in eqn. (\ref{eq:W_gen}) and the D-term (\ref{eq:D-termCompact}) into another form with real-indexed flux components. This is the beauty of simplicity of toroidal models in which one can analytically compute all the relevant data unlike a generic CY case. As we will see later, revisiting the study of a couple of toroidal examples using cohomology-indexed flux components will give us some more insights for invoking non-trivial additional Bianchi identities. 

\section{Converting the first formulation constraints into cohomology ingredients}
\label{sec_convertBIs}
In this section, we consider two explicit examples in type IIB superstring compactification on orientifolds of  ${\mathbb T}^6/{\mathbb Z}_4$ and ${\mathbb T}^6/({\mathbb Z}_2\times{\mathbb Z}_2)$ sixfolds. For these two examples, now our aim is to translate all the complicated flux constraints of the first formulation of Bianchi identities into the ones using cohomology ingredients so that we could compare the two things, and in fact show that the two formulations do not happen to be equivalent. 
\subsection{Model A: Type IIB $\hookrightarrow$ ${\mathbb T}^6/{\mathbb Z}_4$-orientifold}
\subsubsection*{Fixing the conventions} Here we consider the untwisted sector of type IIB superstring compactification on the orientifold of ${\mathbb T}^6/{\mathbb Z}_4$ orbifold.  Let us recall the relevant features of this example from \cite{Robbins:2007yv}. In the toroidal orbifold, we consider the following redefinition of complexified coordinates on ${\mathbb T}^6$
\begin{eqnarray}
\label{eq:coordinates}
& & \hskip-0.5cm z^1 = x^1 + i \, x^2 +e^{i \pi /4}\, (x^3 + i \, x^4) , \quad z^2 = x^3 + i \, x^4 +e^{i 3\, \pi /4}\, (x^1 + i \, x^2) , \, \quad  z^3 = x^5 + i \, x^6 . \nonumber
\end{eqnarray}
The orbifold action ${\mathbb Z}_4$ and the holomorphic involution $\sigma$ are given as under,
\bea
& & \hskip-1.5cm  \Theta({\mathbb Z}_4): (z^1, z^2, z^3) \longrightarrow (i \, z^1, i \, z^2, - z^3), \qquad \sigma : (z^1, z^2, z^3) \longrightarrow (- e^{i \, \pi /4}\, z^1, e^{i \, \pi /4}\, z^2, - i \, z^3). 
\eea
The splittings of the hodge numbers are as: $h^{2,1}=1_+ + 0_-$ and $h^{1,1}= 3_+ + 2_-$. The splitting of various $(p,q)$-cohomology bases are given as under  \cite{Robbins:2007yv},
\vskip0.3cm 
{\bf \hskip-0.5cm Splitting of (1,1)-cohomology bases:}
\begin{eqnarray}
 && \hskip-0.75cm \mu_1 = \frac{i}{4}\, \left(dz^1 \wedge d \ov{z}^1 +dz^2 \wedge d \ov{z}^2 \right) = dx^1 \wedge dx^2 + dx^3 \wedge dx^4, \\
 && \hskip-0.75cm \mu_2 = \frac{i}{2\sqrt{2}}\, \left(dz^1 \wedge d \ov{z}^1 -dz^2 \wedge d \ov{z}^2 \right) = dx^1 \wedge dx^3+dx^1 \wedge dx^4 - dx^2 \wedge dx^3 + dx^2 \wedge dx^4, \nonumber\\
 && \hskip-0.75cm \mu_3 = \frac{i}{2}\, \left(dz^3 \wedge d \ov{z}^3 \right) = dx^5 \wedge dx^6, \nonumber\\
 && \hskip-0.75cm \nu_1 = \frac{1-i}{4}\, \left(dz^1 \wedge d \ov{z}^2 + i \, d\ov{z}^1 \wedge d {z}^2 \right) = dx^1 \wedge dx^3- dx^1 \wedge dx^4 + dx^2 \wedge dx^3 + dx^2 \wedge dx^4, \nonumber\\
 && \hskip-0.75cm \nu_2 = -\frac{e^{-i \pi /4}}{4}\, \left(dz^1 \wedge d \ov{z}^2 - i \, d\ov{z}^1 \wedge d {z}^2 \right) = dx^1 \wedge dx^2 - dx^3 \wedge dx^4. \nonumber
\end{eqnarray}
\vskip0.0cm 
{\bf \hskip-0.5cm Splitting of (2,2)-cohomology bases:}
\begin{eqnarray}
 & & \hskip-2cm  \tilde{\mu}^1 = \mu_1 \wedge \mu_3, \, \, \, \tilde{\mu}^2 = \mu_2 \wedge \mu_3, \, \, \, \tilde{\mu}^3 = \frac{1}{2} \, \mu_1 \wedge \mu_1 , \qquad \quad \tilde{\nu}^1 = \nu_1 \wedge \mu_3, \, \, \, \tilde{\nu}^2 = \nu_2 \wedge \mu_3
\end{eqnarray}
\vskip0.0cm 
{\bf \hskip-0.5cm Splitting of (2,1)-cohomology bases:}
\bea
\label{eq:symplecticBases}
& & \hskip-2cm a_1 = -\frac{i}{2} \left(dz^1\wedge dz^2\wedge d {\ov z}^3-d {\ov z}^1\wedge d {\ov z}^2 \wedge d {z}^3 \right) = \beta^0 + \beta^1 + \beta^2 - \beta^3, \nonumber\\
& & \hskip-2cm b^1 = \frac{1}{2} \left(dz^1\wedge dz^2\wedge d {\ov z}^3+d {\ov z}^1\wedge d {\ov z}^2 \wedge d {z}^3 \right) = \alpha_0 + \alpha_1 + \alpha_2 - \alpha_3,  \\
& & \hskip-2cm A_0 = \frac{1}{2} \left(dz^1\wedge dz^2\wedge d {z}^3+d {\ov z}^1\wedge d {\ov z}^2 \wedge d {\ov z}^3 \right) = \alpha_0 - \alpha_1 - \alpha_2 - \alpha_3, \nonumber\\
& & \hskip-2cm B^0 = -\frac{i}{2} \left(dz^1\wedge dz^2\wedge d {z}^3-d {\ov z}^1\wedge d {\ov z}^2 \wedge d {\ov z}^3 \right) = -\beta^0 + \beta^1 + \beta^2  + \beta^3, \nonumber
\eea
where the following notations have been considered,
\bea
\label{eq:Realbasis}
 & & \alpha_0=1\wedge 3\wedge 5\,, \quad  \alpha_1=1\wedge 4\wedge 6, \quad \alpha_2=2\wedge 3\wedge 6\,, \quad \alpha_3=2\wedge 4\wedge 5\\
 & &  \beta^0=2\wedge 4\wedge 6\, , \quad \beta^1=2\wedge 3\wedge 5\, , \quad \beta^2=1\wedge 4\wedge 5\,, \quad \beta^3=1\wedge 3\wedge 6\,. \nonumber
\eea
In addition, the orientifold even zero-form and the six-form are defined as ${\bf 1}$ and $\Phi_6 = dx^1 \wedge dx^2 \wedge dx^3 \wedge dx^4 \wedge dx^5 \wedge dx^6$ respectively, whereas there are no harmonic 1-forms and their dual five-forms. Moreover, the non-vanishing intersection numbers defined in eqn. (\ref{eq:intersection}) are \cite{Robbins:2007yv},
\bea
\label{eq:intersectionForm}
& & f=\frac{1}{4}, \, \, \hat{d}_\alpha^\beta = diag\left(\frac{1}{2},-1,\frac{1}{4}\right), \, \, \, {d}_a^b = diag\left(-1,-\frac{1}{2}\right)\nonumber\\
& & \left(k_{113}=\frac{1}{2}, \, k_{223}=-1\right)  \, \, \, \, \, {\rm and} \, \, \, \, \, \, \left(\hat{k}_{311}=-1, \, \hat{k}_{322}= -\frac{1}{2}\right).
\eea
As this setup has $h^{2,1}(X) = 1_+ +0_-$, and $h^{1,1}(X) = 3_+ +2_-$, there are three complexified K\"ahler moduli ($T_\alpha$), two complexified odd axions ($G^a$) and no complex structure moduli. Thus we have effectively two independent components for $H_3$ and $F_3$ as well non-geometric $R$-flux as generically one should have $2(h^{2,1}_-+1)$ for the former case while $(2h^{2,1}_+)$ for the later case. Moreover the geometric flux $\omega$ and the non-geometric flux $Q$ have a total of $\left(2 \, h^{11}_+ \, (h^{21}_-+1) + 2 \, h^{11}_- \, h^{21}_+ \right)$ components as we can guess from the collection of flux components surviving under the orientifold action as given in eqn.  (\ref{eq:allowedFluxes}). This way there are 10 independent components for each of geometric-flux ($\omega$) as well as non-geometric flux ($Q$). To be more specific, the total orientifolding induces the following additional restrictions on the various components of the moduli and fluxes,
\begin{itemize}
\item{{\bf $B_{ij}$\,-axion:} $\quad B_{12} = -  B_{34} \equiv b^2, \qquad  B_{13} = -  B_{14} =  B_{23} =  B_{24} \equiv b^1$.}
\item{{\bf $C_{ik}$\,-axion :} $\quad  C_{12} = -  C_{34} \equiv c^2 , \qquad   C_{13} = -  C_{14} =  C_{23} =  C_{24} \equiv c^1$.}
\item{{\bf $C_{ijkl}$\,-axion :} $\quad C_{1256} =  C_{3456} \equiv \rho_1, \quad  C_{1356} =  C_{2456} =-  C_{2356} =  C_{1456} \equiv \rho_2 , \quad   C_{1234} \equiv \rho_3$.}
\item{{\bf $F_{ijk}$\,-flux :}  $\quad F_{135} = -  F_{245} = - F_{146}=- F_{236}, \quad  F_{246} = -  F_{136} = - F_{145}=- F_{235}$. }
\item{{\bf $H_{ijk}$\,-flux :} $\quad H_{135} = -  H_{245} = - H_{146}=- H_{236}, \quad H_{246} = -  H_{136} = - H_{145}=- H_{235}$.}
\item{{\bf $\omega_{ij}^k$\,-flux :} 
\bea
  & & \hskip-1cm \omega_{15}^{1}= - \, \omega_{25}^{2}=- \,\omega_{36}^{3}= \,\omega_{46}^{4}, \, \, \omega_{16}^{1}=- \,\omega_{26}^{2}= \,\omega_{35}^{3}=- \,\omega_{45}^{4}, \omega_{25}^{1}= \,\omega_{15}^{2}=- \omega_{46}^{3}= - \,\omega_{36}^{4}, \nonumber\\
  & &  \hskip-1cm \omega_{26}^{1}= \,\omega_{16}^{2}= \,\omega_{45}^{3}= \,\omega_{35}^{4}, \, \, \omega_{35}^{1}=- \,\omega_{45}^{2}=- \,\omega_{26}^{3}= -\,\omega_{16}^{4}, \, \, \, \,\omega_{36}^{1}=- \,\omega_{46}^{2}= \,\omega_{25}^{3}= \,\omega_{15}^{4}, \nonumber\\
  & & \hskip-1cm \omega_{45}^{1}= \,\omega_{35}^{2}= \,\omega_{16}^{3}=- \,\omega_{26}^{4}, \, \, \, \, \omega_{46}^{1}= \,\omega_{36}^{2}=- \omega_{15}^{3}= \,\omega_{25}^{4}, \, \, \omega_{13}^{5}=- \,\omega_{24}^{5}= \,\omega_{14}^{6}= \,\omega_{23}^{6},\nonumber\\
  & & \hskip3.5cm \omega_{14}^{5}= \,\omega_{23}^{5}=- \,\omega_{13}^{6}= \,\omega_{24}^{6} \nonumber
\eea }
\item{{\bf $Q^{ij}_k$\,-flux : \, }
\bea
& & \hskip-1cm Q^{15}_{1}= - \, Q^{25}_{2}= \,Q^{36}_{3}=- \,Q^{46}_{4}, \,  \, Q^{16}_{1}=- \,Q^{26}_{2}= -\,Q^{35}_{3}= \,Q^{45}_{4}, \, \, Q^{25}_{1}= \,Q^{15}_{2}= Q^{46}_{3}=  \,Q^{36}_{4}, \nonumber\\
& &  \hskip-1cm Q^{26}_{1}= \,Q^{16}_{2}= -\,Q^{45}_{3}= -\,Q^{35}_{4}, \, \, Q^{35}_{1}= -\,Q^{45}_{2}= \,Q^{26}_{3}= \,Q^{16}_{4}, \, \,Q^{36}_{1}=- \,Q^{46}_{2}= - \,Q^{25}_{3}= -\,Q^{15}_{4}, \nonumber\\
& & \hskip-1cm Q^{45}_{1}= \,Q^{35}_{2}= -\,Q^{16}_{3}= \,Q^{26}_{4}, \, \, Q^{46}_{1}= \,Q^{36}_{2}= Q^{15}_{3}= - \,Q^{25}_{4}, \, \, Q^{13}_{5}=- \,Q^{24}_{5}=- \,Q^{14}_{6}=- \,Q^{23}_{6}, \nonumber\\
& & \hskip3.9cm Q^{14}_{5}= \,Q^{23}_{5}= \,Q^{13}_{6}=- \,Q^{24}_{6} \nonumber
\eea}
\item{{\bf $R^{ijk}$\,-flux :} $\qquad R^{135} = -  R^{245} =  R^{146}=  R^{236}, \quad R^{246} = -  R^{136} =  R^{145}=  R^{235}.$}
\end{itemize}
\subsubsection*{Flux conversion relations}
We have the following relations among the various flux parameters in the two formulations,
\bea
\label{eq:fluxconversionB}
& & \hskip-1.3cm H^0 = H_{135}, \quad \quad \quad  H_0 = -H_{246}\, , \nonumber\\
& & \hskip-1.3cm \omega_{a}{}^{0} \equiv \left(\begin{array}{c}-\omega_{15}{}^1 -\omega_{16}{}^1 - \omega_{25}{}^1 + \omega_{26}{}^1\\ -\omega_{36}{}^1 - \omega_{45}{}^1\end{array}\right) \,, \quad \omega_{a0} \equiv \left(\begin{array}{c}\omega_{15}{}^1 - \omega_{16}{}^1 -\omega_{25}{}^1 - \omega_{26}{}^1\\ \omega_{35}{}^1 -\omega_{46}{}^1\end{array}\right), \nonumber\\
& & \hskip-1.4cm \hat{\omega}_{\alpha}{}^{1} \equiv \left(\begin{array}{c}\omega_{35}{}^1 +\omega_{46}{}^1\\ -\omega_{15}{}^1 + \omega_{16}{}^1 -\omega_{25}{}^1 - \omega_{26}{}^1\\ -\omega_{13}{}^5\end{array}\right), \, \hat{\omega}_{\alpha 1} \equiv \left(\begin{array}{c}\omega_{36}{}^1 -\omega_{45}{}^1\\-\omega_{15}{}^1 -\omega_{16}{}^1 + \omega_{25}{}^1 - \omega_{26}{}^1\\ \omega_{14}{}^5\end{array}\right)\, , \\
& & \hskip-1.4cm \hat{Q}^{\alpha 0} \equiv \left(\begin{array}{c} -Q^{15}{}_3 + Q^{16}{}_4 \\ Q^{15}{}_1+ Q^{15}{}_2+ Q^{16}{}_1- Q^{16}{}_2 \\ Q^{13}{}_5\end{array}\right)\, , \quad \hat{Q}^{\alpha}{}_{0} \equiv \left(\begin{array}{c} -Q^{15}{}_4 - Q^{16}{}_3 \\ -Q^{15}{}_1 + Q^{15}{}_2 + Q^{16}{}_1 + Q^{16}{}_2 \\ Q^{13}{}_6\end{array}\right), \,  \nonumber\\
& & \hskip-1.4cm {Q}^{a1} \equiv \left(\begin{array}{c} -Q^{15}{}_1 - Q^{15}{}_2 + Q^{16}{}_1 - Q^{16}{}_2 \\ -Q^{15}{}_4 + Q^{16}{}_3\end{array}\right), \, {Q}^{a}{}_1 \equiv \left(\begin{array}{c} -Q^{15}{}_1+ Q^{15}{}_2- Q^{16}{}_1- Q^{16}{}_2 \\ -Q^{15}{}_3 - Q^{16}{}_4\end{array}\right) \, , \nonumber\\
& & \hskip-1.4cm R^1 = - R^{135}, \quad \quad \quad R_1 =  R^{246} \, . \nonumber
\eea
Let us mention that there is always a bijection between the number of independent real-indexed and cohomology-indexed components for each of the fluxes $F_3, H_3$ and non-geometric $R$-fluxes. However, the same cannot be argued to be generically true for the geometric flux $\omega$ and the non-geometric $Q$-flux. Nevertheless, in some toroidal examples (such as ${\mathbb T}^6/\Gamma$ where $\Gamma$ is a crystallographic action in ${\mathbb Z}_2 \times {\mathbb Z}_2, {\mathbb Z}_3, {\mathbb Z}_3\times {\mathbb Z}_3, {\mathbb Z}_4$ and ${\mathbb Z}_{6-I}$) the bijection appears to hold, e.g. see \cite{Ihl:2007ah, Robbins:2007yv, DeWolfe:2005uu}. Nevertheless, in this current example, there is a bijection in all the fluxes between the two formulations, and so we can explicitly relate the flux components in two formalism by using these conversion relations given in eqn. (\ref{eq:fluxconversionB}). 
Therefore, using the conversion relations given in eqn. (\ref{eq:fluxconversionB}), one can completely invert the five types of Bianchi identities of eqn. (\ref{eq:bianchids1}) given in the first formulation. Let us do it one-by-one. 
\subsubsection*{(i). $(H \omega)$-type identities:} In this case, we find that
\bea
\label{eq:modelBbis1}
& & {H}_{m[\underline{ij}} {\omega}_{\underline{kl}]}{}^{m}=0 \Longrightarrow \biggl\{ H_0 \, \omega_1{}^0 = H^0 \, \omega_{10}\, , \quad H_0 \, \omega_2{}^0 = H^0 \, \omega_{20} \biggr\}\, 
\eea
which is identical to the constraint: $H^\Lambda \, \omega_{a \Lambda} - H_\Lambda \, \omega_a{}^{\Lambda} =0$.
\subsubsection*{(ii). $(HQ + \omega^2)$-type identities:} The identities given in ``$ ({H}_{m[\underline{ij}} \, {Q}^{ml}{}_{\underline{k}]} =  {\omega}_{[\underline{ij}}{}^{m}{}  \, {\omega}_{\underline{k}]m}{}^{l}{})$" translates into many coupled constraints, and a careful reshuffling results in the following two sets of constraints,
\begin{subequations}
\bea
\label{eq:ModelBhq1}
& & H_0 \hat{Q}^{10}-H^0 \hat{Q}^1{}_0=0, \quad  H_0 \hat{Q}^{20}-H^0 \hat{Q}^2{}_0=0, \quad H_0 \hat{Q}^{30}-H^0 \hat{Q}^3{}_0=0, \\
& & \hskip-1.0cm \omega _{20} \omega _1{}^0-\omega _{10} \omega _2{}^0 = 0, \quad \hat{\omega }_{21} \hat{\omega }_1{}^1-\hat{\omega }_{11} \hat{\omega }_2{}^1=0, \quad \hat{\omega }_{31} \hat{\omega }_1{}^1-\hat{\omega }_{11} \hat{\omega }_3{}^1=0, \quad \hat{\omega }_{21} \hat{\omega }_3{}^1-\hat{\omega }_{31} \hat{\omega }_2{}^1=0 \nonumber
\eea
\bea
\label{eq:ModelBhq2}
& & \hskip-0.5cm H^0 Q^1{}_1+H_0 Q^{11}+\hat{\omega }_{31} \omega _1{}^0+\omega _{10} \hat{\omega }_3{}^1 =0, \quad H^0 Q^2{}_1+H_0 Q^{21}+\hat{\omega }_{31} \omega _2{}^0+\omega _{20} \hat{\omega }_3{}^1=0\, , \\
& & \hskip-0.5cm H_0 Q^1{}_1-H^0 Q^{11}-\omega _1{}^0 \hat{\omega }_3{}^1+\omega _{10} \hat{\omega}_{31}=0, \quad H_0 Q^2{}_1-H^0 Q^{21}-\omega _2{}^0 \hat{\omega }_3{}^1+\omega _{20} \hat{\omega}_{31}=0\, . \nonumber\\
& & \hskip-1cm H_0 \hat{Q}^{1}{}_{0}+H^0 \hat{Q}^{10} = 0, \quad H_0 \hat{Q}^{2}{}_{0}+H^0 \hat{Q}^{20} = 0, \quad \hat{\omega }_1{}^1 \hat{\omega }_3{}^1+\hat{\omega }_{11} \hat{\omega }_{31}=0, \quad \hat{\omega }_2{}^1 \hat{\omega }_3{}^1+\hat{\omega }_{21} \hat{\omega }_{31}=0, \nonumber\\
& & \hskip-1cm 8 H_0 \hat{Q}^3{}_0+8 H^0 \hat{Q}^{30}+\left(\omega _1{}^0\right){}^2+2 \left(\omega _2{}^0\right){}^2+2 \left(\hat{\omega }_1{}^1\right){}^2-\left(\hat{\omega }_2{}^1\right){}^2+\omega _{10}^2+2
   \omega _{20}^2+2 \hat{\omega }_{11}^2-\hat{\omega }_{21}^2=0\, .\nonumber
\eea
\end{subequations}
Now it is obvious from the contraction of (2,1)-cohomology indices that the first collection (\ref{eq:ModelBhq1}) corresponds to the constraints of second formulation Bianchi identities in eqn. (\ref{eq:BIs1}) while those in eqn. (\ref{eq:ModelBhq2}) are new constraints. 

\subsubsection*{(iii). $(HR + \omega Q)$-type identities:} This class of identity given as ``$({H}_{ijm} \, {R}^{klm} + {\omega}_{{ij}}{}^{m}{} \, {Q}^{{kl}}{}_{m}{}-
     4\, {\omega}_{m[\underline{i}}{}^{[\underline{k}}{} \, {Q}^{\underline{l}]m}{}_{\underline{j}]})=0$" results in the most complicated and lengthy set of constraints. However after a very tedious attempt, we have managed to rearrange the pieces within each constraint such that the same could be useful and relevant while invoking the second formulation constraints. In fact, the identities translated into cohomology language can be again reshuffled into two sets of constraints, one being the subset of second formulation and the other being the new ones, and we present them both as below,
\begin{subequations}
\bea
\label{eq:omegaQ2a}
& & \omega _{10} \hat{Q}^{10}-\hat{Q}^1{}_0 \omega _1{}^0 = 0,\quad \omega _{10} \hat{Q}^{20}-\hat{Q}^2{}_0 \omega _1{}^0 = 0, \quad \omega _{10} \hat{Q}^{30}-\hat{Q}^3{}_0 \omega _1{}^0 = 0,\\
& & \omega _{20} \hat{Q}^{10}-\hat{Q}^1{}_0 \omega _2{}^0 = 0,\quad \omega _{20} \hat{Q}^{20}-\hat{Q}^2{}_0 \omega _2{}^0 = 0, \quad \omega _{20} \hat{Q}^{30}-\hat{Q}^3{}_0 \omega _2{}^0 = 0,\nonumber\\
& & \hat{\omega }_{11} Q^{11}-Q^1{}_1 \hat{\omega }_1{}^1= 0, \quad \hat{\omega }_{21} Q^{11}-Q^1{}_1 \hat{\omega }_2{}^1= 0, \quad \hat{\omega }_{31} Q^{11}-Q^1{}_1 \hat{\omega }_3{}^1= 0, \nonumber\\
& & \hat{\omega }_{11} Q^{21}-Q^2{}_1 \hat{\omega }_1{}^1= 0, \quad \hat{\omega }_{21} Q^{21}-Q^2{}_1 \hat{\omega }_2{}^1= 0, \quad \hat{\omega }_{31} Q^{21}-Q^2{}_1 \hat{\omega }_3{}^1= 0, \nonumber
\eea
\bea
\label{eq:omegaQ2b}
& & 4 H_0 R_1-\omega _{10} Q^1{}_1-2 \omega _{20} Q^2{}_1+2 \hat{\omega }_{11} \hat{Q}^1{}_0-\hat{\omega }_{21} \hat{Q}^2{}_0+4 \hat{\omega }_{31}
   \hat{Q}^3{}_0 = 0, \nonumber\\
& & 4 H_0 R^1-\omega _{10} Q^{11}-2 \omega _{20} Q^{21} +2  \hat{\omega }_1{}^1 \hat{Q}^1{}_0- \hat{\omega }_2{}^1\hat{Q}^2{}_0+4 \hat{Q}^3{}_0 \hat{\omega }_3{}^1 = 0 ,\\
& & 4 H^0R_1  - \omega _1{}^0 Q^1{}_1 -2  \omega _2{}^0 Q^2{}_1 +2
   \hat{\omega }_{11} \hat{Q}^{10}-\hat{\omega }_{21} \hat{Q}^{20} +4 \hat{\omega }_{31} \hat{Q}^{30}= 0 , \nonumber\\
& & 4 H^0 R^1- \omega _1{}^0 Q^{11} -2  \omega _2{}^0 Q^{21} +2  \hat{\omega }_1{}^1 \hat{Q}^{10}- \hat{\omega }_2{}^1\hat{Q}^{20}+4  \hat{\omega }_3{}^1\hat{Q}^{30} = 0\, . \nonumber
\eea
\bea
\label{eq:omegaQ2c}
& & \hskip-0.75cm \omega _{10} \hat{Q}^1{}_0+\hat{Q}^{10} \omega _1{}^0 = 0, \quad  \omega _{10} \hat{Q}^2{}_0+\hat{Q}^{20} \omega _1{}^0 = 0, \quad  \omega _{20} \hat{Q}^1{}_0+\hat{Q}^{10} \omega _2{}^0 = 0, \quad \omega _{20} \hat{Q}^2{}_0+\hat{Q}^{20} \omega_{2}{}^0= 0 \, ,\nonumber\\
& & \hskip-0.75cm \hat{\omega }_{11} Q^1{}_1+Q^{11} \hat{\omega }_1{}^1=0, \quad \hat{\omega }_{21} Q^1{}_1+Q^{11} \hat{\omega }_2{}^1=0, \quad \hat{\omega }_{11} Q^2{}_1+Q^{21} \hat{\omega }_1{}^1=0, \quad \hat{\omega }_{21} Q^2{}_1+Q^{21} \hat{\omega }_2{}^1=0\, , \nonumber\\
& & \hskip-0.75cm 2 R_1 H^0+2 H_0 R^1-2 \hat{\omega }_{31} \hat{Q}^{30}-2 \hat{Q}^3{}_0 \hat{\omega }_3{}^1 = 0, \quad \quad 2 H_0 R_1-2 H^0 R^1-2 \hat{\omega }_{31} \hat{Q}^3{}_0+2 \hat{Q}^{30} \hat{\omega}_3{}^1=0\,, \nonumber\\
& & \omega_{10}Q^2{}_1+Q^{11} \omega_2{}^0 = \omega_{20} Q^1{}_1 + Q^{21}\omega_1{}^0 , \quad \hat{Q}^{10}\hat{\omega}_2{}^1+\hat{\omega}_{11} \hat{Q}^2{}_0 = \hat{Q}^{20}\hat{\omega}_1{}^1+\hat{\omega}_{21} \hat{Q}^1{}_0 \nonumber\\
& & \omega_{20}Q^{11}+Q^{1}{}_{1} \omega_2{}^0 = \omega_{10} Q^{21} + Q^{2}{}_{1}\omega_1{}^0 , \quad \hat{\omega}_{11} \hat{Q}^{20}+\hat{\omega}_{1}{}^{1} \hat{Q}^2{}_0 = \hat{\omega}_{21}\hat{Q}^{10}+\hat{\omega}_{2}{}^{1} \hat{Q}^1{}_0 \,.
\eea     
\end{subequations}
Now it is worth to emphasize that  first set of 12 constraints in eqn. (\ref{eq:omegaQ2a}) corresponds to  $\omega _{a \Lambda} \hat{Q}^{\alpha \Lambda}=\hat{Q}^\alpha{}_\Lambda \omega _a{}^\Lambda$ and $\hat{\omega} _{\alpha K} {Q}^{a K}={Q}^a{}_K \hat{\omega} _\alpha{}^K$ of the second formulation in eqn. (\ref{eq:BIs1}) while the ones given in eqn. (\ref{eq:omegaQ2b}) correspond to the remaining Bianchi identities of the second formulation in eqn. (\ref{eq:BIs2}). However, all the constraints of eqn. (\ref{eq:omegaQ2c}) are new to the second formulation. 

\subsubsection*{(iv). $(R \omega + Q^2)$-type identities:} When we consider the flux constraint ``$({\omega}_{mi}{}^{[\underline{j}}{}\,\, {R}^{\underline{kl}]m} \, \,- {Q}^{[\underline{jk}}{}_{m}{}  \, {Q}^{\underline{l}]m}{}_{i})=0$" from the Bianchi identities in the first formulation, after some careful reshuffling of the resulting set of constraints, we arrive at the followings
\begin{subequations}
\bea
\label{eq:ModelBromega1}
& & \hskip1cm \hat{\omega }_{11} R^1-R_1 \hat{\omega }_1{}^1=0, \quad \hat{\omega }_{21} R^1-R_1 \hat{\omega }_2{}^1=0, \quad \hat{\omega }_{31} R^1-R_1 \hat{\omega }_3{}^1=0 \\
& & \hskip-0.8cm Q^2{}_1 Q^{11}-Q^1{}_1 Q^{21}=0, \,\, \hat{Q}^2{}_0 \hat{Q}^{10}-\hat{Q}^1{}_0 \hat{Q}^{20}=0, \, \, \hat{Q}^2{}_0 \hat{Q}^{30}-\hat{Q}^3{}_0 \hat{Q}^{20}=0, \, \, \hat{Q}^3{}_0 \hat{Q}^{10}-\hat{Q}^1{}_0 \hat{Q}^{30}=0\, ,\nonumber
\eea
\bea
\label{eq:ModelBromega2}
& & \hskip-1.0cm \hat{Q}^3{}_0 Q^{11}+Q^1{}_1 \hat{Q}^{30}+\omega _{10} R^1+R_1 \omega _1{}^0=0, \quad \hat{Q}^3{}_0\, {Q}^1{}_1-Q^{11} \hat{Q}^{30}+R_1 \omega _{10}-R^1 \omega _1{}^0=0 \,, \\
& & \hskip-1.0cm \hat{Q}^3{}_0 Q^{21}+Q^2{}_1 \hat{Q}^{30}+\omega _{20} R^1+R_1 \omega _2{}^0=0, \quad \hat{Q}^3{}_0 \hat{Q}^2{}_1-Q^{21} \hat{Q}^{30}+R_1 \omega _{20} -R^1 \omega _2{}^0=0\,, \nonumber\\
& &  \hskip-1.2cm \hat{Q}^1{}_0 \hat{Q}^3{}_0+\hat{Q}^{10} \hat{Q}^{30} = 0, \,  \, \hat{Q}^2{}_0 \hat{Q}^3{}_0+\hat{Q}^{20} \hat{Q}^{30} = 0, \quad R^1 \hat{\omega }_1{}^1+R_1 \hat{\omega }_{11}=0, \, \, R^1 \hat{\omega }_2{}^1+R_1 \hat{\omega }_{21}=0\nonumber\\
& & \hskip-0.5cm \left(Q^1{}_1\right)^2+2 \left(Q^2{}_1\right)^2+2 \left(\hat{Q}^1{}_0\right)^2-\left(\hat{Q}^2{}_0\right)^2+\left(Q^{11}\right)^2+2 \left(Q^{21}\right)^2+2
   \left(\hat{Q}^{10}\right)^2-\left(\hat{Q}^{20}\right)^2\nonumber\\
& & \hskip2cm +8 R^1 \hat{\omega }_3{}^1+8 R_1 \hat{\omega }_{31}=0\, .\nonumber
\eea
\end{subequations}
Now it is clearly seen that the first collection (\ref{eq:ModelBromega1}) corresponds to the constraints of second formulation (\ref{eq:BIs1}) while those in eqn. (\ref{eq:ModelBromega2}) are new constraints.

\subsubsection*{(v). $(R Q)$-type identities:}
Finally, in this case we find that
\bea
\label{eq:modelBbis5}
& &  {Q}^{[\underline{ij}}{}_{m}{}\, {R}^{\underline{kl}]m}=0 \Longrightarrow \biggl\{ R^1 \, Q^1{}_1 = R_1 \, Q^{11}\, , \quad R^1 \, Q^2{}_1 = R_1 \, Q^{21} \biggr\}\, 
\eea
We note that this collection represents the same set of flux constraints as the one coming from identity ``$R^K \, Q_{K}^a - R_K \, Q^{aK} =0$'' of the second formulation.

Let us collect an equivalent set of all the NS-NS Bianchi identities obtained by translating all the constraints of first formulation into cohomology indices. We find a total 66 coupled constraints among 24 flux parameters as can be classified into two parts below,\\

{\bf \hskip-0.7cm A subset of constraints equivalent to the second formulation:}
\bea
\label{eq:ModelBbisSecond}
& & \hskip-1cm H_0 \, \omega_1{}^0 = H^0 \, \omega_{10}\, , \quad H_0 \, \omega_2{}^0 = H^0 \, \omega_{20}, \nonumber\\
& & \hskip-1cm H_0 \hat{Q}^{10}=H^0 \hat{Q}^1{}_0, \quad  H_0 \hat{Q}^{20}=H^0 \hat{Q}^2{}_0, \quad H_0 \hat{Q}^{30}=H^0 \hat{Q}^3{}_0, \quad \omega _{20} \omega _1{}^0=\omega _{10} \omega _2{}^0 , \nonumber\\
& & \hskip-1cm \hat{\omega }_{21} \hat{\omega }_1{}^1=\hat{\omega }_{11} \hat{\omega }_2{}^1, \quad \hat{\omega }_{31} \hat{\omega }_1{}^1=\hat{\omega }_{11} \hat{\omega }_3{}^1, \quad \hat{\omega }_{21} \hat{\omega }_3{}^1=\hat{\omega }_{31} \hat{\omega }_2{}^1, \quad \omega _{10} \hat{Q}^{10}=\hat{Q}^1{}_0 \omega _1{}^0, \nonumber\\
& & \hskip-1cm \omega _{10} \hat{Q}^{20}=\hat{Q}^2{}_0 \omega _1{}^0, \quad \omega _{10} \hat{Q}^{30}=\hat{Q}^3{}_0 \omega _1{}^0, \quad \omega _{20} \hat{Q}^{10}=\hat{Q}^1{}_0 \omega _2{}^0,\quad \omega _{20} \hat{Q}^{20}=\hat{Q}^2{}_0 \omega _2{}^0,\nonumber\\
& & \hskip-1cm \omega _{20} \hat{Q}^{30}=\hat{Q}^3{}_0 \omega _2{}^0, \quad \hat{\omega }_{11} Q^{11}=Q^1{}_1 \hat{\omega }_1{}^1, \quad \hat{\omega }_{21} Q^{11}=Q^1{}_1 \hat{\omega }_2{}^1, \quad \hat{\omega }_{31} Q^{11}=Q^1{}_1 \hat{\omega }_3{}^1, \nonumber\\
& & \hskip-1cm \hat{\omega }_{11} Q^{21}=Q^2{}_1 \hat{\omega }_1{}^1, \quad \hat{\omega }_{21} Q^{21}=Q^2{}_1 \hat{\omega }_2{}^1, \quad \hat{\omega }_{31} Q^{21}=Q^2{}_1 \hat{\omega }_3{}^1, \quad \hat{\omega }_{11} R^1=R_1 \hat{\omega }_1{}^1, \nonumber\\
& & \hskip-1cm \hat{\omega }_{21} R^1=R_1 \hat{\omega }_2{}^1, \quad \hat{\omega }_{31} R^1=R_1 \hat{\omega }_3{}^1, \quad Q^2{}_1 Q^{11}=Q^1{}_1 Q^{21}, \quad \hat{Q}^2{}_0 \hat{Q}^{10}=\hat{Q}^1{}_0 \hat{Q}^{20}, \nonumber\\
& & \hskip-1cm \hat{Q}^2{}_0 \hat{Q}^{30}=\hat{Q}^3{}_0 \hat{Q}^{20}, \, \, \hat{Q}^3{}_0 \hat{Q}^{10}=\hat{Q}^1{}_0 \hat{Q}^{30}\, \quad R^1 \, Q^1{}_1 = R_1 \, Q^{11}\, , \quad R^1 \, Q^2{}_1 = R_1 \, Q^{21} ,\nonumber\\
& & \hskip-1cm 4 H_0 R_1-\omega _{10} Q^1{}_1-2 \omega _{20} Q^2{}_1+2 \hat{\omega }_{11} \hat{Q}^1{}_0-\hat{\omega }_{21} \hat{Q}^2{}_0+4 \hat{\omega }_{31}
   \hat{Q}^3{}_0 = 0, \\
& & \hskip-1cm 4 H_0 R^1-\omega _{10} Q^{11}-2 \omega _{20} Q^{21} +2  \hat{\omega }_1{}^1 \hat{Q}^1{}_0- \hat{\omega }_2{}^1\hat{Q}^2{}_0+4 \hat{Q}^3{}_0 \hat{\omega }_3{}^1 = 0 ,\nonumber\\
& & \hskip-1cm 4 H^0R_1  - \omega _1{}^0 Q^1{}_1 -2  \omega _2{}^0 Q^2{}_1 +2
   \hat{\omega }_{11} \hat{Q}^{10}-\hat{\omega }_{21} \hat{Q}^{20} +4 \hat{\omega }_{31} \hat{Q}^{30}= 0 , \nonumber\\
& & \hskip-1cm 4 H^0 R^1- \omega _1{}^0 Q^{11} -2  \omega _2{}^0 Q^{21} +2  \hat{\omega }_1{}^1 \hat{Q}^{10}- \hat{\omega }_2{}^1\hat{Q}^{20}+4  \hat{\omega }_3{}^1\hat{Q}^{30} = 0.\, \nonumber
\eea
{\bf \hskip-0.0cm  A subset of constraints beyond the second formulation:}
\bea
\label{eq:ModelBbisNotSecond}
& & \hskip-1cm H_0 \hat{Q}^{1}{}_{0}+H^0 \hat{Q}^{10} = 0, \, \, H_0 \hat{Q}^{2}{}_{0}+H^0 \hat{Q}^{20} = 0, \, \, \hat{\omega }_1{}^1 \hat{\omega }_3{}^1+\hat{\omega }_{11} \hat{\omega }_{31}=0, \, \, \hat{\omega }_2{}^1 \hat{\omega }_3{}^1+\hat{\omega }_{21} \hat{\omega }_{31}=0, \nonumber\\
& & \hskip-1cm \omega _{10} \hat{Q}^1{}_0+\hat{Q}^{10} \omega _1{}^0 = 0, \, \, \omega _{10} \hat{Q}^2{}_0+\hat{Q}^{20} \omega _1{}^0 = 0, \, \,\omega _{20} \hat{Q}^1{}_0+\hat{Q}^{10} \omega _2{}^0 = 0, \, \, \omega _{20} \hat{Q}^2{}_0+\hat{Q}^{20} \omega_{2}{}^0= 0 \, ,\nonumber\\
& & \hskip-1cm \hat{\omega }_{11} Q^1{}_1+Q^{11} \hat{\omega }_1{}^1=0, \, \, \hat{\omega }_{21} Q^1{}_1+Q^{11} \hat{\omega }_2{}^1=0, \, \, \hat{\omega }_{11} Q^2{}_1+Q^{21} \hat{\omega }_1{}^1=0, \, \, \hat{\omega }_{21} Q^2{}_1+Q^{21} \hat{\omega }_2{}^1=0\, , \nonumber\\
& &  \hskip-1cm \hat{Q}^1{}_0 \hat{Q}^3{}_0+\hat{Q}^{10} \hat{Q}^{30} = 0, \,  \, \hat{Q}^2{}_0 \hat{Q}^3{}_0+\hat{Q}^{20} \hat{Q}^{30} = 0, \, \, R^1 \hat{\omega }_1{}^1+R_1 \hat{\omega }_{11}=0, \, \, R^1 \hat{\omega }_2{}^1+R_1 \hat{\omega }_{21}=0\nonumber\\
& & \hskip-1cm H^0 Q^1{}_1+H_0 Q^{11}+\hat{\omega }_{31} \omega _1{}^0+\omega _{10} \hat{\omega }_3{}^1 =0, \quad H^0 Q^2{}_1+H_0 Q^{21}+\hat{\omega }_{31} \omega _2{}^0+\omega _{20} \hat{\omega }_3{}^1=0\, , \nonumber\\
& & \hskip-1.0cm \hat{Q}^{30} Q^1{}_1+\hat{Q}^3{}_0 Q^{11}+R_1 \omega _1{}^0+\omega _{10} R^1=0, \quad \hat{Q}^{30} Q^2{}_1 +\hat{Q}^3{}_0 Q^{21}+R_1 \omega _2{}^0 +\omega _{20} R^1=0  \,, \\
& & \hskip-1cm H_0 Q^1{}_1-H^0 Q^{11}-\omega _1{}^0 \hat{\omega }_3{}^1+\omega _{10} \hat{\omega}_{31}=0, \quad H_0 Q^2{}_1-H^0 Q^{21}-\omega _2{}^0 \hat{\omega }_3{}^1+\omega _{20} \hat{\omega}_{31}=0\, , \nonumber\\
& & \hskip-1.0cm \hat{Q}^3{}_0\, {Q}^1{}_1-Q^{11} \hat{Q}^{30}-R^1 \omega _1{}^0+R_1 \omega _{10}=0, \quad \hat{Q}^3{}_0 {Q}^2{}_1-Q^{21} \hat{Q}^{30} -R^1 \omega _2{}^0+R_1 \omega _{20}=0\,, \nonumber\\
& & \hskip-1cm 2 R_1 H^0+2 H_0 R^1-2 \hat{\omega }_{31} \hat{Q}^{30}-2 \hat{Q}^3{}_0 \hat{\omega }_3{}^1 = 0, \quad \quad 2 H_0 R_1-2 H^0 R^1-2 \hat{\omega }_{31} \hat{Q}^3{}_0+2 \hat{Q}^{30} \hat{\omega}_3{}^1=0\,, \nonumber\\
& & \omega_{10}Q^2{}_1+Q^{11} \omega_2{}^0 = \omega_{20} Q^1{}_1 + Q^{21}\omega_1{}^0 , \quad \hat{Q}^{10}\hat{\omega}_2{}^1+\hat{\omega}_{11} \hat{Q}^2{}_0 = \hat{Q}^{20}\hat{\omega}_1{}^1+\hat{\omega}_{21} \hat{Q}^1{}_0 \nonumber\\
& & \omega_{20}Q^{11}+Q^{1}{}_{1} \omega_2{}^0 = \omega_{10} Q^{21} + Q^{2}{}_{1}\omega_1{}^0 , \quad \hat{\omega}_{11} \hat{Q}^{20}+\hat{\omega}_{1}{}^{1} \hat{Q}^2{}_0 = \hat{\omega}_{21}\hat{Q}^{10}+\hat{\omega}_{2}{}^{1} \hat{Q}^1{}_0 \,\nonumber\\
& & \hskip-1cm 8 H_0 \hat{Q}^3{}_0+8 H^0 \hat{Q}^{30}+\left(\omega _1{}^0\right){}^2+2 \left(\omega _2{}^0\right){}^2+2 \left(\hat{\omega }_1{}^1\right){}^2-\left(\hat{\omega }_2{}^1\right){}^2+\omega _{10}^2+2 \omega _{20}^2+2 \hat{\omega }_{11}^2-\hat{\omega }_{21}^2=0\, ,\nonumber\\
& & \hskip-0.5cm \left(Q^1{}_1\right)^2+2 \left(Q^2{}_1\right)^2+2 \left(\hat{Q}^1{}_0\right)^2-\left(\hat{Q}^2{}_0\right)^2+\left(Q^{11}\right)^2+2 \left(Q^{21}\right)^2+2
   \left(\hat{Q}^{10}\right)^2-\left(\hat{Q}^{20}\right)^2\nonumber\\
& & \hskip2cm +8 R^1 \hat{\omega }_3{}^1+8 R_1 \hat{\omega }_{31}=0\, .\nonumber
\eea
Now we emphasize that the collections of constraints in eqn. (\ref{eq:ModelBbisNotSecond}) is indeed `effectively new' in the sense that they are non-trivial on top of imposing the identities of eqn. (\ref{eq:ModelBbisSecond}), in the cases otherwise there will be no subtlety in the two formulations. To illustrate the same, one can (at least) consider the `special solution' which we have discussed earlier. In that case, after setting half of the fluxes (say with with upper (2,1) cohomology indices) to zero, one finds that only one constraint in (\ref{eq:ModelBbisSecond}) survives to be non-trivial while there still remain many distinct non-trivial constraints in the collection  (\ref{eq:ModelBbisNotSecond}). This is the quick illustration with the special solutions, and for more general solutions, one may rather expect more non-trivial impact of the `additional' constraints. 

Moreover, let us mention that similar to the case of \cite{Robbins:2007yv}, we have also allowed non-zero values for the fluxes $\omega_{ij}{}^j$ and $Q^{ij}{}_j$ which involve only one free six-dimensional index. However, such flux components have been argued to be set to zero in \cite{Shelton:2005cf}, and so it is interesting to investigate the influence of setting such flux components to zero, and to check if they could wipe out the discrepancy between the two sets of Bianchi identity formulations \footnote{We thank the referee for asking us to investigate on this important point.}. If we demand these fluxes to be zero (i.e. $\omega_{ij}{}^j = 0, \, Q^{ij}{}_j =0$) in the flux conversion relations in eqn. (\ref{eq:fluxconversionB}), we find that 
\bea
\label{eq:fluxconversionB123}
& & \hskip-0.3cm \omega_{a}{}^{0} \equiv \left(\begin{array}{c}  - \omega_{25}{}^1 + \omega_{26}{}^1\\ -\omega_{36}{}^1 - \omega_{45}{}^1\end{array}\right) \,, \qquad \qquad \omega_{a0} \equiv \left(\begin{array}{c} -\omega_{25}{}^1 - \omega_{26}{}^1\\ \omega_{35}{}^1 -\omega_{46}{}^1\end{array}\right), \nonumber\\
& & \hskip-0.4cm \hat{\omega}_{\alpha}{}^{1} \equiv \left(\begin{array}{c}\omega_{35}{}^1 +\omega_{46}{}^1\\  -\omega_{25}{}^1 - \omega_{26}{}^1\\ -\omega_{13}{}^5\end{array}\right), \qquad \qquad \hat{\omega}_{\alpha 1} \equiv \left(\begin{array}{c}\omega_{36}{}^1 -\omega_{45}{}^1\\ \omega_{25}{}^1 - \omega_{26}{}^1\\ \omega_{14}{}^5\end{array}\right)\, , \\
& & \hskip-0.4cm \hat{Q}^{\alpha 0} \equiv \left(\begin{array}{c} -Q^{15}{}_3 + Q^{16}{}_4 \\  Q^{15}{}_2- Q^{16}{}_2 \\ Q^{13}{}_5\end{array}\right), \quad \qquad \hat{Q}^{\alpha}{}_{0} \equiv \left(\begin{array}{c} -Q^{15}{}_4 - Q^{16}{}_3 \\  Q^{15}{}_2  + Q^{16}{}_2 \\ Q^{13}{}_6\end{array}\right), \,  \nonumber\\
& & \hskip-0.4cm {Q}^{a1} \equiv \left(\begin{array}{c}  - Q^{15}{}_2  - Q^{16}{}_2 \\ -Q^{15}{}_4 + Q^{16}{}_3\end{array}\right), \qquad \qquad  {Q}^{a}{}_1 \equiv \left(\begin{array}{c}  Q^{15}{}_2- Q^{16}{}_2 \\ -Q^{15}{}_3 - Q^{16}{}_4\end{array}\right) \,. \nonumber
\eea
This simplification is equivalent to reduce the number of independent flux components (allowed by the orientifold structure) from 10 to 8 for each of the fluxes $\omega$ and $Q$. Moreover, this does not influence the presence of bijection between the flux components counted via real indices and cohomology indices. To be more specific, the effect of setting $\omega_{ij}{}^j = 0$ and  $Q^{ij}{}_j =0$ amounts to have the following additional constraints on the cohomology indexed fluxes,
\bea
\label{eq:wiijQiij}
& & \omega_{15}^1 = 0 = \omega_{16}^1 \quad \Longleftrightarrow \quad \omega_1{}^0 =- \hat{\omega}_{21}, \quad \omega_{10} = \hat{\omega}_2{}^1 \\
& & Q^{15}_1 = 0 =Q^{16}_1 \quad \Longleftrightarrow \quad Q^1{}_1 = \hat{Q}^{20}, \quad Q^{11} = - \hat{Q}^2{}_0 \nonumber
\eea
Now it clear that setting $\omega_{15}^1$ and $\omega_{16}^1$ fluxes to zero reduces the independent components of $\omega_{ij}{}^k$-type fluxes from 10 to 8, which effectively reduces two independent flux component of cohomology indices also. The same thing happens to be true with the $Q$-flux components as well. Although these additional flux constraints in eqn. (\ref{eq:wiijQiij}) simplify the set of identities given in eqns. (\ref{eq:ModelBbisSecond})-(\ref{eq:ModelBbisNotSecond}) a bit further, however the same do not help in removing the extra Bianchi identities of the first formulations which are not covered by the known version of the second formulation. To illustrate the same, let us again consider the `special solution' of Bianchi identities and assume that fluxes with upper (2,1)-cohomology indices are set to zero. Subsequently the identities in  eqns. (\ref{eq:ModelBbisSecond})-(\ref{eq:ModelBbisNotSecond}) after imposing the conditions in eqn. (\ref{eq:wiijQiij}) are reduced into the following constraints, 
\bea
\label{eq:wiijQiijRestBIs}
& & \hskip-0.5cm H_0 R_1=\hat{\omega }_{31} \hat{Q}^3{}_0, \quad R_1 \hat{\omega }_{11}=0, \quad \hat{\omega }_{11} \hat{\omega }_{31}=0, \quad \hat{\omega }_{11} Q^2{}_1=0, \quad H_0
   Q^2{}_1+\omega _{20} \hat{\omega }_{31}=0, \\
& & \hskip-0.5cm H_0 \hat{Q}^1{}_0=0,\omega _{20} \hat{Q}^1{}_0=0, \quad \left(Q^2{}_1\right)^2+ \left(\hat{Q}^1{}_0\right)^2+4 R_1\hat{\omega }_{31}=0, \quad \hat{Q}^1{}_0 \hat{Q}^3{}_0=0,\nonumber\\
& & \hskip-0.5cm 4 H_0 \hat{Q}^3{}_0+ \omega _{20}^2+ \hat{\omega }_{11}^2=0, \quad Q^2{}_1 \hat{Q}^3{}_0+R_1 \omega _{20}=0, \quad 2 H_0 R_1+ \hat{\omega }_{11} \hat{Q}^1{}_0+2 \hat{\omega }_{31} \hat{Q}^3{}_0= \omega _{20} Q^2{}_1\, ,\nonumber
\eea
where except for the last constraint, all the other ones are additionally `new' ones  and do not follow from the known version of the second formulation. This has been the case for (at least) the simplest non-trivial solutions, namely the `special solutions' which serves as a proof for the mismatch, and we expect same results for more complicated structure with the more generic cases. Thus, we show that the mismatch between two formulations still persists even after imposing $\omega_{ij}{}^j = 0$ and $Q^{ij}{}_j =0$. Moreover, this observation will also be supported from our next toroidal example which does not have any allowed flux components of type $\omega_{ij}{}^j$ and $Q^{ij}{}_j$ by the orientifold construction itself, however the mismatch between the two formulations remain intact as we will see later on.

\subsubsection*{Some interesting observations}
Now let us point out some interesting observations from these collection of constraints given in eqns. (\ref{eq:ModelBbisSecond})-(\ref{eq:ModelBbisNotSecond}),
\begin{itemize}
\item{As we have already shown generically in eqns. (\ref{eq:newOrbitsBIs1})-(\ref{eq:newOrbitsBIs2}), the first collection of Bianchi identities given in eqn. (\ref{eq:ModelBbisSecond}) holds with our new flux orbits defined in eqns.  (\ref{eq:OddOrbitA})-(\ref{eq:OddOrbitB}). Moreover, it is interesting to find that the second collection (as a whole as given) in eqn. (\ref{eq:ModelBbisNotSecond}) also holds for our new flux orbits. This can be verified by considering the simplified version of the new flux orbits which are given as under,
\bea
\label{eq:orbitsB2Ex2}
& & \hskip-1.2cm {\mathbb H}_0 = H_0 + (\omega_{01} \, {b}_1+\omega_{02} \, {b}_2) - \hat{Q}^3{}_0 \, \left(2\, b_1^2 + b_2^2\right), \quad  {\mathbb \mho}_{01} = \omega_{01} -4\, \hat{Q}^3{}_0 \, b_1,\,   \\
& & \hskip-1.2cm  {\mathbb \mho}_{02} = \omega_{02} - 2\, \hat{Q}^3{}_0 \, b_2, \qquad {\hat{\mathbb Q}^{1}{}_{0}} = \hat{Q}^{1}{}_{0} , \, {\hat{\mathbb Q}^{2}{}_{0}} = \hat{Q}^{2}{}_{0}, \, {\hat{\mathbb Q}^{3}{}_{0}} = \hat{Q}^{3}{}_{0} \, ,\nonumber\\
& & \hskip-1.2cm {\mathbb R}_1= R_1, \quad  {\mathbb Q}^1_1 = Q^1_1+4\,R_1, \quad {\mathbb Q}^2_1 = Q^2_1 +2\, R_1\, ,  \nonumber\\
& & \hskip-1.2cm  \hat{\mho}_{11} = \hat{\omega}_{11},\, \, \hat{\mho}_{21} = \hat{\omega}_{21}, \, \hat{\mho}_{31} = \hat{\omega}_{31} - \left( Q_1{}^1 b_1 + Q_1{}^2 b_2 \right) - R_1 (2 \, b_1^2 + b_2^2 ) \nonumber
\eea
where the new flux orbits with upper even/odd (2,1)-cohomology indices `0' and `1' can be analogously written. So this analysis serves as another evidence for a very natural relevance of our new generalized flux orbits which were conjectured in \cite{Shukla:2015rua}. Moreover as we will explain later, the Bianchi identities written in terms of new flux orbits will be directly useful for removing many terms from the compact rearrangement of the scalar potential proposed in \cite{Shukla:2015rua, Shukla:2015hpa,Shukla:2016hyy}. Given the coupled and complicated nature of the identities, this inference, about the same being extended to hold even with the new flux-orbits, should not be limited to this particular example, and therefore one may expect that the first formulation identities in eqn. (\ref{eq:bianchids1}) should also be invariant under the generalized flux orbits written in six-dimensional real indices as proposed in \cite{Blumenhagen:2013hva}.
}
\item{There is a symmetry within the set of total constraints in eqns. (\ref{eq:ModelBbisSecond})-(\ref{eq:ModelBbisNotSecond}) which remains unchanged under the following transformations (when applied collectively),
\bea 
& & \hskip-1.5cm H_0 \leftrightarrow R_1, \quad H^0 \leftrightarrow R^1, \quad \omega_{10} \leftrightarrow  Q^1{}_1, \quad\omega_{20} \leftrightarrow Q^2{}_1, \quad \omega_{1}{}^{0} \leftrightarrow  Q^{11}, \quad\omega_{2}{}^{0} \leftrightarrow Q^{21}, \\
& & \hskip-1.5cm \hat{Q}^{1}{}_{0} \leftrightarrow \hat{\omega}_{11}, \quad \hat{Q}^{2}{}_{0} \leftrightarrow \hat{\omega}_{21}, \quad \hat{Q}^{3}{}_{0} \leftrightarrow \hat{\omega}_{31}, \quad \hat{Q}^{10} \leftrightarrow \hat{\omega}_{1}{}^{1}, \quad \hat{Q}^{20} \leftrightarrow \hat{\omega}_{2}{}^{1}, \quad \hat{Q}^{30} \leftrightarrow \hat{\omega}_{3}{}^{1} \nonumber
\eea
}
\end{itemize}
Before we come to the next example, let us make some speculations about a compact version of the `additional' constraints in eqn. (\ref{eq:ModelBbisNotSecond}). A more careful observation shows that using new flux-orbits, we can reshuffle the same into the following constraints with parameters being written into cohomology language, 
\bea 
\label{eq:ModelBbisNotSecondnew}
& & \hskip-0.0cm {\mathbb H}_0 \, \hat{\mathbb Q}^\alpha{}_0+ {\mathbb H}^0 \, \hat{\mathbb Q}^{\alpha0} = 0, \qquad \qquad \hat\mho_{3 1}\, \hat\mho_{\alpha 1} + \hat\mho_3^1\, \hat\mho_\alpha^1 = 0, \qquad \qquad \quad \forall\,  \alpha \in \{1,2\}\, \nonumber\\
& & \hskip-1cm 2\, (\hat{d}^{-1})^3{}_\alpha\, \left({\mathbb H}_0 \, \hat{\mathbb Q}^\alpha{}_0+ {\mathbb H}^0 \, \hat{\mathbb Q}^{\alpha0} \right) + \left({k}_{3 \beta\gamma}\right)^{-1} \left(\hat\mho_{\beta 1}\, \hat\mho_{\gamma 1} + \hat\mho_\beta^1\, \hat\mho_\gamma^1 \right) = \left(\hat{k}_{3 ab}\right)^{-1} \left(\mho_{a0}\, \mho_{b0} + \mho_a^0\, \mho_b^0 \right), \nonumber\\
& & (d^{-1})^a_b \left({\mathbb H}_0\, {\mathbb Q}^{b1}+{\mathbb H}^0\, {\mathbb Q}^b_1\right) + \left(\hat{k}_{\alpha ab}\right)^{-1} \left(\hat{\mho}_{\alpha 1} \mho_{b}^{0} + {\hat\mho}_\alpha^1 \, \mho_{b0}\right) = 0\, , \qquad \forall\,  \, a \nonumber\\
& & (d^{-1})^a_b \left({\mathbb H}^0\, {\mathbb Q}^{b1} - {\mathbb H}_0\, {\mathbb Q}^b_1\right) + \left(\hat{k}_{\alpha ab}\right)^{-1} \left(\hat{\mho}_{\alpha}^{1} \mho_{b}^{0} - {\hat\mho}_{\alpha 1} \, \mho_{b0}\right) = 0\, , \qquad \forall\, \, a \\
& & \nonumber\\
& & {\mathbb R}_1 \, \hat{\mho}_{\alpha 1}+ {\mathbb R}^1 \, \hat{\mho}_{\alpha}^1 = 0, \qquad \qquad \, \, \hat{\mathbb Q}^{3}_0\, \hat{\mathbb Q}^{\alpha}_{0} + \hat{\mathbb Q}^{30}\, \hat{\mathbb Q}^{\alpha0} = 0, \qquad \qquad \quad \forall\,  \alpha \in \{1,2\}\, \nonumber\\
& & \hskip-0.0cm 2\, (f^{-1})\, \left({\mathbb R}_1 \, \hat{\mho}_{3 1}+ {\mathbb R}^1 \, \hat{\mho}_{3}^1 \right) \, - \hat{k}_{3 ab} \, {(d^{-1})}^a_{c}\, {(d^{-1})}^b_{d} \left({\mathbb Q}^{c}_{1}\, {\mathbb Q}^{d}_{1} + {\mathbb Q}^{c1}\, {\mathbb Q}^{d1} \right) \,\nonumber\\
& &  \hskip1.99cm + {k}_{3 \beta\gamma}\, (\hat{d}^{-1})^\beta{}_{\beta'}\, (\hat{d}^{-1})^\gamma{}_{\gamma'} \, \left(\hat{{\mathbb Q}}^{\beta'}_{0}\, \hat{{\mathbb Q}}^{\gamma'}_{0} + \hat{{\mathbb Q}}^{\beta'0}\, \hat{{\mathbb Q}}^{\gamma' 0} \right) = 0\, .\nonumber\\
& & f^{-1} \left({\mathbb R}_1 \mho_{a}^0 + {\mathbb R}^1\mho_{a0}\right) + ({\hat{d}}^{-1})_\alpha^{\alpha'}\, (d^{-1})^b_c \, \, \, \hat{k}_{\alpha' a b} \left(\hat{\mathbb Q}^{\alpha 0} {\mathbb Q}^{c}_1 + \hat{\mathbb Q}^\alpha_0\, {\mathbb Q}^{c1} \right) = 0, \qquad \forall\, \, a \nonumber\\
& & f^{-1} \left({\mathbb R}^1 \mho_{a}^0 - {\mathbb R}_1\mho_{a0}\right) + ({\hat{d}}^{-1})_\alpha^{\alpha'}\, (d^{-1})^b_c \, \, \,\hat{k}_{\alpha' a b} \left(\hat{\mathbb Q}^{\alpha 0} {\mathbb Q}^{c1} - \hat{\mathbb Q}^\alpha_0\, {\mathbb Q}^{c}_{1} \right) = 0, \qquad \forall\, \, a \nonumber
\eea
Given that some of these identities do not follow the same compact version for all $\alpha$'s, the above cohomological representations of identities do not appear to be possibly promoted into a model independent language in a direct way. Nevertheless on that motivation, from eqn. (\ref{eq:ModelBbisNotSecondnew}) we indeed find that the following two identities, which take a kind of rather model independent form, still hold,
\bea 
\label{eq:ModelBbisNotSecondnewV2}
& & \hskip-1cm 2\, (\hat{d}^{-1})^\alpha{}_\beta\, \left({\mathbb H}_0 \, \hat{\mathbb Q}^\beta{}_0+ {\mathbb H}^0 \, \hat{\mathbb Q}^{\beta0} \right) + \left({k}_{\alpha \beta\gamma}\right)^{-1} \left(\hat\mho_{\beta 1}\, \hat\mho_{\gamma 1} + \hat\mho_\beta^1\, \hat\mho_\gamma^1 \right) = \left(\hat{k}_{\alpha ab}\right)^{-1} \left(\mho_{a0}\, \mho_{b0} + \mho_a^0\, \mho_b^0 \right),  \forall \alpha \nonumber\\
& & \hskip-0.0cm 2\, (f^{-1})\, \left({\mathbb R}_1 \, \hat{\mho}_{\alpha 1}+ {\mathbb R}^1 \, \hat{\mho}_{\alpha}^1 \right) \, - \hat{k}_{\alpha ab} \, {(d^{-1})}^a_{c}\, {(d^{-1})}^b_{d} \left({\mathbb Q}^{c}_{1}\, {\mathbb Q}^{d}_{1} + {\mathbb Q}^{c1}\, {\mathbb Q}^{d1} \right) \,\\
& &  \hskip1.99cm + {k}_{\alpha \beta\gamma}\, (\hat{d}^{-1})^\beta{}_{\beta'}\, (\hat{d}^{-1})^\gamma{}_{\gamma'} \, \left(\hat{{\mathbb Q}}^{\beta'}_{0}\, \hat{{\mathbb Q}}^{\gamma'}_{0} + \hat{{\mathbb Q}}^{\beta'0}\, \hat{{\mathbb Q}}^{\gamma' 0} \right) = 0\, , \quad \quad \quad \forall \alpha \,.\nonumber
\eea
This way we have indeed managed to rewrite the additional identities in the cohomological language, however we do not claim these two generic looking identities to be true for arbitrary (rigid) compactifications, and for the time being these speculative relations should be considered to be valid only for this particular example.  Nevertheless, there is some underlying structure within these two identities, and therefore it would be interesting to check for their general validity via a more fundamental route.

\subsection{Model B: Type IIB $\hookrightarrow$ ${\mathbb T}^6/({\mathbb Z}_2\times{\mathbb Z}_2$)-orientifold}
\subsubsection*{Fixing the conventions} Let us briefly revisit the relevant features of a setup within type IIB superstring theory compactified on  $ {\mathbb T}^6 / \left(\mathbb Z_2\times \mathbb Z_2\right)$ orientifold. The complex coordinates $z_i$'s on each of the tori in ${\mathbb T}^6={\mathbb T}^2\times {\mathbb T}^2\times {\mathbb T}^2$ are defined as
\bea
z^1=x^1+ U_1  \, x^2, ~ z^2=x^3+ U_2 \, x^4,~ z^3=x^5+ U_3 \, x^6 ,
\eea
where the three complex structure moduli $U_i$'s can be written as
$U_i= v_i + i\, u_i,\,\,i=1,2,3$. Further, the total orientifold action is given by the two $\mathbb Z_2$ orbifold actions, and an involution $I_6$ being being defined as
\bea
\label{thetaactions}
& & \hskip-1.0cm \theta(z^1,z^2,z^3)= (-z^1,-z^2,z^3), \, \ov\theta(z^1,z^2,z^3)= (z^1,-z^2,-z^3), \, I_6(z^1,z^2,z^3)=(-z^1,-z^2,-z^3),\nonumber
\eea
resulting in a setup with the presence of $O3/O7$-plane. The complex structure moduli dependent pre-potential is given as,
\bea
& & {\cal F} = \frac{{\cal X}^1 \, {\cal X}^2 \, {\cal X}^3}{{\cal X}^0} = U_1 \, U_2 \, U_3\,.
\eea 
Subsequently the holomorphic three-form $\Omega_3=dz^1\wedge dz^2\wedge dz^3$ can be expanded as,
\bea
& & \hskip-1.5cm  \Omega_3\,  = \alpha_0 +  \, U_1 \, \alpha_1 + U_2 \, \alpha_2 + U_3 \alpha_3 \, +\, U_1 \, U_2 \,  U_3 \, \beta^0 -U_2 \, U_3 \, \beta^1- U_1 \, U_3 \, \beta^2 - U_1\, U_2 \, \beta^3 \, .
\eea
Here we have chosen the following basis of closed three-forms
\bea
\label{formbasis}
& &  \hskip-0.5cm \alpha_0=1\wedge 3\wedge 5\,, \, \alpha_1=2\wedge 3\wedge 5\, , \,  \alpha_2=1\wedge 4\wedge 5\, , \, \alpha_3=1\wedge 3\wedge 6, \nonumber\\
& & \hskip-0.5cm \beta^0= 2\wedge 4\wedge 6\, , \, \beta^1= -\, 1\wedge 4\wedge 6\, ,\,   \beta^2=-\, 2\wedge 3\wedge 6\, ,\,   \beta^3=-\, 2\wedge 4\wedge 5 \, , \nonumber
\eea
where $1\wedge 3\wedge 5 = dx^1\wedge dx^3\wedge dx^5$ etc. Now, the basis of orientifold even two-forms and four-forms are as under,
\bea
& & \hskip3cm \mu_1 = dx^1 \wedge dx^2,  \quad \mu_2 = dx^3 \wedge dx^4, \quad  \mu_3 = dx^5 \wedge dx^6 \\
& & \hskip-0.7cm \tilde{\mu}^1 = dx^3 \wedge dx^4 \wedge dx^5 \wedge dx^6,  \quad \tilde{\mu}^2 = dx^1 \wedge dx^2 \wedge dx^5 \wedge dx^6, \quad  \tilde{\mu}^3 = dx^1 \wedge dx^2 \wedge dx^3 \wedge dx^4 \nonumber
\eea
implying that $\hat{d}_\alpha{}^\beta = \delta_\alpha{}^\beta$. The only non-trivial triple intersection number ($\kappa_{\alpha\beta\gamma}$) is given as $\kappa_{123} = 1$ which implies the volume form of the sixfold to be ${\cal V}_E = t_1 \, t_2 t_3$ and so the four cycle volume moduli are given as,
$
\sigma_1 =  \, t_2 \, t_3, \,\,\sigma_2 =  \, t_3 \, t_1, \, \, \,\,\sigma_3 =   \, t_1 \, t_2\, .
$
\subsubsection*{Flux conversion relations}
\bea
\label{eq:fluxconversionA1}
& & H_0 = H_{246}, \, \quad H_1 = -H_{146}, \, \quad H_2 = -H_{236}, \, \quad H_3 = -H_{245}, \\
& & H^0 = H_{135}, \, \quad H^1 = H_{235}, \, \quad H^2 = H_{145}, \, \quad H^3 = H_{136} \, ,\nonumber
\eea
and similarly for $F_3$-flux. In addition we have the following relations for the $Q$-flux components,
\bea
\label{eq:fluxconversionA2}
& & \hat{Q}^1_0 = -Q^{35}{}_2 , \quad  \hat{Q}^1_1 = Q^{35}{}_1, \quad \hat{Q}^1_2 = -Q^{45}{}_2, \quad \hat{Q}^1_3 =-Q^{36}{}_2 , \nonumber\\
& & \hat{Q}^2_0 = -Q^{51}{}_4 , \quad  \hat{Q}^2_1 =- Q^{52}{}_4, \quad \hat{Q}^2_2 = Q^{51}{}_3, \quad \hat{Q}^2_3 =-Q^{61}{}_4 , \nonumber\\
& & \hat{Q}^3_0 = -Q^{13}{}_6 , \quad  \hat{Q}^3_1 = -Q^{23}{}_6, \quad \hat{Q}^3_2 = -Q^{14}{}_6, \quad \hat{Q}^3_3 =Q^{13}{}_5, \\
& & \hat{Q}^{10} = -Q^{46}{}_1 , \quad  \hat{Q}^{11} =- Q^{46}{}_2 \quad \hat{Q}^{12} = Q^{36}{}_1, \quad \hat{Q}^{13} =Q^{45}{}_1 , \nonumber\\
& & \hat{Q}^{20} = -Q^{62}{}_3 , \quad  \hat{Q}^{21} = Q^{61}{}_3, \quad \hat{Q}^{22} = -Q^{62}{}_4, \quad \hat{Q}^{23} =Q^{52}{}_3, \nonumber\\
& & \hat{Q}^{30} = -Q^{24}{}_5 , \quad  \hat{Q}^{31} = Q^{14}{}_5, \quad \hat{Q}^{32} = Q^{23}{}_5, \quad \hat{Q}^{33} =-Q^{24}{}_6 , \nonumber
 \nonumber
\eea
By using the first formulation of Bianchi identities as in eqn. (\ref{eq:bianchids1}), one finds that there are 24 constraints for each of the $HQ$ as well as $QQ$ type. Using the conversion relations (\ref{eq:fluxconversionA1}) and (\ref{eq:fluxconversionA2}) for flux components expressed in the two formulations, we can completely convert these 48 identities coming from eqn. (\ref{eq:bianchids1}), and the same are collated as under,
\subsubsection*{24 Bianchi identities of HQ-type}
\bea
\label{eq:HQFirstFormalism1}
\begin{array}{c}
 \hat{Q}_0{}^2 \, {H}^1+H_0 \, \hat{Q}^{21}=H_3 \hat{Q}_2{}^2+H_2 \hat{Q}_3{}^2 , \quad \hat{Q}_0{}^3 \, {H}^2+H_0 \, \hat{Q}^{32}=H_3 \hat{Q}_1{}^3+H_1 \hat{Q}_3{}^3 \\
 \hat{Q}_0{}^1 \, {H}^3+H_0 \, \hat{Q}^{13}=H_2 \hat{Q}_1{}^1+H_1 \hat{Q}_2{}^1 , \quad \hat{Q}_0{}^3 \, {H}^1+H_0 \, \hat{Q}^{31}=H_3 \hat{Q}_2{}^3+H_2 \hat{Q}_3{}^3 \\
 \hat{Q}_0{}^1 \, {H}^2+H_0 \, \hat{Q}^{12}=H_3 \hat{Q}_1{}^1+H_1 \hat{Q}_3{}^1 , \quad \hat{Q}_0{}^2 \, {H}^3+H_0 \, \hat{Q}^{23}=H_2 \hat{Q}_1{}^2+H_1 \hat{Q}_2{}^2 \\
 \end{array}
\eea
 \bea
\label{eq:HQFirstFormalism2}
\begin{array}{c}
 \hat{Q}_1{}^2 \, {H}^0+H_1 \, \hat{Q}^{20}+\, {H}^3 \, \hat{Q}^{22}+\, {H}^2 \, \hat{Q}^{23}=0 \\
 \hat{Q}_2{}^3 \, {H}^0+H_2 \, \hat{Q}^{30}+\, {H}^3 \, \hat{Q}^{31}+\, {H}^1 \, \hat{Q}^{33}=0 \\
 \hat{Q}_3{}^1 \, {H}^0+H_3 \, \hat{Q}^{10}+\, {H}^2 \, \hat{Q}^{11}+\, {H}^1 \, \hat{Q}^{12}=0 \\
 \hat{Q}_3{}^2 \, {H}^0+H_3 \, \hat{Q}^{20}+\, {H}^2 \, \hat{Q}^{21}+\, {H}^1 \, \hat{Q}^{22}=0 \\
 \hat{Q}_2{}^1 \, {H}^0+H_2 \, \hat{Q}^{10}+\, {H}^3 \, \hat{Q}^{11}+\, {H}^1 \, \hat{Q}^{13}=0 \\
 \hat{Q}_1{}^3 \, {H}^0+H_1 \, \hat{Q}^{30}+\, {H}^3 \, \hat{Q}^{32}+\,{H}^2 \, \hat{Q}^{33}=0 \\
 \end{array}
\eea
 \bea
\label{eq:HQFirstFormalism3}
\begin{array}{c}
\hat{Q}_2{}^1 \, {H}^2=H_0 \, \hat{Q}^{10}+H_1 \, \hat{Q}^{11}+H_3 \, \hat{Q}^{13} , \quad \hat{Q}_0{}^1 \, {H}^0+\hat{Q}_1{}^1 \, {H}^1+\hat{Q}_3{}^1 \, {H}^3=H_2 \,\hat{Q}^{12}   \\
\hat{Q}_1{}^2 \,{H}^1=H_0 \, \hat{Q}^{20}+H_2 \, \hat{Q}^{22}+H_3 \, \hat{Q}^{23} , \quad \hat{Q}_0{}^2 \, {H}^0+\hat{Q}_2{}^2 \, {H}^2+\hat{Q}_3{}^2 \, {H}^3=H_1 \, \hat{Q}^{21}  \\
\hat{Q}_2{}^3 \, {H}^2=H_0 \, \hat{Q}^{30}+H_1 \, \hat{Q}^{31}+H_3 \, \hat{Q}^{33} , \quad \hat{Q}_0{}^3 \, {H}^0+\hat{Q}_1{}^3 \, {H}^1+\hat{Q}_3{}^3 \, {H}^3=H_2 \, \hat{Q}^{32}  \\
\hat{Q}_3{}^1 \, {H}^3=H_0 \, \hat{Q}^{10}+H_1 \, \hat{Q}^{11}+H_2 \, \hat{Q}^{12}, \quad \hat{Q}_0{}^1 \, {H}^0+\hat{Q}_1{}^1 \, {H}^1+\hat{Q}_2{}^1 \, {H}^2=H_3 \, \hat{Q}^{13}  \\
\hat{Q}_3{}^2 \, {H}^3=H_0 \, \hat{Q}^{20}+H_1 \, \hat{Q}^{21}+H_2 \, \hat{Q}^{22}, \quad \hat{Q}_0{}^2 \, {H}^0+\hat{Q}_1{}^2 \, {H}^1+\hat{Q}_2{}^2 \, {H}^2=H_3 \, \hat{Q}^{23}  \\
\hat{Q}_1{}^3 \, {H}^1=H_0 \, \hat{Q}^{30}+H_2 \, \hat{Q}^{32}+H_3 \, \hat{Q}^{33}, \quad \hat{Q}_0{}^3 \, {H}^0+\hat{Q}_2{}^3 \, {H}^2+\hat{Q}_3{}^3 \, {H}^3=H_1 \, \hat{Q}^{31}  \\
\end{array}
\eea
\subsubsection*{24 Bianchi identities of QQ-type}
\bea
\label{eq:QQFirstFormalism1}
\begin{array}{c}
\hskip-0.5cm \hat{Q}_1{}^3   \hat{Q}_2{}^2+ \hat{Q}_1{}^2 \hat{Q}_2{}^3=  \hat{Q}_0{}^3 \hat{Q}^{23}+ \hat{Q}_0{}^2 \hat{Q}^{33} , \quad  \hat{Q}_2{}^3 \hat{Q}_3{}^1+ \hat{Q}_2{}^1  \hat{Q}_3{}^3= \hat{Q}_0{}^3 \hat{Q}^{11}+  \hat{Q}_0{}^1 \hat{Q}^{31} \\
\hskip-0.5cm  \hat{Q}_1{}^2   \hat{Q}_3{}^1+  \hat{Q}_1{}^1  \hat{Q}_3{}^2=  \hat{Q}_0{}^2 \hat{Q}^{12}+  \hat{Q}_0{}^1 \hat{Q}^{22}, \quad \hat{Q}_1{}^3  \hat{Q}_3{}^2+  \hat{Q}_1{}^2  \hat{Q}_3{}^3= \hat{Q}_0{}^3 \hat{Q}^{22}+ \hat{Q}_0{}^2  \hat{Q}^{32} \\
\hskip-0.5cm  \hat{Q}_1{}^3  \hat{Q}_2{}^1+ \hat{Q}_1{}^1 \hat{Q}_2{}^3= \hat{Q}_0{}^3 \hat{Q}^{13}+  \hat{Q}_0{}^1 \hat{Q}^{33}, \quad\hat{Q}_2{}^2  \hat{Q}_3{}^1+  \hat{Q}_2{}^1 \hat{Q}_3{}^2= \hat{Q}_0{}^2 \hat{Q}^{11}+  \hat{Q}_0{}^1 \hat{Q}^{21} \\
\end{array}
\eea
\bea
\label{eq:QQFirstFormalism2}
\begin{array}{c}
  \hat{Q}_3{}^3 \, \, \hat{Q}^{20}+\, \,  \hat{Q}_3{}^2 \, \, \hat{Q}^{30}+\, \,\hat{Q}^{31} \, \, \hat{Q}^{22}+\, \,\hat{Q}^{21} \, \,\hat{Q}^{32}=0 \\
  \hat{Q}_1{}^3\, \, \hat{Q}^{10}+\, \,  \hat{Q}_1{}^1 \, \, \hat{Q}^{30}+\, \,\hat{Q}^{32} \, \,\hat{Q}^{13}+\, \,\hat{Q}^{12}\, \,\hat{Q}^{33}=0 \\
  \hat{Q}_2{}^2\, \, \hat{Q}^{10}+\, \,  \hat{Q}_2{}^1\, \, \hat{Q}^{20}+\, \,\hat{Q}^{21} \, \,\hat{Q}^{13}+\, \,\hat{Q}^{11}\, \,\hat{Q}^{23}=0 \\
  \hat{Q}_2{}^3\, \, \hat{Q}^{20}+\, \,  \hat{Q}_2{}^2 \, \,\hat{Q}^{30}+\, \,\hat{Q}^{31} \, \,\hat{Q}^{23}+\, \,\hat{Q}^{21}\, \,\hat{Q}^{33}=0 \\
  \hat{Q}_3{}^3 \, \,\hat{Q}^{10}+\, \,  \hat{Q}_3{}^1\, \, \hat{Q}^{30}+\, \,\hat{Q}^{31} \, \,\hat{Q}^{12}+\, \,\hat{Q}^{11}\, \,\hat{Q}^{32}=0 \\
  \hat{Q}_1{}^2 \, \,\hat{Q}^{10}+\, \,  \hat{Q}_1{}^1\, \, \hat{Q}^{20}+\, \,\hat{Q}^{22} \, \,\hat{Q}^{13}+\, \,\hat{Q}^{12}\, \,\hat{Q}^{23}=0 \\
  \end{array}
\eea
\bea
\label{eq:QQFirstFormalism3}
\begin{array}{c}
\hskip-1.0cm  \hat{Q}_0{}^3 \hat{Q}^{20}+  \hat{Q}_1{}^3 \hat{Q}^{21}+  \hat{Q}_2{}^3 \hat{Q}^{22}= \hat{Q}_3{}^2 \hat{Q}^{33} , \quad \hat{Q}_0{}^2 \hat{Q}^{30}+  \hat{Q}_1{}^2 \hat{Q}^{31}+ \hat{Q}_2{}^2 \hat{Q}^{32}=  \hat{Q}_3{}^3 \hat{Q}^{23}  \\
\hskip-1.0cm  \hat{Q}_0{}^2 \hat{Q}^{10}+  \hat{Q}_1{}^2 \hat{Q}^{11}+  \hat{Q}_3{}^2 \hat{Q}^{13}=  \hat{Q}_2{}^1 \hat{Q}^{22} , \quad \hat{Q}_0{}^1 \hat{Q}^{20}+  \hat{Q}_1{}^1 \hat{Q}^{21}+  \hat{Q}_3{}^1 \hat{Q}^{23}=  \hat{Q}_2{}^2 \hat{Q}^{12}   \\
\hskip-1.0cm  \hat{Q}_0{}^3 \hat{Q}^{10}+ \hat{Q}_1{}^3 \hat{Q}^{11}+ \hat{Q}_2{}^3 \hat{Q}^{12}= \hat{Q}_3{}^1 \hat{Q}^{33} , \quad \hat{Q}_0{}^1 \hat{Q}^{30}+ \hat{Q}_1{}^1 \hat{Q}^{31}+ \hat{Q}_2{}^1 \hat{Q}^{32}=\hat{Q}_3{}^3 \hat{Q}^{13} \\
\hskip-1.0cm  \hat{Q}_0{}^1 \hat{Q}^{30}+  \hat{Q}_2{}^1 \hat{Q}^{32}+ \hat{Q}_3{}^1 \hat{Q}^{33}= \hat{Q}_1{}^3 \hat{Q}^{11}, \quad  \hat{Q}_0{}^3 \hat{Q}^{10}+  \hat{Q}_2{}^3 \hat{Q}^{12}+ \hat{Q}_3{}^3 \hat{Q}^{13}= \hat{Q}_1{}^1 \hat{Q}^{31} \\
\hskip-1.0cm  \hat{Q}_0{}^2 \hat{Q}^{30}+  \hat{Q}_1{}^2 \hat{Q}^{31}+ \hat{Q}_3{}^2 \hat{Q}^{33}= \hat{Q}_2{}^3 \hat{Q}^{22}, \quad\hat{Q}_0{}^3 \hat{Q}^{20}+  \hat{Q}_1{}^3 \hat{Q}^{21}+  \hat{Q}_3{}^3 \hat{Q}^{23}=  \hat{Q}_2{}^2 \hat{Q}^{32} \\
\hskip-1.0cm  \hat{Q}_0{}^1\hat{Q}^{20}+ \hat{Q}_2{}^1 \hat{Q}^{22}+  \hat{Q}_3{}^1 \hat{Q}^{23}= \hat{Q}_1{}^2 \hat{Q}^{11}, \quad \hat{Q}_0{}^2 \hat{Q}^{10}+  \hat{Q}_2{}^2 \hat{Q}^{12}+ \hat{Q}_3{}^2 \hat{Q}^{13}= \hat{Q}_1{}^1 \hat{Q}^{21} \\
\end{array}
\eea
{\it Just to avoid any confusion among the $h^{11}_+$ indices and $h^{21}_-$ indices of $Q$-flux (as both of these take values as $1,2$ and 3) while appearing in the Bianchi identities given in eqns. (\ref{eq:HQFirstFormalism1})-(\ref{eq:QQFirstFormalism3}), we recall that flux components are written as $\hat{Q}_\Lambda{}^\alpha$ and $\hat{Q}^{\alpha \Lambda}$, and any lower index is always a $h^{21}_-$ index in these constraints.} We have divided the $HQ$ and $QQ$ Bianchi identities in three sets, and it is important to mention that the last set of constraints as given in eqns. (\ref{eq:HQFirstFormalism3}) and (\ref{eq:QQFirstFormalism3}) indeed contains the following Bianchi identities of the second formulation,
\bea
\label{eq:hqmodelA}
& & \hskip-1.0cm H^\Lambda \, \hat{Q}_\Lambda{}^\alpha - H_\Lambda \hat{Q}^{\alpha \Lambda} = 0, \quad \hat{Q}^{\alpha\Lambda} \hat{Q}^\beta{}_{\Lambda} - \hat{Q}^{\beta \Lambda} \hat{Q}^\alpha{}_{\Lambda} = 0. 
\eea 
For example, the two constraints of each of the six lines of eqn. (\ref{eq:HQFirstFormalism3}) produce three $HQ$ identities for $\alpha= 1,2,3$ which corresponds to those of the second formulation mentioned above. Similarly the two constraints in each of six lines of eqn. (\ref{eq:QQFirstFormalism3}) produce $QQ$ identities. However, let us mention that the eqn. (\ref{eq:hqmodelA}) happens to be weaker than the collection in eqns. (\ref{eq:HQFirstFormalism3}) and (\ref{eq:QQFirstFormalism3}), and so the reshuffled version obtained after converting the identities into cohomology language, although captures all the identities of second formulation, is not exactly the same.  Also, given that $h^{2,1}_+(X_6) = 0$ for this example, the Bianchi identities in eqn. (\ref{eq:BIs2}) are trivially satisfied, and therefore dose not appear out of the constraints of the first formulation.

Further, other constraints in which $h^{21}_-$ flux indices are not contracted (in any of $HQ$ or $QQ$ type identities), cannot belong to the second formulation; for examples constraints in eqns. (\ref{eq:HQFirstFormalism1}), (\ref{eq:HQFirstFormalism2}), (\ref{eq:QQFirstFormalism1}) and (\ref{eq:QQFirstFormalism2}) fall in this category. Moreover, we find that these four sets of identities can be written as,
\bea
\label{eq:ModelA95}
& & \hskip-1.50cm H^i\, \hat{Q}^\alpha_0 + H_0 \, \hat{Q}^{\alpha i} - (\hat{l}_{ijk})^{-1}\, H_{j} \,\hat{Q}^\alpha_k = 0, \, H^{0}\, \hat{Q}^\alpha_i + H_i\, \hat{Q}^{\alpha 0} + \hat{l}_{ijk}\, H^j \, \hat{Q}^{\alpha k} = 0, \hskip0.1cm   \forall i \, \, {\rm and} \, \, \alpha \ne i \,, \nonumber\\
& & \hskip-1.70cm k_{\alpha\beta\gamma}\, \left(2\, \hat{Q}^{\beta}_0\, \hat{Q}^{\gamma i} - (\hat{l}_{ijk})^{-1} \hat{Q}^{\beta}_j\, \hat{Q}^\gamma_k \right) = 0, \, \, k_{\alpha\beta\gamma}\, \left(2\, \hat{Q}^{\beta 0}\, \hat{Q}^{\gamma}_{i} + \hat{l}_{ijk} \hat{Q}^{\beta j}\, \hat{Q}^{\gamma k} \right) = 0, \hskip0.1cm   \forall i \, \, {\rm and} \, \, \alpha \ne i \,. 
\eea
Recall that the only non-zero intersection numbers in this example are $\hat{l}_{ijk} = 1$ and $k_{\alpha\beta\gamma} = 1$, and so there is not as much structure apparent from eqn. (\ref{eq:ModelA95}) as it was for the respective invoked identities in the previous example. {\it However, we stress here again that a more fundamental reason or proof is needed to trust these speculative identities for generic backgrounds. } 

\subsection{Summary and observations }
From the analysis of this section about converting the first formulation constraints into cohomology ingredients in two concrete examples, we conclude the followings,
\begin{itemize}
\item{First formulation already has all the flux constraints of the second formulation.}
\item{There are some additional flux constraints in the first formulation which cannot be derived from the known identities of the second formulation. A reason for this mismatch could be the fact that in second formulation, flux-actions utilized in eqns. (\ref{eq:action1})-(\ref{eq:action2}) are defined only for harmonic forms while the ones (given in eqn. (\ref{eq:action0}) which are) utilized in deriving the first formulation are known for arbitrary p-forms \cite{Ihl:2007ah}. In other words, the first formulation is derived via imposing ${\cal D}^2 A_p = 0$ on generic $p$-forms while the second formulation is derived by imposing the nilpotency only on the harmonic-forms. This also supports the first point made above.}
\item{Given that the toroidal-orientifold examples we studied do not have any harmonic one-form (and their dual five-form),  and so an immediate, though naive, expectation could be the possibility that some of the `additional' flux constraints may be non-trivial even for the examples beyond toroidal orientifolds, and hence one may expect them to be relevant for setups with Calabi Yau orientifold compactifications.}
\item{Moreover, as we have demonstrated in two examples by rewriting the cohomological identities into some apparently model independent forms, there should exist a cohomology-indexed-version of the additional identities of the first formulation which (once invoked) could provide a completion of the second formulation for generic complex threefolds.}
\item{If these simple observations hold beyond toroidal models, any attempt for model building based on imposing the identities of only the second formulation for simplifying the scalar potential, for example \cite{Blumenhagen:2015qda, Blumenhagen:2015kja,Blumenhagen:2015jva, Blumenhagen:2015xpa, Li:2015taa}, could possibly remain under-constrained.}
\end{itemize}

\section{Solutions of BIs and implications on moduli stabilization: Model A}
\label{sec_solBIs2}
In the previous section we have performed very detailed computation and reshuffling of Bianchi identities, and now we would discuss how one can exploit those results in the context of moduli stabilization.
\subsection{Some direct implications of `additional' BIs on the scalar potential readjustments}
First let us see what we could gain in simplifying the scalar potential via imposing these `additional' Bianchi identities in some analytic sense. For that, let us mention that the compact version of identities given in eqn. (\ref{eq:ModelBbisNotSecondnewV2}) can be utilized to add/remove certain terms for a well motivated reshuffling of the scalar potential. To make this statement clear, now without going into all the details of computing the K\"ahler potential, the superpotential and the D-terms, which have been studied at a couple of occasions elsewhere \cite{Shukla:2015rua,Shukla:2015hpa}, here we simply collect the explicit expressions for the total scalar potential pieces which are given as under,
\bea
\label{eq:VscalarModelB}
& & \hskip-0.75cm V_{{\mathbb F}{\mathbb F}} =  \frac{{\mathbb F}_0^2+{({\mathbb F}^0)}^2}{4 \, s\,{\cal V}_E^2},  \quad V_{ {\mathbb F} {\mathbb H}} =  \frac{{\mathbb H}_0 {\mathbb F}^0-{\mathbb F}_0 {\mathbb H}^0}{2 \, {\cal V}_E^2}, \, V_{{\mathbb F} \hat{\mathbb Q}} =  \frac{\sigma_\alpha\left({\mathbb F}_0 \hat{\mathbb Q}^{\alpha0} - \hat{\mathbb Q}^\alpha_0 \, {\mathbb F}^0\right)}{2 \, s \, {\cal V}_E^2}, \\
& & \hskip-0.75cm V_{{\mathbb H}{\mathbb H}} =  \frac{s \left({\mathbb H}_0^2 +\, {({\mathbb H}^0)}^2 \right)}{4 \, {\cal V}_E^2}, \quad V_{\hat{\mathbb Q} \hat{\mathbb Q}} =\frac{1}{4 \,s\, {\cal V}_E^2} \biggl[\left(4 \sigma_2^2 - \sigma_1^2\right) \left(\hat{\mathbb Q}^1{}_0 \hat{\mathbb Q}^1{}_0 + \hat{\mathbb Q}^{10} \hat{\mathbb Q}^{10} \right) \nonumber\\
& & + \left(\sigma_1^2 - \sigma_2^2\right) \left(\hat{\mathbb Q}_0{}^2 \hat{\mathbb Q}_0{}^2 + \hat{\mathbb Q}^{20} \hat{\mathbb Q}^{20} \right)  + \sigma_3^2 \left(\hat{\mathbb Q}_0{}^3 \hat{\mathbb Q}_0{}^3 + \hat{\mathbb Q}^{30} \hat{\mathbb Q}^{30} \right) + 2 \, \sigma_1 \sigma_2 \left(\hat{\mathbb Q}_0{}^1 \hat{\mathbb Q}_0{}^2 + \hat{\mathbb Q}^{10} \hat{\mathbb Q}^{20} \right)\nonumber\\
& & - 6 \, \sigma_2 \sigma_3 \left(\hat{\mathbb Q}_0{}^2 \hat{\mathbb Q}_0{}^3 +  \hat{\mathbb Q}^{20} \hat{\mathbb Q}^{30} \right)  - 6 \, \sigma_1 \sigma_3 \left(\hat{\mathbb Q}_0{}^1 \hat{\mathbb Q}_0{}^3 + \hat{\mathbb Q}^{10} \hat{\mathbb Q}^{30} \right) \biggr], \nonumber\\
& & \hskip-0.75cm  V_{{\mathbb H} \hat{\mathbb Q}} =  \frac{3 \sigma_\alpha\left({\mathbb H}^0 \, \hat{\mathbb Q}^{\alpha0} + \hat{\mathbb Q}^{\alpha}_0 \, {\mathbb H}_0 \right)}{2  \, {\cal V}_E^2}, \, V_{{\cal \mho}{\cal \mho}} = \frac{\sigma_3 \bigl[\left(\mho_{01} \mho_{01} + \mho_{1}{}^{0} \mho_{1}{}^{0}\right) + 2\, \left(\mho_{02} \mho_{02} + \mho_{2}{}^{0} \mho_{2}{}^{0}\right)\bigr]}{4 {\cal V}_E^2}, \nonumber\\
& & \hskip-0.75cm V_{{\mathbb R}{\mathbb R}} = \frac{{\mathbb R}_1^2+({\mathbb R}^1)^2}{4\,s^2\, f^2}, \, \, V_{{\hat{\mathbb \mho}}{\hat{\mathbb \mho}}} =  \frac{t^\alpha \, t^\beta \,\left(\hat{{\cal \mho}}_{\alpha1} \hat{{\cal \mho}}_{\beta1} + {\hat{\cal \mho}_\alpha{}^1}  {\hat{\cal \mho}_\beta{}^1}\right)}{4\, \,{\cal V}_E^{2}}, \, \,V_{{\mathbb R}{\hat{\cal \mho}}}=-\, \frac{f^{-1}\, t^\alpha \left( {\mathbb R}_1\,\hat{{\cal \mho}}_{\alpha1}+ {\mathbb R}^1\, {\hat{\cal \mho}_\alpha{}^1}\right)}{2 \,s\, {\cal V}_E}\, .\nonumber
\eea
where all the new flux orbits are defined in eqn. (\ref{eq:orbitsB2Ex2}) except the following RR-flux orbit,
\bea
{\mathbb F}_0 = \biggl[F_0 + (\omega_{01} \, {c}^1 + \omega_{02} \, {c}^2) + \hat{Q}^1{}_0 \,{\rho}_1+ \hat{Q}^2{}_0 \,{\rho}_2+ \hat{Q}^3{}_0\, \left({\rho}_3 + \hat{\kappa}_{3 11} c^1 b^1 + \hat{\kappa}_{3 22} c^2 b^2\right) \biggr] +  c_0 \, {\mathbb H}_0\, , \nonumber
\eea
and also we mention here that the flux orbits with upper (2,1)-cohomology indices can be analogously written. Let us note that last three pieces ($V_{{\mathbb R}{\mathbb R}}, V_{{\hat{\mathbb \mho}}{\hat{\mathbb \mho}}}$ and $V_{{\mathbb R}{\hat{\cal \mho}}}$) in this collection (\ref{eq:VscalarModelB}) arise via $D$-terms while the rest of the contributions are from $F$-terms.

\subsubsection*{Some more reshuffling for making speculations to generic case}
It is interesting to find that one can completely get rid of the $V_{{\cal \mho}{\cal \mho}}$ piece in eqn.  (\ref{eq:VscalarModelB}) via introducing some even-indexed generalized geometric-flux components $\hat{\mathbb \mho}_{\alpha1}$ and ${\mathbb \mho}_\alpha^1$ by using the following identity from (\ref{eq:ModelBbisNotSecondnew}), $$\left({\mathbb \mho} _1{}^0\right){}^2 + {\mathbb \mho} _{10}^2+2 {\mathbb \mho} _{20}^2+ 2\left({\mathbb \mho} _2{}^0\right){}^2 = -2 \left(\hat{{\mathbb \mho} }_1{}^1\right){}^2 - 2 \hat{{\mathbb \mho} }_{11}^2 + \left(\hat{{\mathbb \mho} }_2{}^1\right){}^2+ \hat{{\mathbb \mho}}_{21}^2 - 8 \left({\mathbb H}_0 \hat{{\mathbb Q}}^3{}_0+ {\mathbb H}^0 \hat{{\mathbb Q}}^{30}\right) \,.$$
Subsequently one finds that the two modified pieces mentioned below are reshuffled as under,
\bea
\label{eq:ModelBsymmetric1}
& & \hskip-0.5cm V_{{\mathbb H} \hat{\mathbb Q}}^{new} =  -\frac{{\mathbb H}^0 \, \hat{\mathbb Q}^{0} + \hat{\mathbb Q}_0{} \, {\mathbb H}_0 }{2  \, {\cal V}_E^2}, 
\quad  V_{{\hat{\mathbb \mho}}{\hat{\mathbb \mho}}}^{new} =\frac{1}{4 \, {\cal V}_E^2} \biggl[\left(4 t_2^2 - t_1^2\right) \left(\hat{{\cal \mho}}_{11} \hat{{\cal \mho}}_{11} + {\hat{\cal \mho}_1{}^1}  {\hat{\cal \mho}_1{}^1} \right) \\
& & + \left(t_1^2 - t_2^2\right) \left(\hat{{\cal \mho}}_{21} \hat{{\cal \mho}}_{21} + {\hat{\cal \mho}_2{}^1}  {\hat{\cal \mho}_2{}^1}\right)  + t_3^2 \left(\hat{{\cal \mho}}_{31} \hat{{\cal \mho}}_{31} + {\hat{\cal \mho}_3{}^1}  {\hat{\cal \mho}_3{}^1} \right) + 2 \, t_1 t_2 \left(\hat{{\cal \mho}}_{11} \hat{{\cal \mho}}_{21} + {\hat{\cal \mho}_1{}^1}  {\hat{\cal \mho}_2{}^1}\right)\nonumber\\
& &  -6 \, t_1 t_3 \left(\hat{{\cal \mho}}_{11} \hat{{\cal \mho}}_{31} + {\hat{\cal \mho}_1{}^1}  {\hat{\cal \mho}_3{}^1}\right)- 6  \, t_2 t_3 \left(\hat{{\cal \mho}}_{21} \hat{{\cal \mho}}_{31} + {\hat{\cal \mho}_3{}^1}  {\hat{\cal \mho}_3{}^1}\right)\, , \quad V_{{\cal \mho}{\cal \mho}}^{new} = 0\, , \nonumber
\eea
where we have also utilized some redefinitions as $\hat{\mathbb Q}^0 = \hat{\mathbb Q}^{\alpha0}\, \sigma_\alpha$ and  $\hat{\mathbb Q}_0 = \hat{\mathbb Q}^{\alpha}_{0}\, \sigma_\alpha$ along with some additional identities: ${\mathbb H}^0 \hat{\mathbb Q}^{10} + \hat{\mathbb Q}^{1}_0 {\mathbb H}_0 = 0, \, {\mathbb H}^0 \, \hat{\mathbb Q}^{20} + \hat{\mathbb Q}^{2}_0 {\mathbb H}_0 = 0, \,  \hat{{\cal \mho}}_{11} \hat{{\cal \mho}}_{31} + {\hat{\cal \mho}_1{}^1}  {\hat{\cal \mho}_3{}^1} =0$ and $\hat{{\cal \mho}}_{21} \hat{{\cal \mho}}_{31} + {\hat{\cal \mho}_2{}^1}  {\hat{\cal \mho}_3{}^1} =0$ from collection in eqn. (\ref{eq:ModelBbisNotSecondnew}) in order to add and subtract some terms. The main motivation and advantage of doing this reshuffling can be reflected from the following points,
\begin{itemize}
\item{One good thing about this reshuffling is the fact that now the contributions from the even and odd (2,1)-indexed fluxes are separated out; for example $V_{\mathbb H \mathbb H}, V_{\mathbb H \mathbb Q}^{new}, V_{\hat{\mathbb Q} \hat{\mathbb Q}}$ are given by the `odd' (2,1)-indexed fluxes while those of $V_{\mathbb R \mathbb R}, V_{\mathbb R \hat{\mathbb \mho}}, V_{\hat{\mathbb \mho} \hat{\mathbb \mho}}^{new}$ are given by `even' (2,1)-index flux components. {\it Here we emphasize that this would not mean that the fluxes counted via odd (1,1)-cohomology such as $(\omega_{a \Lambda}, \omega_a^\Lambda$) and $(Q^a_K, Q^{aK})$ would loose their relevance as one should be reminded about their implicit appearance through the definitions of `new' generaalized flux orbits in eqns. (\ref{eq:newOrbitsBIs1})-(\ref{eq:newOrbitsBIs2}).} However, the separate notions of $F$- and $D$-terms are not relevant now due to insertions and removal of some mixing pieces via Bianchi identities.}
\item{Motivated by the studies on the origin of total 4D scalar potential from the dimensional reduction of a couple of kinetic pieces \cite{Blumenhagen:2015lta, Shukla:2015hpa} of a ten-dimensional theory (such as DFT), now we can compactly rewrite the volume moduli dependent pieces $V_{\hat{\mathbb Q} \hat{\mathbb Q}}$ and $V_{\hat{\mathbb \mho} \hat{\mathbb \mho}}^{new}$ as below,
\bea
\label{eq:finalModelBreshuffling}
& & V_{\hat{\mathbb Q} \hat{\mathbb Q}} =\frac{1}{4 \,s\, {\cal V}_E^2} \biggl[16\, {\cal V}_E^2\, \tilde{\cal G}_{\alpha \beta} \left(\hat{\mathbb Q}^\alpha{}_0 \hat{\mathbb Q}^\beta{}_0 + \hat{\mathbb Q}^{\alpha0} \hat{\mathbb Q}^{\beta0} \right) - 3 \, \left(\hat{\mathbb Q}{}_0 \hat{\mathbb Q}{}_0 + \hat{\mathbb Q}^{0} \hat{\mathbb Q}^{0} \right) \biggr]\, , \nonumber\\
& & V_{{\hat{\mathbb \mho}}{\hat{\mathbb \mho}}}^{new} =\frac{1}{4 \, {\cal V}_E^2} \biggl[ \frac{{\cal G}^{\alpha \beta}}{16{\cal V}_E^2} \left(\hat{{\cal \mho}}_{\alpha1} \hat{{\cal \mho}}_{\beta1} + {\hat{\cal \mho}_1{}^1}  {\hat{\cal \mho}_\alpha{}^1} \right)   - 3 \, \left(\hat{{\cal \mho}}_{1} \hat{{\cal \mho}}_{1} + {\hat{\cal \mho}{}^1}  {\hat{\cal \mho}{}^1} \right) \biggr]\,.
\eea
These reshufflings in $V_{\hat{\mathbb Q} \hat{\mathbb Q}}$ and $V_{{\hat{\mathbb \mho}}{\hat{\mathbb \mho}}}$ hold for this example which one may expect to get extended to arbitrary compactifications, specially for the case of rigid Calabi Yaus. Moreover, the reasons for this final rearrangement in eqn. (\ref{eq:finalModelBreshuffling}) holding true are the following relations,
\bea
& & \hskip-1.5cm 16\, {\cal V}_E^2\, \tilde{\cal G}_{\alpha \beta} - 3 \, \sigma_\alpha \sigma_\beta = \left(
 \begin{array}{ccc}
4 \sigma_2^2 -  \sigma_1^2& \sigma_1 \, \sigma_2 & -3 \sigma_1 \, \sigma_3\\
\sigma_1 \, \sigma_2 & \sigma_1^2 -  \sigma_2^2 & -3 \sigma_2 \, \sigma_3 \\
-3 \sigma_1 \, \sigma_3 & -3 \sigma_2 \, \sigma_3 & \sigma_3^2 \\
\end{array}
\right)\,, \\
& & \hskip-0.75cm \frac{{\cal G}^{\alpha \beta}}{16{\cal V}_E^2}  - 3 \, t^\alpha \, t^\beta = \left(
 \begin{array}{ccc}
4 t_2^2 -  t_1^2& t_1 \, t_2 & -3\, t_1 \, t_3\\
t_1 \, t_2 & t_1^2 -  t_2^2 & -3\, t_2 \, t_3 \\
-3\, t_1 \, t_3 & -3 \, t_2 \, t_3 & t_3^2 \,\\
\end{array}
\right)\,.\nonumber
\eea
Here the moduli space metrics are generically defined to be normalized as under \cite{Grimm:2004uq},
 \bea
& & {\cal G}_{\alpha \beta} = -\frac{3}{2} \, \left( \frac{k_{\alpha \beta}}{k_0} - \frac{3}{2} \frac{k_\alpha \,k_\beta}{k_0^2}\right), \quad \quad \quad {\cal G}^{\alpha \beta} = -\frac{2}{3} \, k_0 \, k^{\alpha \beta} + 2 \, t^\alpha \, t^\beta, \nonumber
\end{eqnarray} 
where we have introduced $k_0 = 6 \, {\cal V}_E = k_\alpha\, t^\alpha$, $k_\alpha =k_{\alpha \beta} \, t^\beta$, $k_{\alpha\beta} = k_{\alpha\beta\gamma} \, t^\gamma, \, \hat{k}_{a b} = \hat{k}_{\alpha a b} \, t^\alpha$ and $\tilde{\cal G}_{\alpha \beta}=\left(({\hat{d}^{-1}})_{\alpha}{}^{\alpha'}\, {\cal G}_{\alpha' \beta'}\, ({\hat{d}^{-1}})_{\beta}{}^{\beta'}\right),\,\hat{\kappa}_{\alpha a b} = (\hat{d}^{-1})_\alpha{}^{\beta}\hat{k}_{\beta a b}$ \cite{Shukla:2015hpa}.}
\end{itemize}
On these lines of final reshuffling in eqn. (\ref{eq:finalModelBreshuffling}) using a peculiar form of additional Bianchi identities, it may be worth to mention here that although most of the computations of \cite{Blumenhagen:2015lta} (e.g. the $N=1$ scalar potential) are done in terms of cohomology ingredients, however the Bianchi identities utilized, to connect the pieces with those of ten-dimensional kinetic terms written with (inverse-)metric of the threefold as in \cite{Blumenhagen:2013hva}, are still those of the first formulation (and not in cohomology basis) even though they are shifted from real six-dimensional indices to a set of complex three-dimensional indices. From the observations in our toroidal examples, we expect the relevance of those `additional' Bianchi identities in generic compactifications, and a more complete version of the second formulation should exit to directly see the connection between the two proposals of {\cite{Blumenhagen:2015lta} and \cite{Blumenhagen:2013hva} in a completely cohomological framework.  

\subsection{Relevant scalar potential pieces for K\"ahler moduli stabilization} Now we switch towards looking at the solutions of Bianchi identities and possibility of moduli stabilization in this example. {\it Let us focus on the volume moduli and say that currently we want to see if all the K\"ahler moduli could be stabilized subject to imposing all the Bianchi identities, as this is probably the most attractive feature for looking at the non-geometric setups so that one could stabilize all moduli at tree level.} For that we will keep in mind that all the new flux orbits generically do not have any volume moduli dependence (e.g. see eqn. (\ref{eq:orbitsB2Ex2}) for this particular example), and so they appear only at the places where $\sigma_\alpha$ and $t^\alpha$ are explicit in eqn. (\ref{eq:VscalarModelB}). Based on the above arguments, we find that
\begin{itemize}
\item{The three pieces $V_{\mathbb F \mathbb F}, V_{\mathbb H \mathbb H}$ and $V_{\mathbb R \mathbb R}$ are not relevant for K\"ahler moduli stabilization as these can be written as $V_0 = n_0 + \frac{m_0}{{\cal V}_E^2\, }$, 
where $n_0$ and $m_0$ are flux-dependent functions of anything but the volume moduli, and so the No-scale structure remains intact.}
\item{As there are no RR tadpoles (in the untwisted sector) in this example \cite{Robbins:2007yv}, the second formulation identities simply nullify the pieces in $V_{\mathbb F \mathbb H}$ and $V_{\mathbb F \hat{\mathbb Q}}$.}
\item{As we have seen, in the remaining five pieces $\left(V_{\hat{\mathbb Q} \hat{\mathbb Q}} + V_{{\mathbb H} \hat{\mathbb Q}} + V_{{\cal \mho}{\cal \mho}} + V_{{\hat{\mathbb \mho}}{\hat{\mathbb \mho}}} +V_{{\mathbb R}{\hat{\cal \mho}}} \right)$ relevant for volume moduli stabilization, we can eliminate the $V_{{\cal \mho}{\cal \mho}}$ via a crucial Bianchi identity to get the modified pieces $V_{{\mathbb H} \hat{\mathbb Q}}^{new}$ and $V_{{\hat{\mathbb \mho}}{\hat{\mathbb \mho}}}^{new}$.}
\item{Subsequently, the only relevant pieces for the volume moduli stabilization remain to be the following four pieces,}
\end{itemize}
\bea
\label{eq:ModelBsymmetric12}
& & \hskip-0.5cm V_{{\mathbb H} \hat{\mathbb Q}}^{new} =  -\frac{\sigma_3\left({\mathbb H}^0 \, \hat{\mathbb Q}^{30} + \hat{\mathbb Q}^3{}_0{} \, {\mathbb H}_0 \right)}{2  \, {\cal V}_E^2}, \, \,  V_{\hat{\mathbb Q} \hat{\mathbb Q}} =\frac{1}{4 \,s\, {\cal V}_E^2} \biggl[\left(4 \sigma_2^2 - \sigma_1^2\right) \left(\hat{\mathbb Q}^1{}_0 \hat{\mathbb Q}^1{}_0 + \hat{\mathbb Q}^{10} \hat{\mathbb Q}^{10} \right)  \\
& &+ \left(\sigma_1^2 - \sigma_2^2\right) \left(\hat{\mathbb Q}_0{}^2 \hat{\mathbb Q}_0{}^2 + \hat{\mathbb Q}^{20} \hat{\mathbb Q}^{20} \right)  + \sigma_3^2 \left(\hat{\mathbb Q}_0{}^3 \hat{\mathbb Q}_0{}^3 + \hat{\mathbb Q}^{30} \hat{\mathbb Q}^{30} \right) + 2 \, \sigma_1 \sigma_2 \left(\hat{\mathbb Q}_0{}^1 \hat{\mathbb Q}_0{}^2 + \hat{\mathbb Q}^{10} \hat{\mathbb Q}^{20} \right)\nonumber\\
& & \hskip-0.5cm  V_{{\mathbb R}{\hat{\cal \mho}}}=-\frac{t_3\left({\mathbb R}_1\,\hat{{\cal \mho}}_{31}+ {\mathbb R}^1\, {\hat{\cal \mho}_3{}^1}\right)}{2 \,f\, s\, {\cal V}_E}\,, \,\,  V_{{\hat{\mathbb \mho}}{\hat{\mathbb \mho}}}^{new} =\frac{1}{4 \, {\cal V}_E^2} \biggl[\left(4 t_2^2 - t_1^2\right) \left(\hat{{\cal \mho}}_{11} \hat{{\cal \mho}}_{11} + {\hat{\cal \mho}_1{}^1}  {\hat{\cal \mho}_1{}^1} \right) \nonumber\\
& & + \left(t_1^2 - t_2^2\right) \left(\hat{{\cal \mho}}_{21} \hat{{\cal \mho}}_{21} + {\hat{\cal \mho}_2{}^1}  {\hat{\cal \mho}_2{}^1}\right)  + t_3^2 \left(\hat{{\cal \mho}}_{31} \hat{{\cal \mho}}_{31} + {\hat{\cal \mho}_3{}^1}  {\hat{\cal \mho}_3{}^1} \right) + 2 \, t_1 t_2 \left(\hat{{\cal \mho}}_{11} \hat{{\cal \mho}}_{21} + {\hat{\cal \mho}_1{}^1}  {\hat{\cal \mho}_2{}^1}\right)\nonumber
\eea

\subsection{Switching-off the fluxes with even or odd (2,1)-cohomology index}
As a warm up for investigations on moduli stabilization, let us say we want to look at the relevant identities while we switch-off all the fluxes with even (2,1)-cohomology indices. Subsequently, imposing the flux parameters to be integer-valued, the simplified version of the Bianchi identities given in eqns. (\ref{eq:ModelBbisSecond})-(\ref{eq:ModelBbisNotSecond}) reduces into the only solutions which involve,
\bea
\label{eq:even21zero}
& & \hat{Q}^1{}_0 = 0, \quad \hat{Q}^{10} = 0, \quad \hat{Q}^2{}_0 = 0, \quad \hat{Q}^{20} = 0.
\eea
Similarly if we assume all the fluxes with odd (2,1)-cohomology indices to be set to zero, then the resulting identities have the only integral-flux solutions which involve,
\bea
\label{eq:odd21zero}
& & \hat{\omega}_1{}^1 = 0, \quad \hat{\omega}_{11} = 0, \quad \hat{\omega}_2{}^1 = 0, \quad \hat{\omega}_{21} = 0.
\eea
These two observations stop our current interest of moduli stabilization with such solutions of Bianchi identities as the same will not be able to stabilize the K\"ahler moduli $T_1$ and $T_2$. This can be immediately read-off from the collection of pieces in eqn.(\ref{eq:ModelBsymmetric12}).

This could have been anticipated because in the first case with the absence of $D$-term contributions (when we have discarded the even-indexed fluxes), the condition (\ref{eq:even21zero}) shows that there are no other coupling of $T_1$ and $T_2$ chiral variables in the superpotential to break the no-scale structure along these directions. Similarly when we do not have any odd-(2,1) index fluxes present (i.e. in the absence of superpotential), then the condition (\ref{eq:odd21zero}) shows that remaining $D$-terms do not have any dependence on two-cycle volume moduli $t_1$ and $t_2$ leaving them unfixed. Therefore, we conclude that this is indeed not a good simplification in the flux sampling to follow for moduli stabilization, at least for our current interest of stabilizing all volume moduli at tree level. Nevertheless, these observations can be taken as advantage towards inflationary applications by stabilizing these flat directions at the subleading order, say via non-perturbative effects. 

\subsection{Switching-off half of the NS-NS fluxes: `Special solutions' }
As a less-simple assumption than the previous ones, let us look at the possibilities with what we have called as `special solutions' of Bianchi identities via setting half of the fluxes to zero on top of imposing the additional flux conditions in eqn. (\ref{eq:wiijQiij}). Subsequently the resulting constraint given in eqn. (\ref{eq:wiijQiijRestBIs}) produces the 
following solutions,
\bea
\label{eq:solModelBHalf123}
& & \hskip-0.5cm {\bf S1:} \, \, \hat{Q}^3{}_0=0\,, \hat{Q}^1{}_0=0\,, Q^2{}_1=0\,, \hat{\omega }_{31}=0\,, \hat{\omega }_{11}=0\,, \omega _{20}=0\,, R_1=0\,, H_0=0 \\
& & \hskip-0.5cm {\bf S2:} \, \, \hat{Q}^3{}_0=0\,, \hat{Q}^1{}_0=0\,,Q^2{}_1=0\,, \hat{\omega }_{31}=0\,, \hat{\omega }_{11}=0\,, \omega _{20}=0\,, R_1=0 \nonumber\\
& & \hskip-0.5cm {\bf S3:} \, \, \hat{Q}^3{}_0=0\,, \hat{Q}^1{}_0=0\,, Q^2{}_1=0\,, \hat{\omega }_{31}=0\,, \hat{\omega }_{11}=0\,, \omega_{20}=0\,, R_1\neq 0\,, H_0=0\nonumber\\
& & \hskip-0.5cm {\bf S4:} \, \, \hat{Q}^1{}_0=0\,, Q^2{}_1=0\,, \hat{\omega }_{31}=0\,,\hat{\omega }_{11}=0\,, \omega _{20}=0\,, H_0=0\,, \hat{Q}^3{}_0\neq 0\nonumber\\
& & \hskip-0.5cm {\bf S5:} \, \, \hat{Q}^3{}_0=0\,, \hat{Q}^1{}_0=0\,, Q^2{}_1=0\,,\hat{\omega }_{11}=0\,, \omega _{20}=0\,, R_1=0\,, \hat{\omega }_{31}\neq 0 \nonumber\\
& & \hskip-0.5cm {\bf S6:} \, \, \hat{Q}^1{}_0=0\,, Q^2{}_1=0\,, \hat{\omega }_{31}=0\,, \hat{\omega }_{11}=0\,, R_1=0\,, \hat{Q}^3{}_0\neq 0\,, H_0=-\frac{\omega_{20}^2}{4 \hat{Q}^3{}_0}\,, \omega _{20}\neq 0 \nonumber\\
& & \hskip-0.5cm {\bf S7:} \, \, \hat{Q}^3{}_0=0\,, \hat{Q}^1{}_0=0\,, \hat{\omega }_{11}=0\,, \omega
   _{20}=0\,, \hat{\omega }_{31}\neq 0\,, R_1=-\frac{\left(Q^2_1\right)^2}{4 \hat{\omega }_{31}}\,, H_0=0\,, Q^2{}_1\neq 0 \nonumber\\
& & \hskip-0.5cm {\bf S8:} \, \, \hat{Q}^1{}_0=0\,, Q^2{}_1=0\,, \hat{\omega }_{31}=0\,, R_1=0\,, \hat{Q}^3{}_0\neq 0\,, H_0=\frac{-\omega _{20}^2-\hat{\omega }_{11}^2}{4
   \hat{Q}^3{}_0}\nonumber\\
& & \hskip-0.5cm {\bf S9:} \, \, \hat{Q}^3{}_0=0\,, \hat{\omega }_{11}=0\,, \omega _{20}=0\,, \hat{\omega }_{31}\neq 0\,,
   R_1=\frac{-\left(Q^2_1\right)^2-\left(\hat{Q}^1_0\right)^2}{4 \hat{\omega }_{31}}\,, H_0=0 \nonumber\\
& & \hskip-1.0cm {\bf S10:} \, \, \hat{Q}^1{}_0=0\,, \hat{\omega
   }_{11}=0\,, Q^2{}_1\neq 0, \omega _{20}=\frac{4 \hat{\omega }_{31} \hat{Q}^3{}_0}{Q^2{}_1}, \hat{\omega }_{31}\neq 0\,,
   R_1=-\frac{\left(Q^2_1\right)^2}{4 \hat{\omega }_{31}}\,, \hat{Q}^3{}_0\neq 0\,, H_0=-\frac{\omega _{20}^2}{4 \hat{Q}^3{}_0}.\nonumber
\eea
Note that in this collection of solutions in eqn. (\ref{eq:solModelBHalf123}), the last solution has the possibility of $H_3$ and $R$-flux both being non-zero while in other solutions, at least one or both are zero. In fact, the last solution turns out to be the one in which all the four kinds of NS-NS fluxes, namely $H, \omega, Q$ and $R$, could be turned-on simultaneously. However, we did not find any of these 10 solutions to result into stabilizing {\it all} the K\"ahler moduli and the dilaton. In this regard, it would be interesting to see if the inclusion of non-geometric $P$-flux could help, but for that case the Bianchi identities are not known in any of the two formulations, as we elaborate more on it in the appendix. Finally, we also note here that we have not considered the most generic case as all the generic 66 Bianchi identities in eqns. (\ref{eq:ModelBbisSecond})-(\ref{eq:ModelBbisNotSecond}) could not be simultaneously solved in an analytic manner. 

\section{Solutions of BIs and implications on moduli stabilization: Model B}
\label{sec_solBIs1}
In this section, we perform a detailed analysis for looking at the possible solutions of Bianchi identities for Model B. So far we have just translated the constraints of the first formulation as given in eqn.  (\ref{eq:bianchids1}) into cohomology ingredients as collected in eqns. (\ref{eq:HQFirstFormalism1})-(\ref{eq:QQFirstFormalism3}). These are in total 48 quadratic flux constraints, and now the aim is to look for (some) possible solutions of the same so that one could impose them directly on the scalar potential expressed in cohomology language to perform moduli stabilization.

Before investigating for the possible non-supersymmetric vacua, let us recall that we need to demand the following generic conditions for the solutions to lie within the physical domains of effective field theoretic description,
\bea
& & \hskip-2.0cm Im(U_i) < 0, \quad Im(\tau) > 0, \quad Im(T_\alpha) < 0 \quad \Longrightarrow \{u < 0 , \, s >0, \, \sigma >0\} \quad (\rm{isotropic \, \, case})
\eea
Note that the four-cycle volume part in the definition of $T_\alpha$ appears with a minus sign in our convention in eqn. (\ref{eq:N=1_coords}) and that is why we need $Im(T_\alpha) < 0$. The $F$-term contributions to the scalar potential can be computed from the following expressions of the K\"ahler potential and the generalized flux-induced superpotential,
\bea
\label{eq:KWmain}
& & \hskip-1.6cm K = -\ln\left(-i(\tau-\ov\tau)\right) -\sum_{j=1}^{3} \ln\left(i(U_j - \ov U_j)\right) - \sum_{\alpha=1}^{3} \ln\left(i(T_\alpha - \ov T_\alpha\right)  \\
\label{eq:KWmain}
& & \hskip-1.6cm W = \biggl[\left(F_\Lambda + \tau\, H_\Lambda  + \hat{Q}^{\alpha}{}_{\Lambda} \, T_\alpha\right)\, {\cal X}^\Lambda + \left(F^\Lambda + \tau\, H^\Lambda  + \hat{Q}^{\alpha \Lambda} \, T_\alpha\right) \, {\cal F}_\Lambda \biggr]\, ,
\eea
where $\Lambda = 0, 1, 2, 3$ and $\alpha = 1,2,3$ implying the presence of 8 components for each of three form fluxes $H_3$ and $F_3$ while 24 components of non-geometric $Q$-flux. Moreover, the choice of involution is such that it has 64 $O3$-planes as well as as three $(O7)_I$-planes corresponding to three ${\mathbb T}^4$'s as there are in total three ${\mathbb Z}_2$ actions including the orientifold involution. Therefore, on top of the NS-NS Bianchi identities (\ref{eq:IsoHQ})-(\ref{eq:IsoQQ}), one has to satisfy the tadpole cancellation conditions given as,
\bea
\label{eq:tadpoles2}
& & \hskip-1.5cm N_3 \equiv 32-N_{D3} = -\, \left(H_\Lambda \, F^\Lambda - H^\Lambda \, F_\Lambda\right), \quad  N_{D_7}^\alpha = \left(F^\Lambda \, \hat{Q}_\Lambda{}^\alpha - F_\Lambda \,\hat{Q}^{\alpha\Lambda}  \right), \, \quad \forall \alpha= 1, 2,3.
\eea 
Let us mention that the total $F$-term contribution to the scalar potential results in 2422 number of terms \cite{Blumenhagen:2013hva,Shukla:2015hpa}, and any analytic attempt for moduli stabilization using the full scalar potential which has 40 flux parameters and 14 scalars sounds quite impractical, and so we consider some simplified flux-solutions of Bianchi identities. 
\subsection{Switching-off half of the NS-NS fluxes: `Special solutions' }
Let us seek for the possibilities with the special solutions first, and assume that all the NS-NS fluxes with upper $h^{2,1}_-$ indices are rotated away, and subsequently imposing (\ref{eq:symplecticRotation}) on the HQ and QQ Bianchi identities in eqns. (\ref{eq:HQFirstFormalism1})-(\ref{eq:QQFirstFormalism3}), which are derived from the first formulation, we get,
\bea
\label{eq:HQhalf}
\begin{array}{c}
  H_3 \, \hat{Q}_2{}^2+H_2 \, \hat{Q}_3{}^2 = 0, \quad H_3 \, \hat{Q}_1{}^3+H_1 \, \hat{Q}_3{}^3 = 0, \quad H_2 \, \hat{Q}_1{}^1+H_1 \, \hat{Q}_2{}^1=0 , \\
  H_3 \, \hat{Q}_2{}^3+H_2 \, \hat{Q}_3{}^3 = 0, \quad H_3 \, \hat{Q}_1{}^1+H_1 \, \hat{Q}_3{}^1 =0 , \quad H_2 \, \hat{Q}_1{}^2+H_1 \, \hat{Q}_2{}^2 = 0\\
\end{array}
\eea
\bea
\label{eq:QQhalf}
\begin{array}{c}
 \hat{Q}_1{}^3 \, \,  \hat{Q}_2{}^2+\, \,  \hat{Q}_1{}^2 \, \,  \hat{Q}_2{}^3= 0 , \quad \hat{Q}_2{}^3 \, \,  \hat{Q}_3{}^1+\, \,  \hat{Q}_2{}^1 \, \,  \hat{Q}_3{}^3=0, \quad \hat{Q}_1{}^2 \, \,  \hat{Q}_3{}^1+\, \,  \hat{Q}_1{}^1 \, \,  \hat{Q}_3{}^2=0 ,\\
  \hat{Q}_1{}^3 \, \,  \hat{Q}_3{}^2+\, \,  \hat{Q}_1{}^2 \, \,  \hat{Q}_3{}^3=0, \quad \hat{Q}_1{}^3 \, \,  \hat{Q}_2{}^1+\, \,  \hat{Q}_1{}^1 \, \,  \hat{Q}_2{}^3=0 , \quad \hat{Q}_2{}^2 \, \,  \hat{Q}_3{}^1+\, \,  \hat{Q}_2{}^1 \, \,  \hat{Q}_3{}^2=0 \, .\\
\end{array}
\eea
This is good that complicated Bianchi identities are reduced to a set of quite simple constraints. Again we emphasize that the second formulation constraints given by identities (\ref{eq:BIs1})-(\ref{eq:BIs2}) do not  produce any of these 12 constraints. Now we have reduced the number of NS-NS flux parameters from 32 to 16, out of which 12 flux parameters have to satisfy 12 (non-independent) constraints in eqns. (\ref{eq:HQhalf})-(\ref{eq:QQhalf}) while the remaining four $(H_0, \hat{Q}_0{}^1,\hat{Q}_0{}^2,\hat{Q}_0{}^3)$ remain unconstrained by NS-NS Bianchi identities. Moreover, these 12 flux constraints result in 38 solutions which do not have good relevance to be listed here. This is because of the fact that even after rotating away half of the NS-NS flux parameters, the number of terms in the scalar potential just reduces from 2422 to 522, and in the absence of any trustworthy hierarchy like LARGE volume scenarios \cite{Balasubramanian:2005zx}, it is still quite difficult to perform any analytic study of moduli stabilization and so we take another step of simplification which is the isotropic limit.

\subsection{Most generic solutions of Bianchi identities in isotropic limit}
A more simplified approach of  `isotropic' case is usually adopted for further simplifications in this setup. This corresponds to considering all the three ${\mathbb T}^2$'s in ${\mathbb T}^6 = {\mathbb T}^2 \times {\mathbb T}^2 \times {\mathbb T}^2$ to be identical which reduces the number of real scalars from 14 to 6 while those of flux parameters from 40 to 14. This isotropic limit induces the following simplifications,
\bea
\label{eq:Isotropycond}
& & \hskip-1.0cm {\rm Moduli}: \quad \quad U_i = U \equiv v + i \, u, \quad T_\alpha = T \equiv \rho - i \, \sigma, \quad \quad  \forall \, i, \alpha \in  \{1,2,3 \} \\
& & \hskip-1.0cm {\rm Fluxes}: \quad \quad  F_0 = f_0, \, F_i = f_1, \, F^0 = f^0, \, F^i = f^1 \, \qquad  H_0 = h_0, \, H_i = h_1, \, H^0 = h^0, \, H^i = h^1\, , \nonumber\\
& & \hskip+1.40cm \hat{Q}_0{}^i = q_0{}^1, \quad \hat{Q}_i{}^i  = q_1{}^1, \quad \hat{Q}_i{}^j  = q_1{}^2 \quad \quad \hat{Q}^{i0}=q^{10}, \quad \hat{Q}^{ii}  = q^{11}, \quad \hat{Q}^{ij}  = q^{21} \nonumber
\eea
where $i,j\in\{1,2,3\}$ and $i\ne j$. In this isotropic limit, Bianchi identities (\ref{eq:HQFirstFormalism1})-(\ref{eq:QQFirstFormalism3}) of the first formulation simplify into the following form,
\bea
\label{eq:IsoHQ}
& &  \hskip-1cm q_0{}^1 \, {h}^1+ h_0 \, q^{21}-h_1 \, q_1{}^1- h_1 \, q_1{}^2 = 0, \quad  q_1{}^2 \, {h}^0 + h_1 \, q^{10}+\, {h}^1 \, q^{11}+\, {h}^1 \, q^{21}=0, \\
& & \hskip-1cm  h_0 \, q^{10}+ h_1 \, q^{11}+ h_1 \, q^{21} - q_1{}^2 \, {h}^1= 0, \quad  q_0{}^1 \, {h}^0+ q_1{}^1 \, {h}^1+q_1{}^2 \, {h}^1- h_1 \,q^{21} = 0\nonumber
\eea
and
\bea
\label{eq:IsoQQ}
& & \hskip2.0cm q_1{}^2 \, q_1{}^1+ q_1{}^2 \, q_1{}^2-  q_0{}^1 \, q^{21}- q_0{}^1 \, q^{11} = 0, \\
& &   \quad q_0{}^1 q^{10}+  q_1{}^2 q^{21} = 0, \quad q_1{}^1 \,q^{10}+\,  q_1{}^2 \, q^{10}+\,q^{21} \, q^{11}+\,q^{21} \, q^{21}=0 \nonumber
\eea
We find the following 14 solutions of these simplified seven identities in eqns. (\ref{eq:IsoHQ})-(\ref{eq:IsoQQ}),
\bea
\label{eq:fluxSolution}
& & \hskip-0.5cm {\bf S1:} \quad \, q_1{}^2=0, \,  q_1{}^1=0, \,  q_0{}^1=0, \,  q^{21}=0, \,  q^{11}=0, \,  q^{10}=0 \\
& & \hskip-0.5cm {\bf S2:} \quad \, q_1{}^1=-q_1{}^2, \,  q^{11}=-q^{21}, \,  q_0{}^1\neq 0, \,  q^{10}=-\frac{q^{21}
   q_1{}^2}{q_0{}^1}, \,  q^{21}\neq 0, \,  h_1=\frac{h^0 q_0{}^1}{q^{21}}, \,  h_0=-\frac{h^1 q_0{}^1}{q^{21}} \nonumber\\
& & \hskip-0.5cm {\bf S3:} \quad \,  q_1{}^1=-q_1{}^2, \,  q^{21}=0, \,  q^{11}=0, \,  q^{10}=0, \, 
   h^1=0, \,  h^0=0, \,  q_0{}^1\neq 0 \nonumber\\
& & \hskip-0.5cm {\bf S4:} \quad \, q_1{}^1=-q_1{}^2, \,  q_0{}^1=0, \,  q^{21}=0, \,  q^{10}\neq 0, \,  h_1=\frac{-h^1 q^{11}-h^0 q_1{}^2}{q^{10}}, \,  h_0=\frac{h^1 q_1{}^2-h_1
   q^{11}}{q^{10}}, \,  q_1{}^2\neq 0 \nonumber\\
& & \hskip-0.5cm {\bf S5:} \quad \, q_1{}^2=0, \,  q_0{}^1=0, \,  q^{21}=0, \,  q^{10}=0, \,  h^1=0, \,  h_1=0, \,  q_1{}^1\neq 0 \nonumber\\
& & \hskip-0.5cm {\bf S6:} \quad \,  q_1{}^2=0, \,  q_0{}^1=0, \, 
   q_1{}^1\neq 0, \,  q^{10}=\frac{-\left(q^{21}\right)^2-q^{11} q^{21}}{q_1{}^1}, \,  q^{21}\neq 0, \,  h_1=\frac{h^1 q_1{}^1}{q^{21}}, \,  h_0=\frac{h_1 q_1{}^1}{q^{21}} \nonumber\\
& & \hskip-0.5cm {\bf S7:} \quad \, q_1{}^2=0, \,  q_1{}^1=0, \,  q_0{}^1=0, \,  q^{21}=0, \,  q^{10}\neq 0, \,  h_1=-\frac{h^1 q^{11}}{q^{10}}, \,  h_0=-\frac{h_1 q^{11}}{q^{10}} \nonumber\\
& & \hskip-0.5cm {\bf S8:} \quad \,  q_1{}^2=0, \,  q_1{}^1=0, \, 
   q_0{}^1=0, \,  q^{21}\neq 0, \,  q^{11}=-q^{21}, \,  h_1=0, \,  h_0=0 \nonumber\\
& & \hskip-0.5cm {\bf S9:} \quad \, q_0{}^1\neq 0, \,  q^{11}=\frac{\left(q_1{}^2\right){}^2+q_1{}^1 q_1{}^2-q^{21} q_0{}^1}{q_0{}^1}, \, 
   q_1{}^1+q_1{}^2\neq 0, \,  q^{10}=\frac{-\left(q^{21}\right)^2-q^{11} q^{21}}{q_1{}^1+q_1{}^2}, \,  q^{21}\neq 0, \, \nonumber\\
& & \hskip2cm h_1=\frac{h^0 q_0{}^1+h^1 q_1{}^1+h^1 q_1{}^2}{q^{21}}, \,  h_0=\frac{-h^1
   q_0{}^1+h_1 q_1{}^1+h_1 q_1{}^2}{q^{21}} \nonumber\\ 
& & \hskip-0.5cm {\bf S10:} \quad \, q^{21}=0, \,  q_0{}^1\neq 0, \,  q^{11}=\frac{q_1{}^2 \left(q_1{}^1+q_1{}^2\right)}{q_0{}^1}, \,  q^{10}=0, \,  q_1{}^2\neq 0, \, 
   h^0=-\frac{h^1 q^{11}}{q_1{}^2}, \,  q_1{}^1+q_1{}^2\neq 0, \, \nonumber\\
& & \hskip2cm h_1=\frac{h^1 q_0{}^1}{q_1{}^1+q_1{}^2} \nonumber\\
& & \hskip-0.5cm {\bf S11:} \quad \, q_1{}^1=-q_1{}^2, \,  q_0{}^1=0, \,  q^{21}=0, \,  q^{10}=0, \, 
   q_1{}^2\neq 0, \,  h^0=-\frac{h^1 q^{11}}{q_1{}^2}, \,  q^{11}\neq 0, \,  h_1=\frac{h^1 q_1{}^2}{q^{11}}  \nonumber\\
& & \hskip-0.5cm {\bf S12:} \quad \, q_1{}^1=-q_1{}^2, \,  q_0{}^1=0, \,  q^{21}=0, \,  q^{11}=0, \, 
   q^{10}=0, \,  h^1=0, \,  h^0=0, \,  q_1{}^2\neq 0 \nonumber\\
& & \hskip-0.5cm {\bf S13:} \quad \, q_1{}^2=0, \,  q^{21}=0, \,  q^{11}=0, \,  q^{10}=0, \,  q_0{}^1\neq 0, \,  h^0=-\frac{h^1 q_1{}^1}{q_0{}^1}, \,  q_1{}^1\neq
   0, \,  h_1=\frac{h^1 q_0{}^1}{q_1{}^1} \nonumber\\
& & \hskip-0.5cm {\bf S14:} \quad \, q_1{}^2=0, \,  q_1{}^1=0, \,  q_0{}^1=0, \,  q^{21}=0, \,  q^{10}=0, \,  q^{11}\neq 0, \,  h^1=0, \,  h_1=0 \,.\nonumber
\eea
Now we will use some of these solutions for illustrating their relevance in studying the moduli stabilization.
\subsection{AdS extremum with `special' solutions of Bianchi identities}
Considering the isotropy condition (\ref{eq:Isotropycond}) along with the symplectic rotation (\ref{eq:symplecticRotation}) of half of the NS-NS fluxes reduces the 2422 terms of total $F$-term scalar potential into just 90 terms! Subsequently, it is quite remarkable that one can even think of performing some analytic investigations. Now we remind that special solutions only utilize the fact that half of the fluxes (with upper $h^{2,1}_-$ indices) can be rotated away, and subsequently the NS-NS Bianchi identity constraints of the second formulation are trivially satisfied in this example. Now, there still remain some `additional' flux constraints coming from the Bianchi identities of the first formulation, and those in eqns. (\ref{eq:IsoHQ})-(\ref{eq:IsoQQ}) are simplified into just two constraints given as under,
\bea
\label{eq:NS-NS_Iso}
& & h_1 \,(q_1{}^1+ \, q_1{}^2) = 0, \quad \quad q_1{}^2 \,(q_1{}^1+ \, q_1{}^2) = 0 \, .
\eea
\subsubsection*{(i). Realizing AdS vacua via simplest case with $H \neq 0$ and $Q\neq 0$ simultaneously}
Let us consider one of the simplest kind of solutions of eqn. (\ref{eq:NS-NS_Iso}), i.e. $h_1 = 0, \, q_1{}^1 = 0, \, q_1{}^2 = 0$. Solving the extremization conditions, we find that $\partial_\rho V =0$ and $\partial_{C_0} V =0$ both are satisfied at $(\ov{C_0} h_0+3 \ov \rho _1 q_0{}^1+f_0+3 {\ov v}^2 f^1+3 f_1 \ov v)= {\ov v}^3 f^0$, and one has the following classes of extrema,
\bea
\label{eq:extremasModelB}
& & \hskip-2.0cm {\bf E1}: \, \ov v = \frac{f^1}{f^0}, \, \, \ov s = \mp \frac{2\, \sqrt{\frac{5}{3}}\, \delta^3}{3\, h_0 \, {f^0}^2}, \, \, \ov u = \mp \frac{\sqrt{\frac{5}{3}}\, \delta}{f^0}, \, \, \ov \sigma = \pm\frac{2\, \sqrt{\frac{5}{3}}\, \delta^3}{3\, (f^0)^2\, q_0^1},\nonumber\\
& & \hskip-2.0cm {\bf E2}: \, \ov v = \frac{f^1}{f^0},\,  \ov s = \pm \frac{2 \sqrt[4]{2} \sqrt{5} \, \delta^3}{3^{3/4} h_0 \left(f^0\right)^2}, \, \ov u = \pm \frac{\sqrt{5}\, \delta}{\sqrt[4]{6} f^0}, \, \ov \sigma = \mp \frac{\sqrt{5} \, \delta^3}{6^{3/4} \left(f^0\right)^2 q_0{}^1}, \nonumber\\
& & \hskip-2.0cm {\bf E3}: \, \ov v = \frac{f^1}{f^0} \pm \frac{\delta}{\sqrt{3} f^0},\,  \ov s = \mp\frac{4 \delta ^3}{3 \sqrt{3} h_0 \left(f^0\right)^2}, \, \ov u =  \mp \frac{2 \delta }{\sqrt{3} f^0}, \, \ov \sigma = \pm \frac{4 \delta ^3}{3 \sqrt{3} \left(f^0\right)^2 q_0{}^1}, \\
& & \hskip-2.0cm {\bf E4}: \, \ov v = \frac{f^1}{f^0} \pm \frac{\delta}{3\, \sqrt{11} f^0},\,  \ov s = \mp \frac{196 \delta ^3}{27 \sqrt{11} h_0 \left(f^0\right)^2}, \ov u= \mp \frac{14 \delta }{3 \sqrt{11} f^0}, \ov \sigma = \pm \frac{196 \delta ^3}{99
   \sqrt{11} \left(f^0\right)^2 q_0{}^1}, \nonumber\\
& & \hskip-1.0cm V_0^{(1)} = \frac{3 \, \ov u^3 \, \left(f^0\right)^2}{25 \, \ov s \, \ov \sigma^3}, \quad V_0^{(2)} = \frac{2 \, \ov u^3 \, \left(f^0\right)^2}{25 \, \ov s \, \ov \sigma^3}, \quad V_0^{(3)} = \frac{ \ov u^3 \, \left(f^0\right)^2}{8 \, \ov s \, \ov \sigma^3}, \quad V_0^{(4)} = \frac{33 \, \ov u^3 \, \left(f^0\right)^2}{392 \, \ov s \, \ov \sigma^3}\,, \nonumber
\eea
where $\delta =\sqrt{ \left(f^1\right)^2+f_1 f^0}$, and we denote the value of potential at the respective extremum point as $V_0^{(i)}$ for the four cases. Note that the two identities in eqn. (\ref{eq:NS-NS_Iso}) can be satisfied via $q_1{}^2 = -q_1{}^1$, while there are many simpler flux-solutions to play with; for example,  $\{h_1 = 0, \, q_1{}^2 = -q_1{}^1\}$, $\{h_1 = 0, \, q_1{}^2 = 0\}$, $\{q_1{}^1 = 0, \, q_1{}^2 = 0\}$ and the simplest one is $\{h_1 = 0, \, q_1{}^1 = 0, \, q_1{}^2 = 0\}$ which we are considering for the moment. In addition, one has to take care of the RR-flux constraints given in eqn. (\ref{eq:tadpoles2}) which reduces into the following form,
\bea
\label{eq:RR_Iso}
& & \hskip-2cm N_3 \equiv 32-N_{D3} = - \left(h_0 \, f^0 + 3 h_1 \, f^1\right) \, , \qquad  N_{D_7} = f^0 \, q_0{}^1 + f^1(q_1{}^1+ 2 \, q_1{}^2) 
\eea
Further, one should note that imposing $\{\ov s>0, \ov u<0, \ov \sigma >0 \}$ clearly means that all the solutions adopt negative value at the extremum (\ref{eq:extremasModelB}), and so the point corresponding to minimum would be $AdS$. Moreover one has to fix the signs of fluxes such that one ensures $N_{D_3} = 32 + h_0\, f^0 \ge 0$ and $N_{D_7} = f^0 \, q_0{}^1 \ge 0$ as seen from RR tadpole cancellation conditions given in eqn. (\ref{eq:RR_Iso}). Further, we note that there are the following flux scalings in the stabilized values of the moduli,
\bea
& & \ov s \sim \frac{\left( \left(f^1\right)^2+f_1 f^0\right)^{3/2}}{h_0\, \left(f^0\right)^2}\, \quad \ov u \sim \frac{\sqrt{ \left(f^1\right)^2+f_1 f^0}}{f^0}, \quad \ov \sigma \sim \frac{\left( \left(f^1\right)^2+f_1 f^0\right)^{3/2}}{q_o{}^1\, \left(f^0\right)^2}\,.
\eea
This shows that in order to trust the effective field theory description via restricting the solutions into weak (string) coupling, large volume and large complex structure limit, one needs to choose the parameters $h_0, f_0$ and $q_0{}^1$ as small as possible while setting larger values of flux parameter $f_1$ and $f^1$.  Here we provide a particular flux sampling for which all the four extremum solutions are manifest,
\bea
& & \hskip-1.50cm h_0 = 32, \quad q_0{}^1 = -1, \quad f^0 = -1, \quad f^1 = 5, \quad f_1 = -2,  \quad N_{D_3} = 0, \quad N_{D_7} = 1\, ,
\eea
where we have taken larger value of $h_0$ to make string coupling non-trivial (and of order $0.1$). Now after numerically solving all the extremization conditions simultaneously, and imposing $\{\ov s>0, \ov u<0 ,\, \ov \sigma >0\}$ to rule out unphysical solutions, we indeed get precisely those four analytic solutions which we discussed earlier, and they are given as under, 
\begin{table}[H]
\centering
\begin{tabular}{|l||l|l|l|l|l|l||}
\hline
 S. No. & \quad $\ov v$  & \quad $\ov u$ & \quad $\ov s$  & \quad $\ov \sigma$ & \quad $V_0$ & $\#$ of flat dires.  \\
\hline
\, \, {\bf E1} & -5 & -6.7082 & 3.77336 &120.748 & -5.45299 $\times 10^{-6}$ & 1 in $(C_0, \rho)$-plane \\
\hline
\, \, {\bf E2} & -5 &-7.4239 &10.2289 & 81.8309 &-5.83981$\times 10^{-6}$ & 1 in $(C_0, \rho)$-plane \\
\hline
\, \, {\bf E3} & -8 & -6 & 3.375 & 108 & -6.35066 $\times 10^{-6}$ & 2 \{in $(C_0, \rho)$-plane and \\
                  & -2& -6 & 3.375 & 108 & -6.35066 $\times 10^{-6}$ &   and $(s, \sigma)$-plane\} \\
\hline
\, \, {\bf E4} & -5.52223 & -7.31126 & 9.59603 & 83.7472 & -5.83717 $\times 10^{-6}$ & 1 in $(C_0, \rho)$-plane\\
                 & -4.47777  & -7.31126 & 9.59603 & 83.7472 & -5.83717 $\times 10^{-6}$ &   \\
\hline
\end{tabular}
\caption{Stabilized values of moduli/axions and the potential at the four extremum points. In addition, one of the RR axions are stabilized through: $(\ov{C_0} h_0+3 \ov \rho _1 q_0{}^1+f_0+3 {\ov v}^2 f^1+3 f_1 \ov v)=v^3 f^0$.}
\label{tab_sampling0}
\end{table}
The table [\ref{tab_sampling0}] shows that one can easily have quite large values of overall Einstein-frame volume (${\cal V}_E \simeq \sigma^{3/2}\sim 10^3$) of the threefold along with large complex structure moduli $|\ov u| \sim 7$ and weak string coupling ($g_s \sim 0.1$) for this flux sampling. Moreover, these values at the minimum are realized with integral values of fluxes, satisfying the total set of NS-NS and RR Bianchi identities. Further, we find that the extremum {\bf E3} corresponding to the lowest value can stabilize only 4 out of 6 moduli/axions. Investigating the Hessian shows that there is a direction in the $(s, \sigma)$-plane which remains flat along with the one in the $(C_0, \rho)$-plane. So all non-axionic directions are not fixed. In addition, we find that Hessian has negative eigenvalue in {\bf E1} and {\bf E4} implying those extrema to be saddle points while the second extremum {\bf E2} is a minimum for the five directions still leaving a flat axionic direction in $(C_0, \rho)$-plane. Nevertheless the same may be useful for building axionic inflationary models by generating subleading terms via non-perturbative effects on the lines of \cite{Blumenhagen:2014gta,Gao:2014uha,Blumenhagen:2014nba}. These observations on the no-scale structure being only partially broken demands that one should look at some less simple flux samplings. 

\subsubsection*{(ii). More complicated cases of the special solutions of Bianchi identities}
Now let us take the generic potential in which the isotropy conditions are imposed on top of using `special solutions' of Bianchi identities,  and investigate how far we can go for analytic studies of moduli stabilization. 
\subsubsection*{Axionic extremization conditions}
Though it is still hard to analytically solve the extremization conditions for the saxions, however we find that axion stabilization conditions can be ``collectively" expressed as,
\bea
\label{eq:mainAxionStab}
& & \hskip-0.5cm \ov\rho=\frac{f_0 \, h_1 - f_1 \, h_0 + \ov v^2 \, h_0 \, f^0+ 2 \, \ov v^3 \, h_1\, f^0 - 2 \, \ov v \, h_0\, f^1 - 3 \, \ov v^2 \, h_1 \, f^1}{h_0 \, q_1{}^1 + 2 \, h_0 \, q_1{}^2 - 3 \, h_1 \, q_0{}^1} \, , \\
& & \hskip-0.5cm \ov C_0 = \frac{1}{h_0\, q_1{}^1 + 2 \, h_0 \,q_1{}^2 - 3 \, h_1 \,q_0{}^1} \biggl(3 \, f_1 \,q_0{}^1+ 6 \, \ov v\, f^1 \,q_0{}^1 - f_0 \,q_1{}^1 + 3 \, \ov v^2 \, f^1 \,q_1{}^1  \nonumber\\
& & \hskip1.5cm - 2 \, f_0 \, q_1{}^2 + 6 \, \ov v^2 \, f^1 \, q_1{}^2 - \ov v^2 \, f^0 (3 \, q_0{}^1 + 2 \, \ov v \, q_1{}^1 + 4 \, \ov v \, q_1{}^2)\biggr) \, ,\nonumber\\
& & \hskip-0.5cm \ov v = \frac{\ov u^4 \, f^0 \, f^1 - \ov s^2 \, h_0 \, h_1-\ov \sigma^2 \, q_0{}^1(q_1{}^1+ 2 q_1{}^2)}{3\, \ov s^2 h_1^2 + \ov u^4 (f^0)^2 + \ov \sigma^2 (q_1{}^1+ 2 q_1{}^2)^2} \, .\nonumber
\eea
Here,  ``collectively" refers to the fact that stabilized values $\ov \rho, \ov C_0$ and $\ov v$ are obtained in mutually coupled manner as first we solve for $\ov \rho$ and $\ov C_0$ by considering $\partial_\rho V = 0 =\partial_{C_0} V$. Then we use the two subsequent constraints to get the simplified versions of extremization conditions for complex structure axionic partner $v$ and the saxions $\sigma, u$ as well as the dilaton $s$.
\subsubsection*{Saxionic extremization conditions}
Now we provide the polynomial constraints solving which one will get $\ov u, \ov s$ and $\ov \sigma$. Using the extremization conditions $\partial_\rho V = 0 = \partial_{C_0} V$, one finds that the simplified version of saxion stabilization conditions are given as,
\bea
\label{eq:mainSaxionStab1}
& & \hskip-1cm \partial_\sigma V = 0 \Longrightarrow \qquad 2 h_1 \ov s \, \ov u^2 \left(-3 \ov u f^1+4 \ov\sigma  q_1{}^1+8 \ov\sigma  q_1{}^2\right)+2 h_0 \ov s \left(3 h_1 \ov s\, \ov v-\ov u^3 f^0\right)+4 \ov\sigma \, \ov u^3 f^0 q_0{}^1\nonumber\\
& & \hskip1cm +4 \ov \sigma \, \ov u^3 f^1 q_1{}^1+\ov u^6 \left(f^0\right)^2 +4 \ov\sigma  q_1{}^2 \left(2 \ov u^3 f^1+\ov\sigma  q_1{}^1 \left(\ov v^2-\ov u^2\right)+\ov\sigma  \, \ov v q_0{}^1\right) +3 \ov u^4\, \ov v^2 \left(f^0\right)^2 \nonumber\\
& & \hskip1cm +3 \ov u^4 \left(f^1\right)^2 +3 h_1^2 \ov s^2 \left(\ov u^2+3 \ov v^2\right)+h_0^2 \ov s^2 +\ov\sigma ^2 \left(q_0{}^1\right){}^2+4\ov \sigma ^2
   \left(q_1{}^2\right){}^2 \left(\ov v^2-\ov u^2\right) \nonumber\\
& & \hskip1.0cm +\ov\sigma ^2 \, \ov v^2 \left(q_1{}^1\right){}^2+2 \ov\sigma ^2 \, \ov v q_0{}^1 q_1{}^1 =6 \ov u^4 \, \ov v f^0 f^1+\ov \sigma ^2 \, \ov u^2 \left(q_1{}^1\right){}^2
\eea
\bea
\label{eq:mainSaxionStab2}
& & \hskip-1.3cm \partial_u V = 0 \Longrightarrow \qquad \ov v \left(2 \ov u^4 f^0 f^1+3 \ov \sigma ^2 q_1{}^1 \left(\ov v q_1{}^1+2 q_0{}^1\right)\right)+4 h_1 \ov \sigma \,  \ov s \, \ov u^2 \left(q_1{}^1+2 q_1{}^2\right) +6 h_0 h_1 \ov s^2 \, \ov v \nonumber\\
& & \hskip2.0cm  +h_1^2 \ov s^2 \left(\ov u^2+9 \ov v^2\right) +h_0^2 \ov s^2 +3 \ov \sigma ^2 \left(q_0{}^1\right){}^2+4 \ov\sigma ^2 q_1{}^2 \left(3 \ov v q_0{}^1-q_1{}^1 \left(\ov u^2-3 \ov v^2\right)\right)\nonumber\\
& &  \hskip1.5cm =\ov u^4 \left(\left(f^0\right)^2 \left(\ov u^2+\ov v^2\right)+\left(f^1\right)^2\right)+\ov\sigma ^2 \, \ov u^2 \left(q_1{}^1\right){}^2 +4 \ov\sigma ^2
   \left(q_1{}^2\right){}^2 \left(\ov u^2-3 \ov v^2\right)
\eea
\bea
\label{eq:mainSaxionStab3}
& & \hskip-1.9cm \partial_s V = 0 \Longrightarrow \qquad 6 \ov \sigma \, \ov u^3 f^0 q_0{}^1+\left(f^0\right)^2
   \left(\ov u^6+3 \ov u^4\,  \ov v^2\right)+3 \biggl[2 \ov \sigma  \, \ov u^3 f^1 \left(q_1{}^1+2 q_1{}^2\right)+\ov u^4 \left(f^1\right)^2\nonumber\\
& & \hskip1cm +\ov \sigma ^2 \left(4 \left(q_1{}^2\right){}^2
   \left(\ov v^2-\ov u^2\right)+4 q_1{}^2 \left(q_1{}^1 \left(\ov v^2-\ov u^2\right)+\ov v q_0{}^1\right)+\left(\ov v q_1{}^1+q_0{}^1\right){}^2\right)\biggr]\nonumber\\
& & \hskip1cm =6 \ov u^4 \, \ov v f^0 f^1+3 h_1^2 \ov s^2 \left(\ov u^2+3 \ov v^2\right)+6 h_0 h_1 \ov s^2 \, \ov v+h_0^2 \ov s^2+3 \ov \sigma^2 \, \ov u^2 \left(q_1{}^1\right){}^2
\eea
Note that these saxionic values at the respective minima are also coupled in the axion extremization conditions (\ref{eq:mainAxionStab}) via $\ov v$, and so will implicitly affect the overall moduli stabilization. As it appears from the complicated form of extremization conditions, it is hard to analytically solve these high degree and coupled polynomial constraints. Actually the main concern comes from the complex structure moduli stabilization as they have cubic couplings in the superpotential while axion-dilaton and the complexfied $T$-moduli are linear in the superpotential. Subsequently, it is easier for the later as it involves only quadratic (though highly coupled) polynomials.  However, for a given simplified flux choice one can numerically solve all the constraints, and investigate for physical vacua. 

Let us mention here that after using the minimum values of $\ov\rho$ and $\ov{C_0}$, and subsequently the relation coming from the extremizing condition $\partial_\sigma V = 0$ in eqn. (\ref{eq:mainSaxionStab1}), the total scalar potential can be reshuffled into the following form,
\bea
& & \hskip-1.0cm V_{\rm 0} = -\frac{f^0 \, q_0{}^1 + f^1(q_1{}^1+ 2 \, q_1{}^2)}{2\, \ov s\, \ov \sigma^2} - \frac{h_1(q_1{}^1+ 2 \, q_1{}^2)}{\ov u \, \ov \sigma^2} \\
& & \hskip2.0cm - \frac{\left(q_0{}^1 + \ov v \, (q_1{}^1+ 2 \, q_1{}^2)\right)^2}{2\, \ov s\, \ov u^3 \, \ov \sigma} + \frac{(q_1{}^1+ 2 \, q_1{}^2)^2}{2\, \ov s\, \ov u \, \ov \sigma}\nonumber
\eea
This small size of scalar potential is quite impressive! Also we stress that we have not actually solved the saxion extremization conditions (\ref{eq:mainSaxionStab1})-(\ref{eq:mainSaxionStab3}) in getting $V_{\rm 0}$, and all we did was to use the relation in eqn. (\ref{eq:mainSaxionStab1}) on top of using  $\partial_i V = 0$ for $i =\rho$ and  $C_0$. Therefore, the values of $\ov\sigma, \ov v, \ov s$ and $\ov u$ appearing in  $V_{\rm 0}$ will have to be supplemented by solving their respective extremization conditions. The reason for writing $V_{\rm 0}$ in the above manner is suitable for exploiting the `additional' flux constraints in (\ref{eq:NS-NS_Iso})-(\ref{eq:RR_Iso}) which subsequently results in,
\bea
\label{eq:VminModelAhalf}
& & \hskip-1.0cm V_{\rm 0} = \biggl[- \frac{\left(q_0{}^1 + \ov v \, (q_1{}^1+ 2 \, q_1{}^2)\right)^2}{2\, \ov s\, \ov u^3 \, \ov \sigma}\biggr] - \frac{h_1 \, q_1{}^2}{\ov u \, \ov \sigma^2}  + \biggl[\frac{(q_1{}^1+ 2 \, q_1{}^2)^2}{2\, \ov s\, \ov u \, \ov \sigma} -\frac{N_{D_7}}{2\, \ov s\, \ov \sigma^2} \biggr] \, .
\eea
Now given that for any physical solution one needs $\{\ov s >0, \, \ov u <0, \, \ov \sigma >0\}$, one finds that the first piece of (\ref{eq:VminModelAhalf}) is positive semidefinite and the last two pieces are negative semidefinite while the second piece can be of any sign, and therefore there may exist some de-Sitter solution when the three terms suitably compete, and more importantly in case when fluxes also manage to satisfy all the minimization conditions. Now one can directly check for the consistency of the following inequalities,
\bea
\label{eq:dSinequality}
& & \ov s >0, \, \ov u <0, \, \ov \sigma >0, \, V_{\rm min} >0, \, \quad {\rm conditions} \, \, (\ref{eq:mainSaxionStab1})-(\ref{eq:mainSaxionStab3}), 
\eea
and if they are not compatible, it would mean that no consistent de-Sitter solutions are possible within the simplifications we have imposed on the scalar potential. For some flux choices, this strategy appears to be numerically faster ; for example,  by considering another solution of Bianchi identities as $\{h_1 \neq 0, \, q_1{}^1 = 0, \, q_1{}^2 = 0\}$, we find that the six extremization conditions  are incompatible with $V_0 > 0$ implying that a consistent de-Sitter vacua cannot be obtained. This we can see just by checking the compatibility of inequalities $\{\ov s >0, \ov u <0 , \ov\sigma>0, V_0 >0 \}$ along with conditions in eqns. (\ref{eq:mainSaxionStab1})-(\ref{eq:mainSaxionStab3}), and without exactly solving them. However this strategy could not produce any conclusion for the $q_1{}^2= - q_1{}^1 \neq0$ case of the special solutions.

In addition, we considered a couple of solutions of Bianchi identities from eqn. (\ref{eq:fluxSolution}) which correspond to the most generic isotropic flux constraints, and hence are more generic to the previous analysis with isotropic special solutions of Bianchi indetities. However we always ended up having either AdS solution or else tachyonic de-Sitter solutions; for example, we find several unstable de-Sitter vacua for the solution of Bianchi identities given in ${\bf S4}$ of  (\ref{eq:fluxSolution}). {\it However, with integral value of flux parameters satisfying the Bianchi identities, we could not find any stable de-Sitter solution.} This observation is consistent with the unusual rarity of finding de-Sitter solutions as investigated through deep numerical analysis in  \cite{Danielsson:2012by,Blaback:2013ht, Blaback:2015zra}.  

Finally, it is worth to mention that we have just investigated some of the flux-solutions among the ones given in (\ref{eq:fluxSolution}) which are obtained from the generic isotropic flux constraints, and it would be interesting to perform a systematic scan using each of the other flux-solutions of Bianchi identities, and see if the de-Sitter solutions could be realized for other solutions. Moreover, considering the inclusion of non-geometric $P$-flux could be an interesting aspect to further extend the investigations for searching de-Sitter vacua in these setups.

\section{Conclusions and discussions}
\label{sec_conclusion}
The main goal of this article has been two-fold. First we have shown that the known versions of the two formulations for representing the NS-NS Bianchi identities in a given non-geometric flux compactification scenario are not equivalent. To illustrate this argument we considered two toroidal examples in the context of type IIB superstring compactification on the orientifolds of  ${\mathbb T}^6/{\mathbb Z}_4$ and ${\mathbb T}^6/{({\mathbb Z}_2 \times {\mathbb Z}_2)}$ orbifolds.  Subsequently, explicit computations have been done for both of the formulations within each of the two setups so that one could compare the resulting flux constraints arising from the two formulations. In particular, for these two examples, first we have translated the first formulation identities completely into cohomology indices which after more careful and tedious reshuffling of pieces have led us to some interesting observation such as,
\begin{itemize}
\item{The set of Bianchi identities of the first formulation already has all the flux constraints arising from the second formulation.}
\item{There are some additional flux constraints in the first formulation which cannot be derived from the known version of the identities of the second formulation.}
\end{itemize}
Both of the formulations can be derived by imposing the nilpotency of the twisted differential operator ${\cal D}$, and a careful observation shows that one reason for this mismatch could be the fact that in second formulation, flux-actions utilized in eqns. (\ref{eq:action1})-(\ref{eq:action2}) are defined only for harmonic forms while in the first formulation the flux action in eqn. (\ref{eq:action0}) is defined for an arbitrary p-form \cite{Ihl:2007ah}. In other words, the first formulation is derived via imposing ${\cal D}^2 A_p = 0$ on arbitrary forms while the second formulation is derived by imposing the nilpotency only on the harmonic-forms. This should be the reason why we recovered all the second formulation identities via some tedious reshuffling of the flux constraints arising from the first formulation.

Given that the toroidal-orientifold examples we studied do not have harmonic one-form (and five-form),  and so after rewriting everything (e.g. scalar potential and the Bianchi identities) in terms of cohomology ingredients, an immediate (though naive,) expectation could be the extension of (the some of) these `additional' flux constraints into beyond toroidal backgrounds such as Calabi Yaus. As we have demonstrated in two examples, there should exist a cohomology-indexed-version of the `additional' identities of the first formulation which (once invoked) could provide a completion of the second formulation for generic complex threefolds. For example, we have managed to write down the following identities in a model independent manner,
\bea 
& & \hskip-1cm 2\, (\hat{d}^{-1})^\alpha{}_\beta\, \left({\mathbb H}_0 \, \hat{\mathbb Q}^\beta{}_0+ {\mathbb H}^0 \, \hat{\mathbb Q}^{\beta0} \right) + \left({k}_{\alpha \beta\gamma}\right)^{-1} \left(\hat\mho_{\beta 1}\, \hat\mho_{\gamma 1} + \hat\mho_\beta^1\, \hat\mho_\gamma^1 \right) = \left(\hat{k}_{\alpha ab}\right)^{-1} \left(\mho_{a0}\, \mho_{b0} + \mho_a^0\, \mho_b^0 \right),  \forall \alpha \nonumber\\
& & \hskip-0.0cm 2\, (f^{-1})\, \left({\mathbb R}_1 \, \hat{\mho}_{\alpha 1}+ {\mathbb R}^1 \, \hat{\mho}_{\alpha}^1 \right) \, - \hat{k}_{\alpha ab} \, {(d^{-1})}^a_{c}\, {(d^{-1})}^b_{d} \left({\mathbb Q}^{c}_{1}\, {\mathbb Q}^{d}_{1} + {\mathbb Q}^{c1}\, {\mathbb Q}^{d1} \right) \,\\
& &  \hskip1.99cm + {k}_{\alpha \beta\gamma}\, (\hat{d}^{-1})^\beta{}_{\beta'}\, (\hat{d}^{-1})^\gamma{}_{\gamma'} \, \left(\hat{{\mathbb Q}}^{\beta'}_{0}\, \hat{{\mathbb Q}}^{\gamma'}_{0} + \hat{{\mathbb Q}}^{\beta'0}\, \hat{{\mathbb Q}}^{\gamma' 0} \right) = 0\, , \quad \quad \quad \forall \alpha \,, \nonumber
\eea
which are true for Model A. We have shown how these are helpful in a decoupling of scalar potential pieces in such a way that pieces involving even and odd (2,1)-indexed fluxes are separated out. It would be also interesting to investigate some more toroidal examples to convert the additional first formulation identities into cohomology ingredients, and subsequently invoke some generic structure to seek the desired additional constraints for the second formulation.

Moreover, on the same lines of arguments, let us point out that in both of the examples we considered, we find another peculiar identification among the identities of the two formulations, 
\bea
\label{eq:additionalBIs}
& & \hskip-1cm H_{k l [\underline i} \, Q^{kl}{}_{\underline j]} = 0 \quad \Longleftrightarrow  \quad H^\Lambda \hat{Q}^\alpha{}_\Lambda=H_\Lambda \hat{Q}^{\alpha \Lambda}, \, \quad {\rm for \, \, Model \, A \,\, and \, \, Model \, B} \nonumber\\
& & \hskip-1cm \omega_{k l}{}^{[\underline{i}} \,R^{k l \underline{j}]} = 0, \quad \Longleftrightarrow \quad \hat{\omega }_{\alpha K} R^K=R_K \hat{\omega }_\alpha{}^K,  \, \quad {\rm for \, \, Model \, A}\,. \nonumber
\eea
The left ones are embedded into the identities (\ref{eq:bianchids1additional}) which unlike the ones in first formulation in eqn. (\ref{eq:bianchids1}), involve contraction of two real indices, and moreover are directly related to the identities of second formulation in eqn. (\ref{eq:BIs1}). For the time being it is not clear if this observation is just accidental for these particular examples, or it is true for generic orientifold compactifications. It will be interesting to investigate on these lines, which if known would help in invoking a compact symplectic version of the `additional' constraints. 

Without being critical, we point out that if these simple observations hold beyond toroidal models, any previous attempts for model building based on simplifying the scalar potential by imposing only the identities of the second formulation, for example \cite{Blumenhagen:2015qda, Blumenhagen:2015kja,Blumenhagen:2015jva, Blumenhagen:2015xpa, Li:2015taa}, could possibly be under-constrained.
 
Later on, we have discussed the possible solutions of Bianchi identities in those two toroidal examples, and have investigated some applications towards moduli stabilization and search of de-Sitter vacua etc. On these lines, we have made the following observations, 
\begin{itemize}
\item{In the first toroidal example, we have found that imposing all the Bianchi identities can very significantly restrict the available flux-parameter space, and may be up to an extent that it is difficult to stabilize {\it all} moduli at the tree level which is supposed  to be among the most attractive features of model building with non-geometric fluxes. So things may not be as supportive in terms of flux tuning as one could have naively thought of to be feasible in the presence of many superpotential flux-couplings with various moduli and axions. }
\item{There are a couple of simple easy-to-get non-trivial solutions which we call `special' solutions of Bianchi identities, in which half of the fluxes could be set to zero. These solutions are motivated by the second formulation in which the set of constraints could be viewed as an orthogonal set of symplectic vectors, and subsequently one can rotate away half of those integral fluxes via appropriate symplectic transformations. This could not have been so much easy-to-guess via taking the route of the first formulation where fluxes mix in a more complicated manner in the identities. Moreover these `special' solutions can reduce the size of scalar potential very significantly, and one can even do some analytic study as we showed in Model A (and in Model B with the isotropic limit). }
\item{Extending the investigations towards non-special solutions of Bianchi identities, we have attempted to search for stable de-Sitter vacua, and though we find some de-Sitter extrema but those are tachyonic, and we have not realized any stable de-Sitter solution on top of satisfying all Bianchi identities.}
\end{itemize}
As a concluding remark, let us mention that we have not intended to provide an exhaustive study on moduli stabilization and/or the search of de-Sitter vacua in this work, and rather we have aimed to show the influence (and clash) of the two formulations of Bianchi identities as a cautionary/guiding remark, which we hope that our analysis would have conveyed.

\section*{Acknowledgments}
I am grateful to Michele Cicoli and Fernando Quevedo for their kind support and encouragements. I am very thankful to Michele Cicoli and Francesco Muia for earlier collaboration, and for sharing some useful conversations with G. Dibitetto and E. Plauschinn to whom I take the opportunity to thank again. I also thank David Andriot, Gustavo A. Duran  and Damian M. Pena for some helpful discussions. I would like to thank Ralph Blumenhagen for many useful discussions and enlightening learning of the subject during earlier collaborations. Finally I  thank the High Energy Physics group at University of Bologna and INFN-Bologna, Bologna (Italy), for their kind hospitality during short visits to the centers where some parts of this work have been done.


\appendix
\numberwithin{equation}{section}
\section{Derivation of Bianchi identities in the second formulation}
\label{sec_BIs1}
Let us consider the twisted differential operator involving all the NS-NS fluxes to be given as
\bea
& & {\cal D} = d + H \wedge.  + \omega \triangleleft . + Q \triangleright. + R \bullet \, \, .\nonumber
\eea
The action of the flux operations $\triangleleft, \triangleright$ and $\bullet$ on a $p$-form changes the same into a $(p+1)$-form, a $(p-1)$-form and a $(p-3)$-form respectively, and can be given as under, 
\bea
&&{\cal D}\, A_p = d \, A_p + H \wedge\, A_p  + \omega \triangleleft \, A_p + Q \triangleright \, A_p + R \bullet \, A_p \, ,
\eea
Subsequently, we find that $({\cal D}^2 \, A_p)$ has seven types of pieces written as $(p+i)$-forms where $i \in \{ 6, 4, 2 ,0, -2 , -4 , -6\}$. Each of these pieces has to vanish individually for ensuring ${\cal D}^2 \, A_p = 0$. In the expansion, the seven pieces are collected as under,
\bea
\label{eq:D2forms}
& \quad {\bf (p+6)}:= & H \wedge (H \wedge A_p) \\
 {\rm (I).} &\quad  {\bf (p+4)}:= & d (H\wedge A_p) + H \wedge (d A_p) + H \wedge (\omega \triangleleft A_p) + \omega \triangleleft(H \wedge A_p) \nonumber\\
 {\rm (II).} &\quad{\bf (p+2)}:= &d^2 \, A_p + d \,(\omega \triangleleft A_p) + \omega \triangleleft  (d\, A_p) \nonumber\\
& & \hskip1cm + H \wedge (Q \triangleright A_p) + Q \triangleright(H \wedge A_p) + \omega \triangleleft(\omega \triangleleft A_p)\nonumber\\
 {\rm (III).} & \quad {\bf (p)}:= & d\, (Q \triangleright A_p)+Q \triangleright(d\, A_p) + Q \triangleright(\omega \triangleleft A_p) +\omega \triangleleft(Q \triangleright A_p) \nonumber\\
& & \hskip1cm + H \wedge (R\bullet A_p) + R \bullet(H \wedge A_p)\nonumber\\
 {\rm (IV).} & \quad {\bf (p-2)}:= & d \, (R \bullet A_p) + R\bullet (d A_p) \nonumber\\
 & & \hskip1cm + \omega \triangleleft (R\bullet A_p) + Q \triangleright (Q \triangleright A_p) + R \bullet (\omega \triangleleft A_p) \nonumber\\
 {\rm (V).} & \quad {\bf (p-4)}:=&  Q \triangleright (R \bullet A_p) + R \bullet (Q \triangleright A_p) \nonumber\\
& \quad {\bf (p-6)}:=& R\bullet (R\bullet A_p) \nonumber
\eea
Now considering various flux-actions on the different  even/odd bases allowed under the total orientifold action, we will derive the resulting constraints following from eqn. (\ref{eq:D2forms}). Given that the internal background is a real six-dimensional manifold,  one can observe that the first and last expressions of eqn. (\ref{eq:D2forms}) can be relevant only for $A_p$ being zero-form ${\bf 1}$ and six-form $\Phi_6$ respectively. However, the same leads to trivial constraints as,
\bea
& & H \wedge (H \wedge {\bf 1}) = 0, \quad R\bullet (R\bullet \Phi_6) = R^K \, (R\bullet a_K) + R_K (R\bullet b^K) = 0\ .
\eea
For simplifying the remaining five type of terms in eqn. (\ref{eq:D2forms}), we will assume that all fluxes are constant parameters \footnote{For non-constant fluxes, the Bianchi identity constraints (\ref{eq:bianchids1}) gets more complicated as can be seen from flux formulation in a DFT analysis \cite{Blumenhagen:2013hva}. However, for our current purpose, we assume the fluxes to be constant parameters throughout.}. Moreover, one observation is very straight that mixing of fluxes in the remaining five constraints are of $H\omega, (\omega^2+HQ), (HR+Q\omega), (Q^2+ \omega R)$ and ($QR$) types. This is quite motivating from the point of view of the first formulation of Bianchi identities as given in eqn. (\ref{eq:bianchids1}).  However, for the second formulation our aim is to compute Bianchi identities with fluxes written in various cohomology bases and not in the real six-dimensional indices. Let us take each constraint one-by-one via considering the flux actions in eqn. (\ref{eq:action1})-(\ref{eq:action2}). 
\subsubsection*{(i). Terms with $(p+4)$-form:}
Using the fact that $H_3$ is a three-form constant flux, we have $d (H\wedge A_p) + H \wedge (d A_p) = 0$, and subsequently we get,
\bea
& & H \wedge (\omega \triangleleft A_p) + \omega \triangleleft(H \wedge A_p) = 0.
\eea
The relevant $A_p$-forms for expecting non-trivial relations correspond to $p= 0, 1$ and $2$. In the absence of non-trivial one-form (and its dual five-form), we find,
\bea
& & H \wedge (\omega \triangleleft A_2) =0, \quad \quad \quad  \omega \triangleleft(H \wedge A_0) = 0
\eea
Considering $A_p$ as the bases of zero-form (${\bf 1}$), the above results in following Bianchi identities,
\bea
& & \hskip-1.1cm 0=\omega \triangleleft(H \wedge {\bf 1}) =H_\Lambda \,  (\omega \triangleleft {\cal B}^\Lambda) +H^\Lambda  \, (\omega \triangleleft {\cal A}_\Lambda) = \left({d}^{-1}\right)_a{}^b \,\tilde{\nu}^a \left(H^\Lambda \, {\omega}_{b \Lambda}- H_\Lambda \, {\omega}_{b}{}^ \Lambda \right)
\eea
If we consider the $p$-form $A_p$ as bases of odd two-forms ($\nu_a$), we get the same constraint,
\bea
H^\Lambda \, {\omega}_{b \Lambda}- H_\Lambda \, {\omega}_{b}{}^ \Lambda = 0.
\eea
\subsubsection*{(ii). Terms with $(p+2)$-form}
\bea
&(II). &d^2 \, A_p + d \,(\omega \triangleleft A_p) + \omega \triangleleft  (d\, A_p) \nonumber\\
& & \hskip1cm + H \wedge (Q \triangleright A_p) + Q \triangleright(H \wedge A_p) + \omega \triangleleft(\omega \triangleleft A_p)\nonumber
\eea
Of course $d^2 = 0$, and again assuming that fluxes are constant parameters, one can get the following flux constraints,
\bea
& & \hskip-1.0cm H^\Lambda \, \hat{Q}_\Lambda{}^\alpha - H_\Lambda \hat{Q}^{\alpha \Lambda} = 0, \quad  \omega_{a}{}^{\Lambda} \omega_{b \Lambda} - \omega_{b}{}^{\Lambda} \omega_{a \Lambda} =0, \quad  \hat{\omega}_{\alpha}{}^{K} \hat{\omega}_{\beta K} - \hat{\omega}_{\beta}{}^{K} \hat{\omega}_{\alpha K} = 0
\eea
Here, the first one comes from $A_p =\{{\bf 1}, {\cal A}_\Lambda, {\cal B}^\Lambda, \tilde{\mu}^\alpha\} $ while the second and third ones arise from using $A_p = \nu_a$ and $A_p = \mu_\alpha$ respectively.

\subsubsection*{(iii). Terms with $(p)$-form}
\bea
& (III). & d\, (Q \triangleright A_p)+Q \triangleright(d\, A_p) + Q \triangleright(\omega \triangleleft A_p) +\omega \triangleleft(Q \triangleright A_p) \nonumber\\
& & \hskip1cm + H \wedge (R\bullet A_p) + R \bullet(H \wedge A_p)\nonumber
\eea
For $A_p = a_K$, this leads to the quadratic flux constraints with mixed even/odd (2,1)-cohomology index given as under,
\bea
& & \hskip1.0cm f^{-1} \, H_\Lambda \, R_K + (d^{-1})_a{}^b \, \omega_{b \Lambda} \, Q^a{}_K + (\hat{d}^{-1})_\alpha{}^\beta \, \hat{Q}^\alpha{}_\Lambda \, \hat{\omega}_{\beta K} = 0 \\
& & \hskip1cm f^{-1} \, H^\Lambda \, R_K + (d^{-1})_a{}^b \, \omega_{b}{}^{ \Lambda} \, Q^a{}_K + (\hat{d}^{-1})_\alpha{}^\beta \, \hat{Q}^{\alpha{}\Lambda} \, \hat{\omega}_{\beta K} = 0\nonumber
\eea
and for $A_p =b^K$, one gets
\bea
& & \hskip1.0cm f^{-1} \, H_\Lambda \, R^K + (d^{-1})_a{}^b \, \omega_{b \Lambda} \, Q^{a{}K} + (\hat{d}^{-1})_\alpha{}^\beta \, \hat{Q}^\alpha{}_\Lambda \, \hat{\omega}_{\beta}{}^{K} = 0 \\
& & \hskip1.0cm f^{-1} \, H^\Lambda \, R^K + (d^{-1})_a{}^b \, \omega_{b}{}^{ \Lambda} \, Q^{a K} + (\hat{d}^{-1})_\alpha{}^\beta \, \hat{Q}^{\alpha{}\Lambda} \, \hat{\omega}_{\beta}{}^{K} = 0\nonumber
\eea
For $A_p = \{{\cal A}_\Lambda, {\cal B}^\Lambda \}$, one gets the same four constraints though in a different set of combinations. Moreover, for $A_p=\{\mu_\alpha, \nu_a, \tilde{\mu}^\alpha, \tilde{\nu}^a\}$, we get
\bea
& & \hskip-1.0cm \omega_{a \Lambda} \hat{Q}^{\alpha \Lambda} - \omega_{a}{}^{\Lambda} \hat{Q}^\alpha{}_{\Lambda} = 0, \quad  Q^{a K} \hat{\omega}_{\alpha K} - Q^{a}{}_{K} \hat{\omega}_{\alpha}^{K} = 0
\eea 
\subsubsection*{(iv). Terms with $(p-2)$-form}
\bea
& (IV). &  d \, (R \bullet A_p) + R\bullet (d A_p) + \omega \triangleleft (R\bullet A_p) + Q \triangleright (Q \triangleright A_p) + R \bullet (\omega \triangleleft A_p) \nonumber
\eea
This results in following Bianchi identities,
\bea
& & \hskip0.0cm R^K \, \hat{\omega}_{\alpha K} - R_K \hat{\omega}_{\alpha}{}^{K} = 0, \quad  \hat{Q}^{\alpha\Lambda} \hat{Q}^\beta{}_{\Lambda} - \hat{Q}^{\beta \Lambda} \hat{Q}^\alpha{}_{\Lambda} = 0 , \quad Q^{a K} \hat{\omega}_{\alpha K} - Q^{a}{}_{K} \hat{\omega}_{\alpha}^{K} = 0 \nonumber
\eea
Here, the first one comes from $A_p =\{\Phi_6, a_K, b^K, {\mu}_\alpha\} $ while the second and third ones arise from using $A_p = \tilde{\mu}^\alpha$ and $A_p = \tilde{\nu}^a$ respectively.
\subsubsection*{(v). Terms with $(p-4)$-form}
\bea
& (V).&  Q \triangleright (R \bullet A_p) + R \bullet (Q \triangleright A_p) \nonumber
\eea
Similar to the case with $(p+4)$-type terms leading to $(H\omega)$-type identities, we have
\bea
R^K \, {Q}^{a K}- R_K \, {Q}_{a}{}^K = 0.
\eea
as seen from considering $A_p =\{\Phi_6, \tilde{\nu}^\alpha\} $.
\subsubsection*{Summary of Bianchi identities in the second formulation:}
Combining everything together, we have the following set of quadratic flux constraints,
\bea
& & \hskip1.0cm H^\Lambda \, \hat{Q}_\Lambda{}^\alpha - H_\Lambda \hat{Q}^{\alpha \Lambda} = 0, \quad \quad \quad H_\Lambda \, \omega_{a}{}^{\Lambda} - H^\Lambda \, \omega_{\Lambda a} = 0, \\
& & \hskip-1cm \hat{Q}^{\alpha\Lambda} \hat{Q}^\beta{}_{k} - \hat{Q}^{\beta \Lambda} \hat{Q}^\alpha{}_{\Lambda} = 0, \quad  \omega_{a}{}^{\Lambda} \omega_{b \Lambda} - \omega_{b}{}^{\Lambda} \omega_{a \Lambda} =0 , \quad \omega_{a \Lambda} \hat{Q}^{\alpha \Lambda} - \omega_{a}{}^{\Lambda} \hat{Q}^\alpha{}_{\Lambda} = 0, \nonumber\\
& & \hskip1.0cm R^K \, \hat{\omega}_{\alpha K} - R_K \hat{\omega}_{\alpha}{}^{K} = 0, \quad \quad \quad R_K \, Q^{a K} - R^K \, Q^{a}{}_{K} = 0, \nonumber\\
& & \hskip-1cm \hat{\omega}_{\alpha}{}^{K} \hat{\omega}_{\beta K} - \hat{\omega}_{\beta}{}^{K} \hat{\omega}_{\alpha K} = 0, \quad  Q^{a K} Q^{b}{}_{K} -Q^{b K} Q^{a}{}_{K} =0 , \quad Q^{a K} \hat{\omega}_{\alpha K} - Q^{a}{}_{K} \hat{\omega}_{\alpha}^{K} = 0, \nonumber
\eea
and
\bea
& & \hskip1.0cm f^{-1} \, H_\Lambda \, R_K + (d^{-1})_a{}^b \, \omega_{b \Lambda} \, Q^a{}_K + (\hat{d}^{-1})_\alpha{}^\beta \, \hat{Q}^\alpha{}_\Lambda \, \hat{\omega}_{\beta K} = 0, \\
& & \hskip1cm f^{-1} \, H^\Lambda \, R_K +(d^{-1})_a{}^b \, \omega_{b}{}^{ \Lambda} \, Q^a{}_K + (\hat{d}^{-1})_\alpha{}^\beta \, \hat{Q}^{\alpha{}\Lambda} \, \hat{\omega}_{\beta K} = 0, \nonumber\\
& & \hskip1.0cm f^{-1} \, H_\Lambda \, R^K + (d^{-1})_a{}^b \, \omega_{b \Lambda} \, Q^{a{}K} + (\hat{d}^{-1})_\alpha{}^\beta \, \hat{Q}^\alpha{}_\Lambda \, \hat{\omega}_{\beta}{}^{K} = 0, \nonumber\\
& & \hskip1.0cm f^{-1} \, H^\Lambda \, R^K + (d^{-1})_a{}^b \, \omega_{b}{}^{ \Lambda} \, Q^{a K} + (\hat{d}^{-1})_\alpha{}^\beta \, \hat{Q}^{\alpha{}\Lambda} \, \hat{\omega}_{\beta}{}^{K} = 0 \,.\nonumber
\eea

\section{On inclusion of non-geometric $P$-flux}
\label{sec_Pflux-inclusion}
In the absence of S-dual $P$-fluxes, we have seen that the two formulations are not completely equivalent, and moreover a ``complete form" of Bianchi identities in the second formulation, i.e. using fluxes with cohomology bases, is not known for a generic compactification background. Also, here we note that while considering the S-dual $P$-fllux, even the Bianchi identities of the first formulation are only known for the orientifolds without odd axions, and also without non-geometric $R$-flux \cite{Aldazabal:2006up,Aldazabal:2008zza, Guarino:2008ik}. The modified Bianchi identities in the first formulation are known to take the following form,
\begin{subequations}
\bea
\label{BIsForm1a}
& & \hskip-2.8cm {\bf (a).} \quad Q^{[ab}{}_p \, Q^{c]p}{}_l = 0, \quad P^{[ab}{}_p \, \, P^{c]p}{}_l = 0, \quad Q^{[ab}{}_p \, P^{c]p}{}_l = 0, \quad P^{[ab}{}_p \, Q^{c]p}{}_l = 0,
\eea
\bea
\label{BIsForm1b}
& & \hskip-5.7cm {\bf (b).} \quad Q^{[ab}{}_p \, \tilde{F}^{c]lp} + \tilde{F}^{p[ab} \, Q^{c]l}{}_p = 0 \hskip1cm \Longleftrightarrow \quad \quad Q\, F_3 = 0
\eea
\bea
\label{BIsForm1c}
& & \hskip-5.5cm {\bf (c).} \quad P^{[ab}{}_p \, \tilde{H}^{c]lp} + \tilde{H}^{p[ab} \, P^{c]l}{}_p = 0 \hskip1cm \Longleftrightarrow \quad \quad P\, H_3 = 0
\eea
\bea
\label{BIsForm1d}
& & \hskip-0.7cm {\bf (d).} \quad \biggl\{ Q^{l[a}{}_p \, \tilde{H}^{bc]p} \, -\, P^{[ab}{}_p \, \tilde{F}^{c]lp} = 0, \quad \quad  P^{l[a}{}_p \, \tilde{F}^{bc]p} \, -\, Q^{[ab}{}_p \, \tilde{H}^{c]lp} = 0 \biggr\} \\
& & \hskip-1.0cm \Longleftrightarrow \quad \biggl\{ (P\, F_3 + Q\, H_3)= 0, \quad Q^{l[a}{}_p \, \tilde{H}^{bc]p} +  Q^{[ab}{}_p \, \tilde{H}^{c]lp} - P^{l[a}{}_p \, \tilde{F}^{bc]p} \, -\, P^{[ab}{}_p \, \tilde{F}^{c]lp} = 0 \biggr\}\,. \nonumber
\eea
Here indices within bracket [] are defined to be anti-symmetrized, and the definitions as $\tilde{H}^{ijk} = \frac{1}{3!} \, \epsilon^{ijklmn} \, {H}_{lmn}$ as well as $\tilde{F}^{ijk} = \frac{1}{3!} \, \epsilon^{ijklmn} \, {F}_{lmn}$ have been used. On top of these flux constraints, one has to satisfy the following additional constraints arising from demanding the antisymmetry of the commutators,
\bea
\label{BIsForm1e}
& & \hskip-2.5cm {\bf (e).} \quad Q^{ab}{}_p \, P^{pc}{}_m - P^{ab}{}_p \, Q^{pc}{}_m = 0, \quad Q^{ab}{}_p \, \tilde{H}^{clp} - P^{ab}{}_p \, \tilde{F}^{clp} - \tilde{H}^{pab}\, Q^{cl}{}_p \, + \, \tilde{F}^{pab}\, P^{cl}{}_p = 0\,.
\eea
\end{subequations}
As we have mentioned earlier the second formulation identities are not known with the inclusion of $P$-flux. Nevertheless, based on modular completion arguments one may have the following identities \cite{Shukla:2016hyy},
\bea
\label{eq:BIs1new}
& & \hskip-1.0cm H^\Lambda \, \hat{Q}_\Lambda{}^\alpha - H_\Lambda \hat{Q}^{\alpha \Lambda} = 0, \quad \quad  F^\Lambda \, \hat{P}_\Lambda{}^\alpha - F_\Lambda \hat{P}^{\alpha \Lambda} =0, \quad \quad  H_\Lambda \, \omega_{a}{}^{\Lambda} - H^\Lambda \, \omega_{a k} = 0, 
\\
& & \hskip-1cm \hat{Q}^{\alpha \Lambda} \hat{Q}^\beta{}_{\Lambda} - \hat{Q}^{\beta \Lambda} \hat{Q}^\alpha{}_{\Lambda} = 0, \, \, \quad \omega_{a}{}^{\Lambda} \omega_{b \Lambda} - \omega_{b}{}^{\Lambda} \omega_{a k} =0, \quad \quad \, \, \hat{P}^{\alpha \Lambda} \hat{P}^\beta{}_{\Lambda} - \hat{P}^{\beta \Lambda} \hat{P}^\alpha{}_{\Lambda} = 0 \nonumber\\
& & \hskip-1cm  \omega_{a \Lambda} \hat{Q}^{\alpha \Lambda} - \omega_{a}{}^{\Lambda} \hat{Q}^\alpha{}_{\Lambda} = 0, \quad \quad \omega_{a \Lambda} \hat{P}^{\alpha \Lambda} - \omega_{a}{}^{\Lambda} \hat{P}^\alpha{}_{\Lambda} = 0, \quad \quad \hat{P}^{\alpha \Lambda} \hat{Q}^{\beta}{}_{\Lambda} - \hat{Q}^{\beta \Lambda} \hat{P}^{\alpha}{}_{\Lambda} = 0\nonumber\\
& & \hskip-1cm Q^{a K} Q^{b}{}_{K} -Q^{b K} Q^{a}{}_{K} =0 , \quad Q^{a K} \hat{\omega}_{\alpha K} - Q^{a}{}_{K} \hat{\omega}_{\alpha}^{K} = 0, \quad \hat{\omega}_{\alpha}{}^{K} \hat{\omega}_{\beta K} - \hat{\omega}_{\beta}{}^{K} \hat{\omega}_{\alpha K} = 0, \nonumber\\
& & \hskip-1cm  P^{a K} P^{b}{}_{K} -P^{b K} P^{a}{}_{K} =0 , \quad P^{a K} \hat{\omega}_{\alpha K} - P^{a}{}_{K} \hat{\omega}_{\alpha}^{K} = 0, \quad P^{a K} Q^{b}{}_{K} -Q^{a K} P^{b}{}_{K} =0,\nonumber
\eea
where we set non-geometric $R$-flux to zero \cite{Shukla:2015rua}. A part of these identities will be verified in one of the current toroidal example where we can still translate the known Bianchi identities of first formulations \cite{Aldazabal:2006up,Aldazabal:2008zza, Guarino:2008ik} into the desired cohomology-indexed form.  

\subsection{Cohomology indexed flux constraints in the isotropic limit for Model B}
The conversion relations for  24 non-geometric $P$-flux components can be written similar to the Q-flux conversion relations in eqn. (\ref{eq:fluxconversionA2}). Now, using these $P$-flux relations along with the ones in eqns.  (\ref{eq:fluxconversionA1})-(\ref{eq:fluxconversionA2}), we have converted all the first formulation Bianchi identities given in eqns. (\ref{BIsForm1a})-(\ref{BIsForm1e}) which are too complicated to deserve listing here. However, it is  worth to mention that the class of Bianchi identities given in eqn. (\ref{BIsForm1d}) produces all the identities of second formulation which are of types: $(H_\Lambda \, \hat{Q}^{\alpha\Lambda} - H^\Lambda \, \hat{Q}^\alpha{}_\Lambda) = 0 $ and $(F_\Lambda \, \hat{P}^{\alpha\Lambda} - F^\Lambda \, \hat{P}^\alpha{}_\Lambda) = 0 $ while the ones of $QQ$ and $PP$ type arise from the respective constraints in the collection (\ref{BIsForm1a}). Subsequently, one observes that these can be trivially satisfied once we choose $H^\Lambda = 0, F^\Lambda =0, \hat{Q}^{\alpha \Lambda} =0$ and $\hat{P}^{\alpha \Lambda} =0$ as an extended version of the `special solutions'. 

Let us mention that the complete scalar potential having 9661 number of terms involves 64 flux parameters and 14 real variables \cite{Shukla:2016hyy}. As we did earlier, a pragmatic step is to take the isotropic limit which reduces the number of flux parameters from 64 to 20 and real moduli/axions from 14 to 6. In this limit, although it is still quite lengthy, for the sake of illustration and completion within the isotropic limit, the various classes of Bianchi identities given in {\bf  (i)}-{\bf (vii)} can be mentioned as under,
\bea
\label{eq:IsoBIs1}
& & \hskip-2.5cm {\bf {(i).}}\, \hskip3cm \left(\hat{Q}^{11}+\hat{Q}^{21}\right) \hat{Q}_0{}^1=\hat{Q}_1{}^2 \left(\hat{Q}_1{}^1+\hat{Q}_1{}^2\right), \\
& & \hat{Q}^{10} \hat{Q}_0{}^1+\hat{Q}^{21}
   \hat{Q}_1{}^2=0,\quad \left(\hat{Q}^{21}\right)^2+\hat{Q}^{11} \hat{Q}^{21}+\hat{Q}^{10} \left(\hat{Q}_1{}^1+\hat{Q}_1{}^2\right)=0 \nonumber
\eea
\bea
\label{eq:IsoBIs2}
& &  \hskip-2.7cm {\bf {(ii).}}\, \hskip3.5cm \left(\hat{P}^{11}+\hat{P}^{21}\right) \hat{P}_0{}^1=\hat{P}_1{}^2 \left(\hat{P}_1{}^1+\hat{P}_1{}^2\right),\\
& &  \hat{P}^{10} \hat{P}_0{}^1+\hat{P}^{21}
   \hat{P}_1{}^2=0, \quad \left(\hat{P}^{21}\right)^2+\hat{P}^{11} \hat{P}^{21}+\hat{P}^{10} \left(\hat{P}_1{}^1+\hat{P}_1{}^2\right)=0 \nonumber
\eea
\bea
\label{eq:IsoBIs3}
& & \hskip-1.4cm {\bf {(iii).}} \qquad  \hat{P}_1{}^2 \hat{Q}^{11}=\hat{P}^{10} \hat{Q}_0{}^1+\left(\hat{P}^{11}+\hat{P}^{21}\right) \hat{Q}_1{}^2, \quad  \hat{P}_0{}^1 \hat{Q}^{10}+\left(\hat{P}_1{}^1+\hat{P}_1{}^2\right)
   \hat{Q}^{21}=\hat{P}^{21} \hat{Q}_1{}^1, \\
& & \hat{P}_0{}^1 \hat{Q}^{11}+\hat{P}^{21} \hat{Q}_0{}^1=\left(\hat{P}_1{}^1+\hat{P}_1{}^2\right) \hat{Q}_1{}^2, \quad \hat{P}^{11} \hat{Q}^{21}+\hat{P}^{21}
   \hat{Q}^{21}+\hat{P}_1{}^2 \hat{Q}^{10}+\hat{P}^{10} \hat{Q}_1{}^1=0 \nonumber
\eea
\bea
\label{eq:IsoBIs4}
& & \hskip-1.4cm {\bf {(iv).}}\, \qquad  \hat{P}_0{}^1 \hat{Q}^{10}+\hat{P}_1{}^2
   \left(\hat{Q}^{11}+\hat{Q}^{21}\right)=\hat{P}^{11} \hat{Q}_1{}^2, \quad \hat{P}_1{}^1 \hat{Q}^{21}=\hat{P}^{10} \hat{Q}_0{}^1+\hat{P}^{21} \left(\hat{Q}_1{}^1+\hat{Q}_1{}^2\right), \\
& & \hat{P}_0{}^1 \hat{Q}^{21}+\hat{P}^{11} \hat{Q}_0{}^1=\hat{P}_1{}^2 \left(\hat{Q}_1{}^1+\hat{Q}_1{}^2\right), \quad  \hat{P}^{21}
   \left(\hat{Q}^{11}+\hat{Q}^{21}\right)+\hat{P}_1{}^1 \hat{Q}^{10}+\hat{P}^{10} \hat{Q}_1{}^2=0 \nonumber
\eea
\bea
\label{eq:IsoBIs5}
& & \hskip-2.0cm {\bf {(v).}}\quad \qquad F_1 \hat{P}^{10}+F^1 \left(\hat{P}^{11}+\hat{P}^{21}\right)+F^0 \hat{P}_1{}^2=H_1 \hat{Q}^{10}+H^1 \left(\hat{Q}^{11}+\hat{Q}^{21}\right)+H^0 \hat{Q}_1{}^2, \\
& & F^1 \hat{P}_1{}^2+H_0
   \hat{Q}^{10}+H_1 \left(\hat{Q}^{11}+\hat{Q}^{21}\right)=F_0 \hat{P}^{10}+F_1 \left(\hat{P}^{11}+\hat{P}^{21}\right)+H^1 \hat{Q}_1{}^2, \nonumber\\
& & F_0 \hat{P}^{10}+F_1
   \left(\hat{P}^{11}+\hat{P}^{21}\right)+H^1 \hat{Q}_1{}^2=F^1 \hat{P}_1{}^2+H_0 \hat{Q}^{10}+H_1 \left(\hat{Q}^{11}+\hat{Q}^{21}\right), \nonumber\\
   & & F_0 \hat{P}^{21}+F^1 \hat{P}_0{}^1+H_1 \left(\hat{Q}_1{}^1+\hat{Q}_1{}^2\right)=F_1 \left(\hat{P}_1{}^1+\hat{P}_1{}^2\right)+H_0 \hat{Q}^{21}+H^1 \hat{Q}_0{}^1, \nonumber\\
& & F_1
   \hat{P}^{21}+H^0 \hat{Q}_0{}^1+H^1 \left(\hat{Q}_1{}^1+\hat{Q}_1{}^2\right)=F^0 \hat{P}_0{}^1+F^1 \left(\hat{P}_1{}^1+\hat{P}_1{}^2\right)+H_1 \hat{Q}^{21}, \nonumber\\
& & F_0 \hat{P}^{10}+F_1 \left(\hat{P}^{11}+2
   \hat{P}^{21}\right)+H^0 \hat{Q}_0{}^1+H^1 \hat{Q}_1{}^1+2 H^1 \hat{Q}_1{}^2\nonumber\\
& & \hskip1.5cm =F^0 \hat{P}_0{}^1+F^1 \hat{P}_1{}^1+2 F^1 \hat{P}_1{}^2+H_0 \hat{Q}^{10}+H_1 \left(\hat{Q}^{11}+2
   \hat{Q}^{21}\right), \nonumber
\eea
\bea
\label{eq:IsoBIs6}
& & \hskip-1.2cm {\bf {(vi).}}\, \hskip3.5cm \hat{P}_0{}^1 \hat{Q}^{21}=\hat{P}^{21}
   \hat{Q}_0{}^1, \quad \quad \quad \hat{P}_1{}^2 \hat{Q}^{10}=\hat{P}^{10}
   \hat{Q}_1{}^2, \\
& & \hat{P}_0{}^1 \hat{Q}^{11}+\hat{P}_1{}^2 \hat{Q}_1{}^1=\hat{P}^{11}
   \hat{Q}_0{}^1+\hat{P}_1{}^1 \hat{Q}_1{}^2 , \quad \quad \hat{P}^{21} \hat{Q}^{11}+\hat{P}_1{}^1 \hat{Q}^{10}=\hat{P}^{11} \hat{Q}^{21}+\hat{P}^{10} \hat{Q}_1{}^1,\nonumber\\
& & \hat{P}_0{}^1
   \hat{Q}^{10}+\hat{P}_1{}^2 \hat{Q}^{11}=\hat{P}^{10} \hat{Q}_0{}^1+\hat{P}^{11} \hat{Q}_1{}^2, \quad \quad \hat{P}_0{}^1
   \hat{Q}^{10}+\hat{P}_1{}^1 \hat{Q}^{21}=\hat{P}^{10} \hat{Q}_0{}^1+\hat{P}^{21} \hat{Q}_1{}^1, \nonumber\\
& & \left(\hat{P}^{11}+\hat{P}^{21}\right)
   \hat{Q}_1{}^2 =\hat{P}_1{}^2 \left(\hat{Q}^{11}+\hat{Q}^{21}\right), \quad \quad \left(\hat{P}_1{}^1+\hat{P}_1{}^2\right) \hat{Q}^{21}=\hat{P}^{21} \left(\hat{Q}_1{}^1+\hat{Q}_1{}^2\right),\nonumber\\
& & \left(\hat{P}^{11}+\hat{P}^{21}\right) \hat{Q}^{10}=\hat{P}^{10} \left(\hat{Q}^{11}+\hat{Q}^{21}\right), \quad \quad \left(\hat{P}_1{}^1+\hat{P}_1{}^2\right)
   \hat{Q}_0{}^1=\hat{P}_0{}^1 \left(\hat{Q}_1{}^1+\hat{Q}_1{}^2\right), \nonumber
\eea
and finally we have,
\bea
\label{eq:IsoBIs7}
& & \hskip-1.5cm {\bf {(vii).}}\, \hskip0.5cm F_0 \hat{P}^{10}+F_1 \hat{P}^{11}+F^0 \hat{P}_0{}^1+F^1 \hat{P}_1{}^1=H_0 \hat{Q}^{10}+H_1 \hat{Q}^{11}+H^0 \hat{Q}_0{}^1+H^1 \hat{Q}_1{}^1, \\
& & F_1 \hat{P}^{10}+F^1
   \left(\hat{P}^{11}+\hat{P}^{21}\right)+F^0 \hat{P}_1{}^2=H_1 \hat{Q}^{10}+H^1 \left(\hat{Q}^{11}+\hat{Q}^{21}\right)+H^0 \hat{Q}_1{}^2, \nonumber\\
& & F_0 \hat{P}^{21}+F^1 \hat{P}_0{}^1+H_1
   \left(\hat{Q}_1{}^1+\hat{Q}_1{}^2\right)=F_1 \left(\hat{P}_1{}^1+\hat{P}_1{}^2\right)+H_0 \hat{Q}^{21}+H^1 \hat{Q}_0{}^1 \nonumber
\eea
Note that, one has to ensure the $3/7$-brane S-duality invariant tadpole cancellations by,
\bea
\label{eq:tadpoles1}
& & \hskip2cm N_3 = -\, \left(H_\Lambda \, F^\Lambda - H^\Lambda \, F_\Lambda\right) \\
& & \hskip-1cm N_{D7}^\alpha = \left(F^\Lambda \, \hat{Q}_\Lambda{}^\alpha - F_\Lambda \,\hat{Q}^{\alpha\Lambda}  \right), \, \quad N_{NS7}^\alpha = \left(F^\Lambda \, \hat{Q}_\Lambda{}^\alpha - F_\Lambda \,\hat{Q}^{\alpha\Lambda}  \right), \nonumber\\
& & \hskip-1cm N_{I7}^\alpha = \left(H^\Lambda \, \hat{Q}_\Lambda{}^\alpha + F^\Lambda \, \hat{P}_\Lambda{}^\alpha  \right) - \left(H_\Lambda \,\hat{Q}^{\alpha\Lambda} + F_\Lambda \,\hat{P}^{\alpha\Lambda}  \right) \quad \quad \quad  \forall \alpha \in \{1, .., h^{1,1}_+ (CY)\}.\nonumber
\eea 
where $N_3$ and $N_{D7}, N_{NS7}, N_{I7}$ are resultant charges of 3 and 7 brane/orientifolds. 
\subsection{`Special solutions' by setting half of the fluxes to zero}
We have attempted to simplifying the most generic isotropic flux constraints given in eqns. (\ref{eq:IsoBIs1})-(\ref{eq:IsoBIs7}) and we found that there are 75 possible solutions which are too lengthy to deserve a listing here. Mowever, let us mention that the total scalar potential still remains too huge to perform any analytic study of moduli stabilization even with the isotropic simplification, and so one can take the isotropic limit of the `special solutions' for which we get,
\bea
\label{eq:IsoHalfBIs}
& & \hskip-2.0cm \hat{Q}_1{}^2 \left(\hat{Q}_1{}^1+\hat{Q}_1{}^2\right)=0, \quad \hat{P}_1{}^2 \left(\hat{P}_1{}^1+\hat{P}_1{}^2\right)=0, \quad \hat{Q}_1{}^2 \left(\hat{P}_1{}^1+\hat{P}_1{}^2\right) \, =0, \nonumber\\
& & \hskip-2.0cm  \hat{P}_1{}^2 \left(\hat{Q}_1{}^1+\hat{Q}_1{}^2\right)=0, \quad \left(\hat{P}_1{}^1+\hat{P}_1{}^2\right) \hat{Q}_0{}^1=\hat{P}_0{}^1 \left(\hat{Q}_1{}^1+\hat{Q}_1{}^2\right), \, \\
& & \hskip-1.0cm  \hat{P}_1{}^2 \hat{Q}_1{}^1 = \hat{Q}_1{}^2 \hat{P}_1{}^1, \quad F_1 \left(\hat{P}_1{}^1+\hat{P}_1{}^2\right)=H_1 \left(\hat{Q}_1{}^1+\hat{Q}_1{}^2\right)\,. \nonumber
\eea
This is indeed a huge simplification that the original Bianchi identities which could occupy several pages are now reduced into these 7 coupled quadratic-flux relations. Moreover, the aforementioned simplified version of the Bianchi identities results in some particular cases given as under,
\bea
& (i). & \hat{Q}_1{}^1= - \hat{Q}_1{}^2, \quad \quad \quad \hat{P}_1{}^1=-\hat{P}_1{}^2 \\
& (ii). & \hat{Q}_1{}^2=0, \quad  \hat{Q}_1{}^1=0, \quad  \hat{P}_1{}^2=0, \quad \hat{P}_1{}^1=0,  \quad \hat{Q}_0{}^1\neq 0 \nonumber\\
& (iii). & \hat{Q}_1{}^2=0, \quad  \hat{Q}_1{}^1=0, \quad  \hat{Q}_0{}^1=0, \quad \hat{P}_1{}^2=0,  \quad \hat{P}_1{}^1=0 \nonumber\\
& (iv). & \hat{Q}_1{}^2=0, \quad \hat{P}_1{}^2=0, \quad  \hat{Q}_1{}^1\neq 0, \quad  \hat{P}_0{}^1=\frac{\hat{P}_1{}^1 \hat{Q}_0{}^1}{\hat{Q}_1{}^1} , \quad
   \hat{P}_1{}^1\neq 0, \quad F_1=\frac{H_1 \hat{Q}_1{}^1}{\hat{P}_1{}^1} \nonumber\\
& (v). & \hat{Q}_1{}^2=0, \quad \hat{Q}_1{}^1=0, \quad \hat{Q}_0{}^1=0, \quad \hat{P}_1{}^2=0, \quad \hat{P}_1{}^1\neq 0, \quad F_1=0 . \nonumber\\
& (vi). & \hat{Q}_1{}^2=0, \quad  \hat{P}_1{}^2=0, \quad  \hat{P}_1{}^1=0, \quad \hat{P}_0{}^1=0,  \quad H_1=0,  \quad \hat{Q}_1{}^1\neq 0 \nonumber
\eea
It would be interesting to use these solutions and perform a systematic study of moduli stabilization and the search for de-Sitter vacua.

\bibliographystyle{utphys}
\bibliography{reference}

\providecommand{\href}[2]{#2}\begingroup\raggedright\begin{thebibliography}{10}

\bibitem{Ihl:2007ah}
M.~Ihl, D.~Robbins, and T.~Wrase, ``{Toroidal orientifolds in IIA with general
  NS-NS fluxes},'' {\em JHEP} {\bf 0708} (2007) 043,
\href{http://www.arXiv.org/abs/0705.3410}{{\tt 0705.3410}}.

\bibitem{Robbins:2007yv}
D.~Robbins and T.~Wrase, ``{D-terms from generalized NS-NS fluxes in type
  II},'' {\em JHEP} {\bf 0712} (2007) 058,
\href{http://www.arXiv.org/abs/0709.2186}{{\tt 0709.2186}}.

\bibitem{Derendinger:2004jn}
J.-P. Derendinger, C.~Kounnas, P.~M. Petropoulos, and F.~Zwirner,
  ``{Superpotentials in IIA compactifications with general fluxes},'' {\em
  Nucl.Phys.} {\bf B715} (2005) 211--233,
\href{http://www.arXiv.org/abs/hep-th/0411276}{{\tt hep-th/0411276}}.

\bibitem{Grana:2012rr}
M.~Gra\~{n}a and D.~Marques, ``{Gauged Double Field Theory},'' {\em JHEP} {\bf
  1204} (2012) 020,
\href{http://www.arXiv.org/abs/1201.2924}{{\tt 1201.2924}}.

\bibitem{Dibitetto:2012rk}
G.~Dibitetto, J.~Fernandez-Melgarejo, D.~Marques, and D.~Roest, ``{Duality
  orbits of non-geometric fluxes},'' {\em Fortsch.Phys.} {\bf 60} (2012)
  1123--1149,
\href{http://www.arXiv.org/abs/1203.6562}{{\tt 1203.6562}}.

\bibitem{Danielsson:2012by}
U.~Danielsson and G.~Dibitetto, ``{On the distribution of stable de Sitter
  vacua},'' {\em JHEP} {\bf 1303} (2013) 018,
\href{http://www.arXiv.org/abs/1212.4984}{{\tt 1212.4984}}.

\bibitem{Blaback:2013ht}
J.~Blåbäck, U.~Danielsson, and G.~Dibitetto, ``{Fully stable dS vacua from
  generalised fluxes},'' {\em JHEP} {\bf 1308} (2013) 054,
\href{http://www.arXiv.org/abs/1301.7073}{{\tt 1301.7073}}.

\bibitem{Damian:2013dq}
C.~Damian, L.~R. Diaz-Barron, O.~Loaiza-Brito, and M.~Sabido, ``{Slow-Roll
  Inflation in Non-geometric Flux Compactification},'' {\em JHEP} {\bf 1306}
  (2013) 109,
\href{http://www.arXiv.org/abs/1302.0529}{{\tt 1302.0529}}.

\bibitem{Damian:2013dwa}
C.~Damian and O.~Loaiza-Brito, ``{More stable de Sitter vacua from S-dual
  nongeometric fluxes},'' {\em Phys.Rev.} {\bf D88} (2013), no.~4, 046008,
\href{http://www.arXiv.org/abs/1304.0792}{{\tt 1304.0792}}.

\bibitem{Hassler:2014mla}
F.~Hassler, D.~Lust, and S.~Massai, ``{On Inflation and de Sitter in
  Non-Geometric String Backgrounds},''
\href{http://www.arXiv.org/abs/1405.2325}{{\tt 1405.2325}}.

\bibitem{deCarlos:2009qm}
B.~de~Carlos, A.~Guarino, and J.~M. Moreno, ``{Complete classification of
  Minkowski vacua in generalised flux models},'' {\em JHEP} {\bf 1002} (2010)
  076,
\href{http://www.arXiv.org/abs/0911.2876}{{\tt 0911.2876}}.

\bibitem{Danielsson:2009ff}
U.~H. Danielsson, S.~S. Haque, G.~Shiu, and T.~Van~Riet, ``{Towards Classical
  de Sitter Solutions in String Theory},'' {\em JHEP} {\bf 09} (2009) 114,
\href{http://www.arXiv.org/abs/0907.2041}{{\tt 0907.2041}}.

\bibitem{Blaback:2015zra}
J.~Blåbäck, U.~H. Danielsson, G.~Dibitetto, and S.~C. Vargas, ``{Universal dS
  vacua in STU-models},'' {\em JHEP} {\bf 10} (2015) 069,
\href{http://www.arXiv.org/abs/1505.04283}{{\tt 1505.04283}}.

\bibitem{Dibitetto:2011qs}
G.~Dibitetto, A.~Guarino, and D.~Roest, ``{Vacua Analysis in Extended
  Supersymmetry Compactifications},'' {\em Fortsch. Phys.} {\bf 60} (2012)
  987--990,
\href{http://www.arXiv.org/abs/1112.1306}{{\tt 1112.1306}}.

\bibitem{Shelton:2005cf}
J.~Shelton, W.~Taylor, and B.~Wecht, ``{Nongeometric flux compactifications},''
  {\em JHEP} {\bf 0510} (2005) 085,
\href{http://www.arXiv.org/abs/hep-th/0508133}{{\tt hep-th/0508133}}.

\bibitem{Aldazabal:2006up}
G.~Aldazabal, P.~G. Camara, A.~Font, and L.~Ibanez, ``{More dual fluxes and
  moduli fixing},'' {\em JHEP} {\bf 0605} (2006) 070,
\href{http://www.arXiv.org/abs/hep-th/0602089}{{\tt hep-th/0602089}}.

\bibitem{Aldazabal:2008zza}
G.~Aldazabal, P.~G. Camara, and J.~Rosabal, ``{Flux algebra, Bianchi identities
  and Freed-Witten anomalies in F-theory compactifications},'' {\em Nucl.Phys.}
  {\bf B814} (2009) 21--52,
\href{http://www.arXiv.org/abs/0811.2900}{{\tt 0811.2900}}.

\bibitem{Font:2008vd}
A.~Font, A.~Guarino, and J.~M. Moreno, ``{Algebras and non-geometric flux
  vacua},'' {\em JHEP} {\bf 0812} (2008) 050,
\href{http://www.arXiv.org/abs/0809.3748}{{\tt 0809.3748}}.

\bibitem{Guarino:2008ik}
A.~Guarino and G.~J. Weatherill, ``{Non-geometric flux vacua, S-duality and
  algebraic geometry},'' {\em JHEP} {\bf 0902} (2009) 042,
\href{http://www.arXiv.org/abs/0811.2190}{{\tt 0811.2190}}.

\bibitem{Hull:2004in}
C.~Hull, ``{A Geometry for non-geometric string backgrounds},'' {\em JHEP} {\bf
  0510} (2005) 065,
\href{http://www.arXiv.org/abs/hep-th/0406102}{{\tt hep-th/0406102}}.

\bibitem{Kumar:1996zx}
A.~Kumar and C.~Vafa, ``{U manifolds},'' {\em Phys.Lett.} {\bf B396} (1997)
  85--90,
\href{http://www.arXiv.org/abs/hep-th/9611007}{{\tt hep-th/9611007}}.

\bibitem{Hull:2003kr}
C.~M. Hull and A.~Catal-Ozer, ``{Compactifications with S duality twists},''
  {\em JHEP} {\bf 0310} (2003) 034,
\href{http://www.arXiv.org/abs/hep-th/0308133}{{\tt hep-th/0308133}}.

\bibitem{Kachru:2003aw}
S.~Kachru, R.~Kallosh, A.~D. Linde, and S.~P. Trivedi, ``{De Sitter vacua in
  string theory},'' {\em Phys. Rev.} {\bf D68} (2003) 046005,
\href{http://www.arXiv.org/abs/hep-th/0301240}{{\tt hep-th/0301240}}.

\bibitem{Balasubramanian:2005zx}
V.~Balasubramanian, P.~Berglund, J.~P. Conlon, and F.~Quevedo, ``{Systematics
  of moduli stabilisation in Calabi-Yau flux compactifications},'' {\em JHEP}
  {\bf 03} (2005) 007,
\href{http://www.arXiv.org/abs/hep-th/0502058}{{\tt hep-th/0502058}}.

\bibitem{Grana:2005jc}
M.~Grana, ``{Flux compactifications in string theory: A Comprehensive
  review},'' {\em Phys. Rept.} {\bf 423} (2006) 91--158,
\href{http://www.arXiv.org/abs/hep-th/0509003}{{\tt hep-th/0509003}}.

\bibitem{Blumenhagen:2006ci}
R.~Blumenhagen, B.~Kors, D.~Lust, and S.~Stieberger, ``{Four-dimensional String
  Compactifications with D-Branes, Orientifolds and Fluxes},'' {\em Phys.Rept.}
  {\bf 445} (2007) 1--193,
\href{http://www.arXiv.org/abs/hep-th/0610327}{{\tt hep-th/0610327}}.

\bibitem{Douglas:2006es}
M.~R. Douglas and S.~Kachru, ``{Flux compactification},'' {\em Rev. Mod. Phys.}
  {\bf 79} (2007) 733--796,
\href{http://www.arXiv.org/abs/hep-th/0610102}{{\tt hep-th/0610102}}.

\bibitem{Denef:2005mm}
F.~Denef, M.~R. Douglas, B.~Florea, A.~Grassi, and S.~Kachru, ``{Fixing all
  moduli in a simple f-theory compactification},'' {\em Adv. Theor. Math.
  Phys.} {\bf 9} (2005) 861--929,
\href{http://www.arXiv.org/abs/hep-th/0503124}{{\tt hep-th/0503124}}.

\bibitem{Blumenhagen:2007sm}
R.~Blumenhagen, S.~Moster, and E.~Plauschinn, ``{Moduli Stabilisation versus
  Chirality for MSSM like Type IIB Orientifolds},'' {\em JHEP} {\bf 01} (2008)
  058,
\href{http://www.arXiv.org/abs/0711.3389}{{\tt 0711.3389}}.

\bibitem{Burgess:2003ic}
C.~P. Burgess, R.~Kallosh, and F.~Quevedo, ``{De Sitter string vacua from
  supersymmetric D terms},'' {\em JHEP} {\bf 10} (2003) 056,
\href{http://www.arXiv.org/abs/hep-th/0309187}{{\tt hep-th/0309187}}.

\bibitem{Westphal:2006tn}
A.~Westphal, ``{de Sitter string vacua from Kahler uplifting},'' {\em JHEP}
  {\bf 03} (2007) 102,
\href{http://www.arXiv.org/abs/hep-th/0611332}{{\tt hep-th/0611332}}.

\bibitem{Burgess:2006cb}
C.~P. Burgess, J.~M. Cline, K.~Dasgupta, and H.~Firouzjahi, ``{Uplifting and
  Inflation with D3 Branes},'' {\em JHEP} {\bf 03} (2007) 027,
\href{http://www.arXiv.org/abs/hep-th/0610320}{{\tt hep-th/0610320}}.

\bibitem{Achucarro:2006zf}
A.~Achucarro, B.~de~Carlos, J.~A. Casas, and L.~Doplicher, ``{De Sitter vacua
  from uplifting D-terms in effective supergravities from realistic strings},''
  {\em JHEP} {\bf 06} (2006) 014,
\href{http://www.arXiv.org/abs/hep-th/0601190}{{\tt hep-th/0601190}}.

\bibitem{Parameswaran:2006jh}
S.~L. Parameswaran and A.~Westphal, ``{de Sitter string vacua from perturbative
  Kahler corrections and consistent D-terms},'' {\em JHEP} {\bf 10} (2006) 079,
\href{http://www.arXiv.org/abs/hep-th/0602253}{{\tt hep-th/0602253}}.

\bibitem{Cremades:2007ig}
D.~Cremades, M.~P. Garcia~del Moral, F.~Quevedo, and K.~Suruliz, ``{Moduli
  stabilisation and de Sitter string vacua from magnetised D7 branes},'' {\em
  JHEP} {\bf 05} (2007) 100,
\href{http://www.arXiv.org/abs/hep-th/0701154}{{\tt hep-th/0701154}}.

\bibitem{Krippendorf:2009zza}
S.~Krippendorf and F.~Quevedo, ``{Metastable SUSY Breaking, de Sitter Moduli
  Stabilisation and Kahler Moduli Inflation},'' {\em JHEP} {\bf 11} (2009) 039,
\href{http://www.arXiv.org/abs/0901.0683}{{\tt 0901.0683}}.

\bibitem{Louis:2012nb}
J.~Louis, M.~Rummel, R.~Valandro, and A.~Westphal, ``{Building an explicit de
  Sitter},'' {\em JHEP} {\bf 10} (2012) 163,
\href{http://www.arXiv.org/abs/1208.3208}{{\tt 1208.3208}}.

\bibitem{Cicoli:2012fh}
M.~Cicoli, A.~Maharana, F.~Quevedo, and C.~P. Burgess, ``{De Sitter String
  Vacua from Dilaton-dependent Non-perturbative Effects},'' {\em JHEP} {\bf 06}
  (2012) 011,
\href{http://www.arXiv.org/abs/1203.1750}{{\tt 1203.1750}}.

\bibitem{Cicoli:2013cha}
M.~Cicoli, D.~Klevers, S.~Krippendorf, C.~Mayrhofer, F.~Quevedo, and
  R.~Valandro, ``{Explicit de Sitter Flux Vacua for Global String Models with
  Chiral Matter},'' {\em JHEP} {\bf 05} (2014) 001,
\href{http://www.arXiv.org/abs/1312.0014}{{\tt 1312.0014}}.

\bibitem{Blaback:2013qza}
J.~Blåbäck, D.~Roest, and I.~Zavala, ``{De Sitter Vacua from Nonperturbative
  Flux Compactifications},'' {\em Phys. Rev.} {\bf D90} (2014), no.~2, 024065,
\href{http://www.arXiv.org/abs/1312.5328}{{\tt 1312.5328}}.

\bibitem{Kallosh:2014wsa}
R.~Kallosh and T.~Wrase, ``{Emergence of Spontaneously Broken Supersymmetry on
  an Anti-D3-Brane in KKLT dS Vacua},'' {\em JHEP} {\bf 12} (2014) 117,
\href{http://www.arXiv.org/abs/1411.1121}{{\tt 1411.1121}}.

\bibitem{Braun:2015pza}
A.~P. Braun, M.~Rummel, Y.~Sumitomo, and R.~Valandro, ``{De Sitter vacua from a
  D-term generated racetrack potential in hypersurface Calabi-Yau
  compactifications},'' {\em JHEP} {\bf 12} (2015) 033,
\href{http://www.arXiv.org/abs/1509.06918}{{\tt 1509.06918}}.

\bibitem{Retolaza:2015nvh}
A.~Retolaza and A.~Uranga, ``{De Sitter Uplift with Dynamical Susy Breaking},''
\href{http://www.arXiv.org/abs/1512.06363}{{\tt 1512.06363}}.

\bibitem{Guarino:2015gos}
A.~Guarino and G.~Inverso, ``{Single-step de Sitter vacua from nonperturbative
  effects with matter},'' {\em Phys. Rev.} {\bf D93} (2016), no.~6, 066013,
\href{http://www.arXiv.org/abs/1511.07841}{{\tt 1511.07841}}.

\bibitem{Bergshoeff:2015jxa}
E.~A. Bergshoeff, K.~Dasgupta, R.~Kallosh, A.~Van~Proeyen, and T.~Wrase, ``{$
  \overline{\mathrm{D}3} $ and dS},'' {\em JHEP} {\bf 05} (2015) 058,
\href{http://www.arXiv.org/abs/1502.07627}{{\tt 1502.07627}}.

\bibitem{Cicoli:2015ylx}
M.~Cicoli, F.~Quevedo, and R.~Valandro, ``{De Sitter from T-branes},'' {\em
  JHEP} {\bf 03} (2016) 141,
\href{http://www.arXiv.org/abs/1512.04558}{{\tt 1512.04558}}.

\bibitem{Garcia-Etxebarria:2015lif}
I.~García-Etxebarria, F.~Quevedo, and R.~Valandro, ``{Global String Embeddings
  for the Nilpotent Goldstino},'' {\em JHEP} {\bf 02} (2016) 148,
\href{http://www.arXiv.org/abs/1512.06926}{{\tt 1512.06926}}.

\bibitem{deCarlos:2009fq}
B.~de~Carlos, A.~Guarino, and J.~M. Moreno, ``{Flux moduli stabilisation,
  Supergravity algebras and no-go theorems},'' {\em JHEP} {\bf 01} (2010) 012,
\href{http://www.arXiv.org/abs/0907.5580}{{\tt 0907.5580}}.

\bibitem{Blumenhagen:2015xpa}
R.~Blumenhagen, C.~Damian, A.~Font, D.~Herschmann, and R.~Sun, ``{The
  Flux-Scaling Scenario: De Sitter Uplift and Axion Inflation},''
\href{http://www.arXiv.org/abs/1510.01522}{{\tt 1510.01522}}.

\bibitem{Blumenhagen:2013hva}
R.~Blumenhagen, X.~Gao, D.~Herschmann, and P.~Shukla, ``{Dimensional Oxidation
  of Non-geometric Fluxes in Type II Orientifolds},'' {\em JHEP} {\bf 1310}
  (2013) 201,
\href{http://www.arXiv.org/abs/1306.2761}{{\tt 1306.2761}}.

\bibitem{Gao:2015nra}
X.~Gao and P.~Shukla, ``{Dimensional oxidation and modular completion of
  non-geometric type IIB action},'' {\em JHEP} {\bf 1505} (2015) 018,
\href{http://www.arXiv.org/abs/1501.07248}{{\tt 1501.07248}}.

\bibitem{Shukla:2015rua}
P.~Shukla, ``{On modular completion of generalized flux orbits},'' {\em JHEP}
  {\bf 11} (2015) 075,
\href{http://www.arXiv.org/abs/1505.00544}{{\tt 1505.00544}}.

\bibitem{Shukla:2015bca}
P.~Shukla, ``{Implementing odd-axions in dimensional oxidation of 4D
  non-geometric type IIB scalar potential},'' {\em Nucl. Phys.} {\bf B902}
  (2016) 458--482,
\href{http://www.arXiv.org/abs/1507.01612}{{\tt 1507.01612}}.

\bibitem{Shukla:2015hpa}
P.~Shukla, ``{A symplectic rearrangement of the four dimensional non-geometric
  scalar potential},'' {\em JHEP} {\bf 11} (2015) 162,
\href{http://www.arXiv.org/abs/1508.01197}{{\tt 1508.01197}}.

\bibitem{Shukla:2016hyy}
P.~Shukla, ``{Reading off the non-geometric scalar potentials via the
  topological data of the compactifying CYs},''
\href{http://www.arXiv.org/abs/1603.01290}{{\tt 1603.01290}}.

\bibitem{Andriot:2013xca}
D.~Andriot and A.~Betz, ``{$\beta$-supergravity: a ten-dimensional theory with
  non-geometric fluxes, and its geometric framework},'' {\em JHEP} {\bf 1312}
  (2013) 083,
\href{http://www.arXiv.org/abs/1306.4381}{{\tt 1306.4381}}.

\bibitem{Andriot:2011uh}
D.~Andriot, M.~Larfors, D.~Lust, and P.~Patalong, ``{A ten-dimensional action
  for non-geometric fluxes},'' {\em JHEP} {\bf 1109} (2011) 134,
\href{http://www.arXiv.org/abs/1106.4015}{{\tt 1106.4015}}.

\bibitem{Blumenhagen:2015lta}
R.~Blumenhagen, A.~Font, and E.~Plauschinn, ``{Relating Double Field Theory to
  the Scalar Potential of N=2 Gauged Supergravity},''
\href{http://www.arXiv.org/abs/1507.08059}{{\tt 1507.08059}}.

\bibitem{Villadoro:2005cu}
G.~Villadoro and F.~Zwirner, ``{N=1 effective potential from dual type-IIA
  D6/O6 orientifolds with general fluxes},'' {\em JHEP} {\bf 0506} (2005) 047,
\href{http://www.arXiv.org/abs/hep-th/0503169}{{\tt hep-th/0503169}}.

\bibitem{Andriot:2012wx}
D.~Andriot, O.~Hohm, M.~Larfors, D.~Lust, and P.~Patalong, ``{A geometric
  action for non-geometric fluxes},'' {\em Phys.Rev.Lett.} {\bf 108} (2012)
  261602,
\href{http://www.arXiv.org/abs/1202.3060}{{\tt 1202.3060}}.

\bibitem{Andriot:2012an}
D.~Andriot, O.~Hohm, M.~Larfors, D.~Lust, and P.~Patalong, ``{Non-Geometric
  Fluxes in Supergravity and Double Field Theory},'' {\em Fortsch.Phys.} {\bf
  60} (2012) 1150--1186,
\href{http://www.arXiv.org/abs/1204.1979}{{\tt 1204.1979}}.

\bibitem{Andriot:2014qla}
D.~Andriot and A.~Betz, ``{Supersymmetry with non-geometric fluxes, or a
  $\beta$-twist in Generalized Geometry and Dirac operator},''
\href{http://www.arXiv.org/abs/1411.6640}{{\tt 1411.6640}}.

\bibitem{Blair:2014zba}
C.~D.~A. Blair and E.~Malek, ``{Geometry and fluxes of SL(5) exceptional field
  theory},''
\href{http://www.arXiv.org/abs/1412.0635}{{\tt 1412.0635}}.

\bibitem{Derendinger:2005ph}
J.-P. Derendinger, C.~Kounnas, P.~Petropoulos, and F.~Zwirner, ``{Fluxes and
  gaugings: N=1 effective superpotentials},'' {\em Fortsch.Phys.} {\bf 53}
  (2005) 926--935,
\href{http://www.arXiv.org/abs/hep-th/0503229}{{\tt hep-th/0503229}}.

\bibitem{Dall'Agata:2009gv}
G.~Dall'Agata, G.~Villadoro, and F.~Zwirner, ``{Type-IIA flux compactifications
  and N=4 gauged supergravities},'' {\em JHEP} {\bf 0908} (2009) 018,
\href{http://www.arXiv.org/abs/0906.0370}{{\tt 0906.0370}}.

\bibitem{Aldazabal:2011yz}
G.~Aldazabal, D.~Marques, C.~Nunez, and J.~A. Rosabal, ``{On Type IIB moduli
  stabilization and N = 4, 8 supergravities},'' {\em Nucl.Phys.} {\bf B849}
  (2011) 80--111,
\href{http://www.arXiv.org/abs/1101.5954}{{\tt 1101.5954}}.

\bibitem{Aldazabal:2011nj}
G.~Aldazabal, W.~Baron, D.~Marques, and C.~Nunez, ``{The effective action of
  Double Field Theory},'' {\em JHEP} {\bf 1111} (2011) 052,
\href{http://www.arXiv.org/abs/1109.0290}{{\tt 1109.0290}}.

\bibitem{Geissbuhler:2011mx}
D.~Geissbuhler, ``{Double Field Theory and N=4 Gauged Supergravity},'' {\em
  JHEP} {\bf 1111} (2011) 116,
\href{http://www.arXiv.org/abs/1109.4280}{{\tt 1109.4280}}.

\bibitem{Blumenhagen:2015qda}
R.~Blumenhagen, A.~Font, M.~Fuchs, D.~Herschmann, and E.~Plauschinn, ``{Towards
  Axionic Starobinsky-like Inflation in String Theory},''
\href{http://www.arXiv.org/abs/1503.01607}{{\tt 1503.01607}}.

\bibitem{Blumenhagen:2015kja}
R.~Blumenhagen, A.~Font, M.~Fuchs, D.~Herschmann, E.~Plauschinn, {\em et al.},
  ``{A Flux-Scaling Scenario for High-Scale Moduli Stabilization in String
  Theory},''
\href{http://www.arXiv.org/abs/1503.07634}{{\tt 1503.07634}}.

\bibitem{Blumenhagen:2015jva}
R.~Blumenhagen, A.~Font, M.~Fuchs, D.~Herschmann, and E.~Plauschinn, ``{Large
  field inflation and string moduli stabilization},'' in {\em {18th
  International Conference From the Planck Scale to the Electroweak Scale
  (Planck 2015) Ioannina, Greece, May 25-29, 2015}}.
\newblock 2015.
\newblock
\href{http://www.arXiv.org/abs/1510.04059}{{\tt 1510.04059}}.
\newblock

\bibitem{Li:2015taa}
T.~Li, Z.~Li, and D.~V. Nanopoulos, ``{Helical Phase Inflation via
  Non-Geometric Flux Compactifications: from Natural to Starobinsky-like
  Inflation},'' {\em JHEP} {\bf 10} (2015) 138,
\href{http://www.arXiv.org/abs/1507.04687}{{\tt 1507.04687}}.

\bibitem{Ceresole:1995ca}
A.~Ceresole, R.~D'Auria, and S.~Ferrara, ``{The Symplectic structure of N=2
  supergravity and its central extension},'' {\em Nucl.Phys.Proc.Suppl.} {\bf
  46} (1996) 67--74,
\href{http://www.arXiv.org/abs/hep-th/9509160}{{\tt hep-th/9509160}}.

\bibitem{D'Auria:2007ay}
R.~D'Auria, S.~Ferrara, and M.~Trigiante, ``{On the supergravity formulation of
  mirror symmetry in generalized Calabi-Yau manifolds},'' {\em Nucl. Phys.}
  {\bf B780} (2007) 28--39,
\href{http://www.arXiv.org/abs/hep-th/0701247}{{\tt hep-th/0701247}}.

\bibitem{Taylor:1999ii}
T.~R. Taylor and C.~Vafa, ``{R R flux on Calabi-Yau and partial supersymmetry
  breaking},'' {\em Phys.Lett.} {\bf B474} (2000) 130--137,
\href{http://www.arXiv.org/abs/hep-th/9912152}{{\tt hep-th/9912152}}.

\bibitem{Grimm:2004uq}
T.~W. Grimm and J.~Louis, ``{The Effective action of N = 1 Calabi-Yau
  orientifolds},'' {\em Nucl.Phys.} {\bf B699} (2004) 387--426,
\href{http://www.arXiv.org/abs/hep-th/0403067}{{\tt hep-th/0403067}}.

\bibitem{Benmachiche:2006df}
I.~Benmachiche and T.~W. Grimm, ``{Generalized N=1 orientifold
  compactifications and the Hitchin functionals},'' {\em Nucl.Phys.} {\bf B748}
  (2006) 200--252,
\href{http://www.arXiv.org/abs/hep-th/0602241}{{\tt hep-th/0602241}}.

\bibitem{Hosono:1994av}
S.~Hosono, A.~Klemm, and S.~Theisen, ``{Lectures on mirror symmetry},''
  \href{http://www.arXiv.org/abs/hep-th/9403096}{{\tt hep-th/9403096}}.
[Lect. Notes Phys.436,235(1994)].

\bibitem{Arends:2014qca}
M.~Arends, A.~Hebecker, K.~Heimpel, S.~C. Kraus, D.~Lust, C.~Mayrhofer,
  C.~Schick, and T.~Weigand, ``{D7-Brane Moduli Space in Axion Monodromy and
  Fluxbrane Inflation},'' {\em Fortsch. Phys.} {\bf 62} (2014) 647--702,
\href{http://www.arXiv.org/abs/1405.0283}{{\tt 1405.0283}}.

\bibitem{Blumenhagen:2012pc}
R.~Blumenhagen, A.~Deser, E.~Plauschinn, and F.~Rennecke, ``{Bianchi Identities
  for Non-Geometric Fluxes - From Quasi-Poisson Structures to Courant
  Algebroids},'' {\em Fortsch. Phys.} {\bf 60} (2012) 1217--1228,
\href{http://www.arXiv.org/abs/1205.1522}{{\tt 1205.1522}}.

\bibitem{Geissbuhler:2013uka}
D.~Geissbuhler, D.~Marques, C.~Nunez, and V.~Penas, ``{Exploring Double Field
  Theory},'' {\em JHEP} {\bf 06} (2013) 101,
\href{http://www.arXiv.org/abs/1304.1472}{{\tt 1304.1472}}.

\bibitem{Andriot:2014uda}
D.~Andriot and A.~Betz, ``{NS-branes, source corrected Bianchi identities, and
  more on backgrounds with non-geometric fluxes},'' {\em JHEP} {\bf 07} (2014)
  059,
\href{http://www.arXiv.org/abs/1402.5972}{{\tt 1402.5972}}.

\bibitem{Shelton:2006fd}
J.~Shelton, W.~Taylor, and B.~Wecht, ``{Generalized Flux Vacua},'' {\em JHEP}
  {\bf 02} (2007) 095,
\href{http://www.arXiv.org/abs/hep-th/0607015}{{\tt hep-th/0607015}}.

\bibitem{Grana:2006hr}
M.~Grana, J.~Louis, and D.~Waldram, ``{SU(3) x SU(3) compactification and
  mirror duals of magnetic fluxes},'' {\em JHEP} {\bf 04} (2007) 101,
\href{http://www.arXiv.org/abs/hep-th/0612237}{{\tt hep-th/0612237}}.

\bibitem{DeWolfe:2005uu}
O.~DeWolfe, A.~Giryavets, S.~Kachru, and W.~Taylor, ``{Type IIA moduli
  stabilization},'' {\em JHEP} {\bf 07} (2005) 066,
\href{http://www.arXiv.org/abs/hep-th/0505160}{{\tt hep-th/0505160}}.

\bibitem{Blumenhagen:2014gta}
R.~Blumenhagen and E.~Plauschinn, ``{Towards Universal Axion Inflation and
  Reheating in String Theory},'' {\em Phys.Lett.} {\bf B736} (2014) 482--487,
\href{http://www.arXiv.org/abs/1404.3542}{{\tt 1404.3542}}.

\bibitem{Gao:2014uha}
X.~Gao, T.~Li, and P.~Shukla, ``{Combining Universal and Odd RR Axions for
  Aligned Natural Inflation},'' {\em JCAP} {\bf 1410} (2014), no.~10, 048,
\href{http://www.arXiv.org/abs/1406.0341}{{\tt 1406.0341}}.

\bibitem{Blumenhagen:2014nba}
R.~Blumenhagen, D.~Herschmann, and E.~Plauschinn, ``{The Challenge of Realizing
  F-term Axion Monodromy Inflation in String Theory},''
\href{http://www.arXiv.org/abs/1409.7075}{{\tt 1409.7075}}.

\end{thebibliography}\endgroup

\end{document}